\documentclass[useAMS,usenatbib]{mnras} 
\usepackage{color}
\usepackage{times,psfig,epsfig} 
\usepackage{rotating, graphicx, adjustbox}
\usepackage{amssymb, amsmath}
\usepackage[T1]{fontenc} 
\usepackage{aecompl}
\usepackage{gensymb}
\usepackage{esint} 

\newcommand{\vb}[1]{\mathbf{#1}}
\newcommand{\dv}[2]{\frac{\mathrm{d}#1}{\mathrm{d}#2}}
\newcommand{\dd}{\mathrm{d}}

\title[On the accretion history of galaxy clusters]{On the accretion history of galaxy clusters: temporal and spatial distribution} 
\author[Vall\'es-P\'erez, Planelles \& Quilis]
 {David Vall\'es-P\'erez$^{1}$, Susana Planelles$^{1}$ and Vicent Quilis$^{1,2}$\\ 
  $^1$Departament d'Astronomia i Astrof\'{\i}sica, Universitat de
  Val\`encia, E-46100 Burjassot (Val\`encia), Spain\\ 
  $^{2}$Observatori Astron\`omic, Universitat de Val\`encia, E-46980
  Paterna (Val\`encia), Spain} 
\date{Released \today}

\pagerange{\pageref{firstpage}--\pageref{lastpage}} \pubyear{2020}

\def\LaTeX{L\kern-.36em\raise.3ex\hbox{a}\kern-.15em
    T\kern-.1667em\lower.7ex\hbox{E}\kern-.125emX}

\begin{document}

\label{firstpage}

\maketitle


\begin{abstract}
We analyse the results of an Eulerian AMR cosmological simulation in order to quantify the mass growth of galaxy clusters, exploring the differences between dark matter and baryons. We have determined the mass assembly histories (MAHs) of each of the mass components and computed several proxies for the instantaneous mass accretion rate (MAR). The mass growth of both components is clearly dominated by the contribution of major mergers, but high MARs can also occur during smooth accretion periods. We explored the correlations between MARs, merger events and clusters' environments, finding the mean densities in $1 \leq r/R_{200m} \leq 1.5$ to correlate strongly with $\Gamma_{200m}$ in massive clusters which undergo major mergers through their MAH. From the study of the dark matter velocity profiles, we find a strong anticorrelation between the MAR proxies $\Gamma_{200m}$ and $\alpha_{200m}$. Last, we present a novel approach to study the angularly-resolved distribution of gas accretion flows in simulations, which allows to extract and interpret the main contributions to the accretion picture and to assess systematic differences between the thermodynamical properties of each of these contributions using multipolar analysis. We have preliminarily applied the method to the best numerically-resolved cluster in our simulation. Amongst the most remarkable results, we find that the gas infalling through the cosmic filaments has systematically lower entropy compared to the isotropic component, but we do not find a clear distinction in temperature.
 \end{abstract}



\section{Introduction}
\label{s:intro}

Galaxy clusters, as the largest and most massive virialised structures in the Universe, are essential pieces, both, for constraining the cosmological parameters and testing the cosmological model and for improving our understanding of structure formation and evolution on galactic scales \citep{Allen_2011}. A precise understanding of the physics of galaxy clusters, with special focus on the baryonic component, is of utmost importance for these purposes (see, for instance, \citealp{Kravtsov_2012} and \citealp{Planelles_2015} for general reviews on galaxy cluster formation).

In particular, the outskirts of galaxy clusters are dynamically active regions, where the infall of baryons and dark matter (DM) feeds the cluster, giving rise to rather complex physical processes of the gaseous component, such as bulk motions, turbulence, clumping, etc. (see \citealp{Walker_2019} for a recent review, and references therein). The mass growth of cluster-sized DM haloes and their baryonic counterparts is usually split into two contributions, namely mergers and smooth accretion. Merger events have already been studied as a source of energetic feedback to the intracluster medium (ICM; e.g., \citealp{Planelles_2009}), introducing deviations with respect to the X-ray \citep{Markevitch_2001, Nagai_2007} and Sunyaev-Zel'dovich (SZ; \citealp{Yu_2015}) scaling relations and, thus, potentially biasing mass estimations (see \citealp{Pratt_2019} for a recent review). As for mass accretion flows, we briefly review the current observational and numerical state of affairs in the following paragraphs.

Gas bulk velocities (along the line of sight) can be directly measured from X-ray line shifts for a small number of nearby clusters (e.g., \citealp{Tamura_2011}, who used Suzaku data to constrain the bulk motions in Abell 2256). Recently, \cite{Sanders_2020} have used XMM-Newton to map the bulk ICM flows in Perseus and Coma clusters. However, these measurements are still restricted to the central regions of clusters, where enough X-ray photons can be collected. The increased sensitivity of ongoing (e.g., eROSITA\footnote{\url{https://www.mpe.mpg.de/eROSITA/}}) and planned (e.g., ATHENA\footnote{\url{https://www.the-athena-x-ray-observatory.eu/}}) facilities will likely extend the X-ray observations to outer regions, thus being able to probe the dynamics of gas in cluster outskirts.

Gas motions can also be inferred from microwave observations through the kinetic SZ (kSZ) effect. Recently, \cite{Adam_2017} obtained the first resolved map of kSZ in a galaxy cluster. Future, high-resolution kSZ observations will likely provide strong constraints on the dynamics in the outskirts of the ICM. We refer the interested reader to \cite{Simionescu_2019} for an extensive review on the possibilities of ICM velocity measurements with X-ray and kSZ.

On the other hand, accretion onto galaxy clusters has been triggering increasing attention in the numerical cosmology community, especially during the last 5 years. For the dark component, \cite{Diemer_2014} found a sharp drop in DM density profiles, corresponding to the so-called ``splashback'' radius, which is generated by recently accreted DM particles in their first apocentric passage, as shown in analytical works by \cite{Adhikari_2014} and \cite{Shi_2016}. Further works by \cite{More_2015}, \cite{Diemer_2017} and \cite{Mansfield_2017} have analysed the relation between the splashback radius and mass accretion rates (MAR), showing that faster accreting DM haloes have generally smaller splashback radii. \cite{Chen_2020} have used statistical techniques to reduce the dimensionality of the mass assembly histories (MAHs) of DM haloes, finding correlations with halo concentrations and other parameters.

Similarly, several works have studied the imprint of mass accretion onto the ICM. In particular, \cite{Lau_2015} have found that the radial profiles of thermodynamical quantities of the ICM (pressure, entropy, temperature, etc.) depend on the MAR. It has been further shown that faster accreting clusters tend to have smaller accretion shock radii \citep{Lau_2015, Shi_2016_shock}, higher ellipticities (\citealp{Chen_2019}; see also \citealp{Lau_2020}, who explore the connection between DM haloes triaxial shapes and several formation history parameters) and more negative residuals with respect to the $T_X-M$ relation \citep{Chen_2019}. Several works have also reported that dynamically disturbed systems display larger hydrostatic mass biases and higher levels of gas clumping in outer cluster regions (see, e.g., \citealp{Biffi_2016} and \citealp{Planelles_2017}, respectively).

It has been seen in $N$-Body simulations that mergers occur primarily through the filament connecting the cluster with its nearest massive neighbour, which is in turn aligned with the major axis of the cluster (e.g., \citealp{Lee_2007, Lee_2008}). However, the accretion pattern of the gaseous component has not yet been extensively covered in the literature. 

In this paper, we examine a small sample of clusters from a hydrodynamical, Eulerian AMR cosmological simulation including cooling and heating, star formation and supernova feedback. The main aim has been studying and characterising the accretion processes on these clusters, paying special attention to the gaseous component. In that sense, we have characterised their MAHs and computed several proxies for the instantaneous MAR (namely, $\Gamma_{200m}$ and $\alpha_{200m}$; see their definitions in Sec. \ref{s:results_mar.mar} and \ref{s:results_mar.velocity_profiles.compare}, respectively), their relation to clusters' environments and merging histories, and the imprint of accretion on the radial density profiles for massive clusters and for low-mass clusters or groups, highlighting the differences between these two classes. Besides, the angularly-resolved distribution of gas accretion flows has been quantified by means of a simple algorithm proposed in this work. Using multipolar analysis, we have been able to extract the main contributions to the accretion picture and to quantify the differences in their thermodynamical quantities.

The manuscript is organised as follows. In Sec. \ref{s:simulation}, we present the numerical details about the simulation and the cluster sample. In Sec. \ref{s:results_mar}, we study and compare several MAR definitions and relate them to the merging history and surrounding densities, while in Sec. \ref{s:results_angular} we present a novel method to evaluate and represent the angular distribution of mass accretion flows. We summarise our main findings and conclusions in Sec. \ref{s:conclusion}. Appendices \ref{s:appendix.symlog} and \ref{s:appendix.realspharm} discuss in more detail some technical issues regarding the symmetric logarithmic scale, used in some representations in this paper, and the real spherical harmonic basis.

\section{The simulation}
\label{s:simulation}
In this section, the details of the simulation we analyse and the cluster sample extracted from it are briefly covered. The simulation has already been employed in previous works (e.g., \citealp{Quilis_2017, Planelles_2018}).

\subsection{Simulation setup}
\label{s:simulation.setup}
The results presented in this paper correspond to the outputs of a cosmological simulation carried out with the Eulerian, adaptive mesh refinement (AMR) code \textsc{MASCLET} \citep{Quilis04}. \textsc{MASCLET} combines a multigrid particle mesh $N$-Body implementation for the description of DM dynamics with \textit{high-resolution shock capturing} techniques for the evolution of the gaseous component.

The background cosmology is set by a spatially flat $\Lambda$-cold dark matter ($\Lambda$CDM) model, assuming a Hubble parameter $h \equiv H_0 / (100 \, \mathrm{km \, s^{-1} \, \mathrm{Mpc^{-1}}}) =  0.678$ and an energetic content given by the density parameters $\Omega_m = 0.31$, $\Omega_\Lambda \equiv \Lambda / 3 H_0^2 = 0.69$, $\Omega_b = 0.048$. A spectral index $n_s = 0.96$ and a normalisation $\sigma_8 = 0.82$ characterise the spectrum of the primordial density fluctuations. These parameters are consistent with the latest results by \cite{Planck_2018}.

The simulation domain corresponds to a cubic box of comoving side length $40 \, \mathrm{Mpc}$, which is discretised in a coarse grid with $128^3$ cells, thus providing a harsh resolution of $\sim 310 \, \mathrm{kpc}$ at the base level ($\ell = 0$). The initial conditions are set up at $z = 100$ using a CDM transfer function \citep{Eisenstein_1998}, with a constrained realisation aimed to produce a massive cluster in the center of the computational domain \citep{Hoffman_1991}. A tentative, low-resolution run from the initial conditions until present time is first performed in order to pick the initially refined regions at the AMR levels $\ell = 1, 2 \, \mathrm{and} \; 3$, that will get their DM mass distribution sampled by particles 8, 64 and 512 times lighter than the ones used in the base level. 

During the evolution of cosmic inhomogeneities, different regions can get refined under a criterion based on the local gaseous and DM densities. The ratio between cells' side lengths at two consecutive AMR levels is set to $\Delta x_{\ell} / \Delta x_{\ell+1} = 2$. Up to $n_{\ell} = 9$ refinement levels are allowed in this run, yielding a best spatial comoving resolution of $\sim 610 \, \mathrm{pc}$. The lightest DM particles have a mass of $\sim 2 \times 10^6 M_\odot$. This provides a peak mass resolution equivalent to having the domain filled with $1024^3$ of such particles.

Our simulation accounts for several cooling processes, such as atomic and molecular cooling for primordial gases, inverse Compton and free-free cooling. Heating by a UV background of radiation is also included, according to the prescriptions of \cite{Haardt_1996}. The abundances are computed under the assumption of the gas being optically thin, in ionization equilibrium, but not in thermal equilibrium \citep{Katz_1996, Theuns_1998}. Tabulated, metallicity-dependent cooling rates from \cite{Sutherland_1993} are employed, with the cooling curves truncated below a temperature threshold of $10^4 \, \mathrm{K}$.

Star formation is parametrised following the ideas of \cite{Yepes_1997} and \cite{Springel_2003}. For more details on the particular implementation in this simulation, we refer the interested reader to \cite{Quilis_2017}. Massive stars produce type-II supernova (SN)  feedback. Feedback mechanisms from type-Ia SN or active galactic nuclei (AGN) are not present in this run. Nevertheless, even though the lack of a central source of energetic feedback (as AGN) could bias the thermal distributions in the inner regions of clusters (see, e.g., \citealp{Planelles_2014, Rasia_2015}), it is not expected to have a noticeable impact on their outskirts, which are the focus of this work. 

\subsection{Structure identification and cluster sample}
\label{s:simulation.structure_identification}

In order to identify galaxy clusters in our simulation, we have made use of the dark matter halo finder \textsc{ASOHF} \citep{Planelles_2010, Knebe_2011}. \textsc{ASOHF} is a spherical overdensity halo finder especially designed to take advantage of the AMR structure of \textsc{MASCLET} outputs. Using \textsc{ASOHF}, we identify a total of 8 DM haloes with virial masses $M_\mathrm{DM} > 10^{13} M_\odot$ by $z = 0$. Two of these haloes have been discarded, as they lie close to the domain boundary and are not faithfully resolved. The remaining six DM haloes and their baryonic counterparts constitute our cluster sample. Two of them have masses above $10^{14} M_\odot$ and can be fully regarded as galaxy clusters. The rest correspond to low-mass clusters or groups. Their main properties at $z = 0$ are summarised in Table \ref{tab:cluster_properties}.

\begin{table*}
	\centering
	\caption{Summary of the main properties of the selected clusters at $z=0$. $x$, $y$ and $z$ refer to the DM center of mass comoving coordinates. The masses $M_\mathrm{DM}$ and $M_\mathrm{gas}$ are measured inside the virial radius, $R_\mathrm{vir}$, defined according to Eqs. (\ref{eq:SO_mass}) and (\ref{eq:bryan_norman_overdensity}). Temperatures and entropies are computed inside $R_\mathrm{vir}$, assuming hydrostatic equilibrium (equations 59 and 64 in \citealp{Voit_2005}) and taking a mean molecular weight $\mu = 0.6$.}
	\begin{tabular}{c|cccccccc}
		\hline
			cluster & $x$ & $y$ & $z$ & $R_\mathrm{vir}$ & $M_\mathrm{DM}$ & $M_\mathrm{gas}$ & $k_B T_\mathrm{vir}$ & $K_{e,\mathrm{vir}}$ \\
			& (Mpc) & (Mpc) & (Mpc) & (Mpc) & ($10^{13} M_\odot$) & ($10^{13} M_\odot$) & (keV) & ($\mathrm{keV \, cm^2}$) \\
			\hline
			CL01 & 0.1 & 0.0 & 0.1 & 1.99 & 42.9 & 4.56 & 3.27 & 1230 \\
			CL02 & -3.2 & 4.9 & -14.9 & 1.26 & 10.9 & 1.33 & 1.34 & 520 \\
			CL03 & 17.0 & -3.0 & 9.4 & 0.96 & 4.8 & 0.66 & 0.77 & 290 \\
			CL04 & 10.7 & -2.5 & 2.0 & 0.95 & 4.6 & 0.44 & 0.74 & 279 \\
			CL06 & -14.6 & -1.0 & -11.0 & 0.71 & 1.9 & 0.22 & 0.43 & 161 \\
			CL08 & -15.0 & -4.6 & 1.9 & 0.61 & 1.2 & 0.12 & 0.41 & 117 \\
			\hline
	\end{tabular}
	\label{tab:cluster_properties}
\end{table*}

In addition to a list of objects for each temporal snapshot, \textsc{ASOHF} also provides, given a halo at some code output, a complete list of its \textit{progenitor haloes} (i.e., the haloes at the previous code output which have DM particles in common with it). We refer to this information as the \textit{full merger tree} of a halo. In order to quantify the accretion phenomena, the \textit{reduced merger tree}, containing only the main progenitor of each halo, needs to be built from the previous information. 

Amongst all the progenitor haloes, the main progenitor has been picked as the one which contributes the most to the descendant halo mass (i.e., the one which gives the most mass). This strategy is also followed in, e.g., \cite{Tormen_2004}. Additionally, we have tested other options described in the literature, like tracing the most bound particles back in time (e.g., \citealp{Planelles_2010}). These alternative definitions yield remarkably similar results, pointing out the robustness of the reconstructed reduced merger trees. 
 
We follow the objects back in time from $z=0$ up to $z=1.5$, for a total of 41 snapshots. No resimulations have been performed.

\section{Accretion rates and their relation to mergers and clusters' environments}
\label{s:results_mar}
Through this section, we cover several topics related to the determination of instantaneous accretion rates (Sec. \ref{s:results_mar.mar}), their relation to clusters' environments and merging histories (Sec. \ref{s:results_mar.mergers_surroundings}) and the impact of accretion on the evolution of the density profiles (Sec. \ref{s:results_mar.density_profiles}). In Sec. \ref{s:results_mar.velocity_profiles} we look at the accretion phenomena from a more dynamically-motivated perspective, by studying the radial velocity profiles.

\subsection{Determination of the mass accretion rates}
\label{s:results_mar.mar}
For each of the DM haloes described in Sec. \ref{s:simulation.structure_identification}, we determine their boundaries and enclosed masses according to the usual spherical overdensity definition \citep{Lacey_1994} with respect to the background matter density:

\begin{equation}
	M_\mathrm{DM}(<R_{\Delta_m}) = \frac{4\pi}{3} R_{\Delta_m}^3 \Delta_m \rho_B
	\label{eq:SO_mass}
\end{equation}

\noindent with $\rho_B(z) = \Omega_m(z) \rho_\mathrm{c}(z)$ and $\rho_\mathrm{c}(z) = \frac{3 H(z)^2}{8\pi G}$. Along this manuscript we focus on $\Delta_m = 200$, as well as on the virial radius, $R_\mathrm{vir}$, defined as the radius enclosing an overdensity \citep{Bryan_1998}:

\begin{equation}
	\Delta_\mathrm{vir, m} = \frac{18 \pi^2 + 82 x - 39 x^2}{\Omega_m}
	\label{eq:bryan_norman_overdensity}
\end{equation}

\noindent being $x \equiv \Omega_m (z) - 1$ and $\Omega_m (z) = \frac{\Omega_m (1+z)^3}{\Omega_m (1+z)^3 + \Omega_\Lambda}$. The computation of the stellar, gaseous and DM masses for each snapshot constitutes the MAH of the different material components of the cluster.

Aiming to quantify the strength of accretion onto the objects in our sample, we define the mass accretion rate as the logarithmic slope of the $M(a)$ curve, where $a$ is the scale factor of the Universe:

\begin{equation}
	\Gamma_\Delta(a) = \frac{\mathrm{d}\log M_\Delta}{\mathrm{d}\log a}
	\label{eq:mar_derivative}
\end{equation}

This definition is more extended in theoretical works (e.g., \citealp{Adhikari_2014}), while most numerical studies opt for replacing the derivative by a quotient difference over a fixed, wide interval $[a_1, a_0]$, in order to avoid the contaminating effects of the intrinsically noisy nature of the MAH of clusters \citep{Diemer_2014, More_2015, Mansfield_2017, Chen_2019}. In these cases, the resulting MAR oughts to be interpreted as an average over several Gyr in the accretion history of the cluster. We shall denote this definition of the MAR as $\Gamma_\Delta^{[a_1, a_0]}$.

Even though the average MAR accounts for the global impact of accretion on the dynamical state of clusters (as shown in \citealp{Chen_2019}), it is not as suitable for analysing a number of aspects, such as the relation of mass infall and clusters' surroundings or punctual events like mergers, as the accretion rates get overly smoothed. In order to measure the \textit{instantaneous} accretion rate through a cluster's history while getting rid of statistical noise, we implement the computation of the derivatives using Savitzky-Golay filters \citep{Savitzky_1964, Press_1990}, whose parameters have been tuned to offer a compromise between smoothness and locality of the computed derivatives.

The definition of MAR in Eq. (\ref{eq:mar_derivative}) can be applied independently to each material component (i.e., DM and baryons\footnote{Baryons account for both, gas and stars. As the total gas mass is not conserved due to star formation, the MAR of gas would be biased low. This effect may be almost negligible in massive clusters, but definitely noticeable in low-mass clusters and groups, where stellar fractions tend to be higher (e.g., \citealp{Planelles_2013}).}). It is also worth mentioning, however, that this is not the only proxy for the MAR of a cluster (see Sec. \ref{s:results_mar.velocity_profiles} for the comparison with an alternative definition). 

\subsubsection{Baryonic and DM MARs}

\begin{figure*}
	\centering
	{\includegraphics[width=0.5\textwidth]{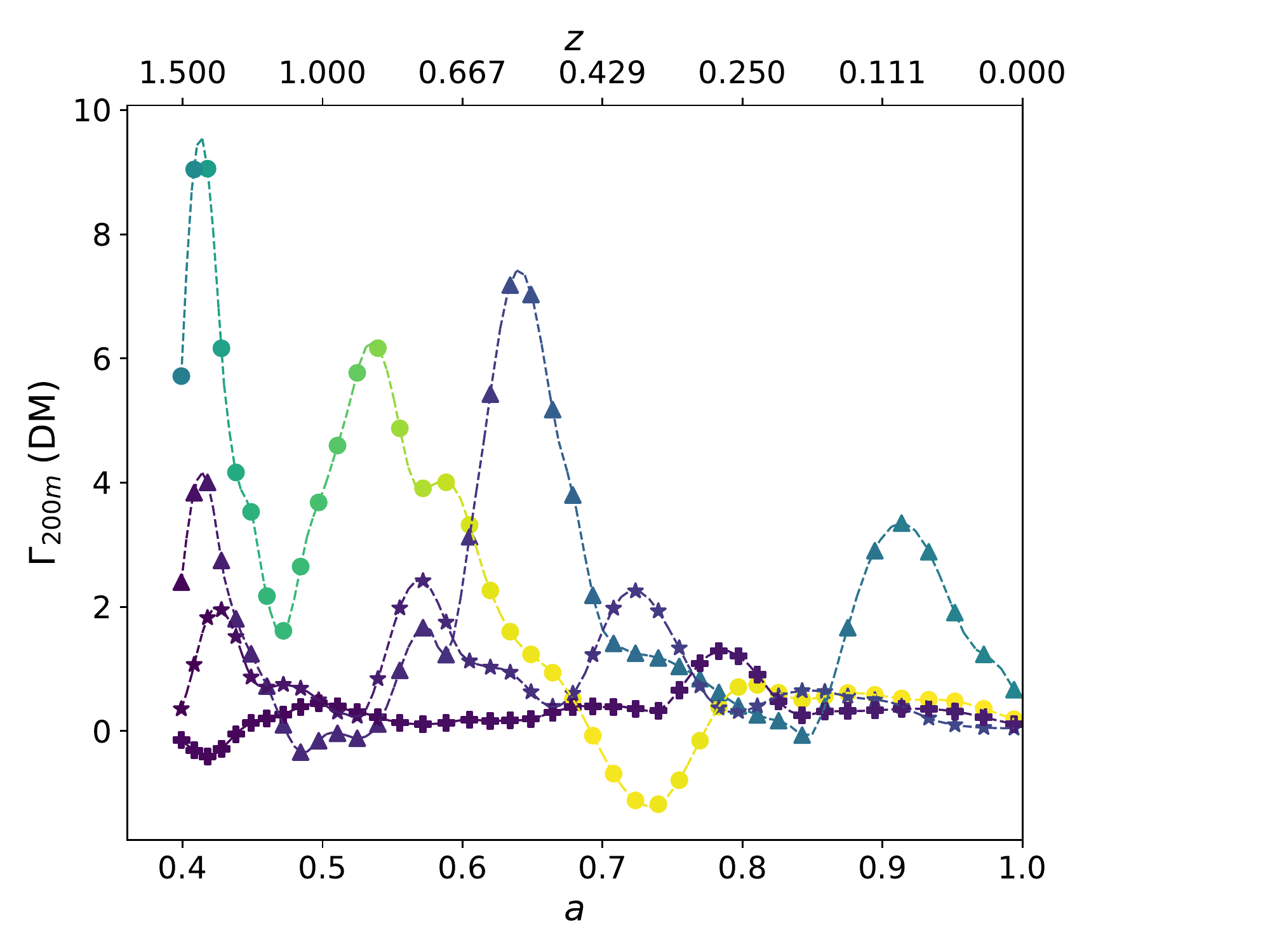}}~ \hspace{-0.1\textwidth} ~
	{\includegraphics[width=0.5\textwidth]{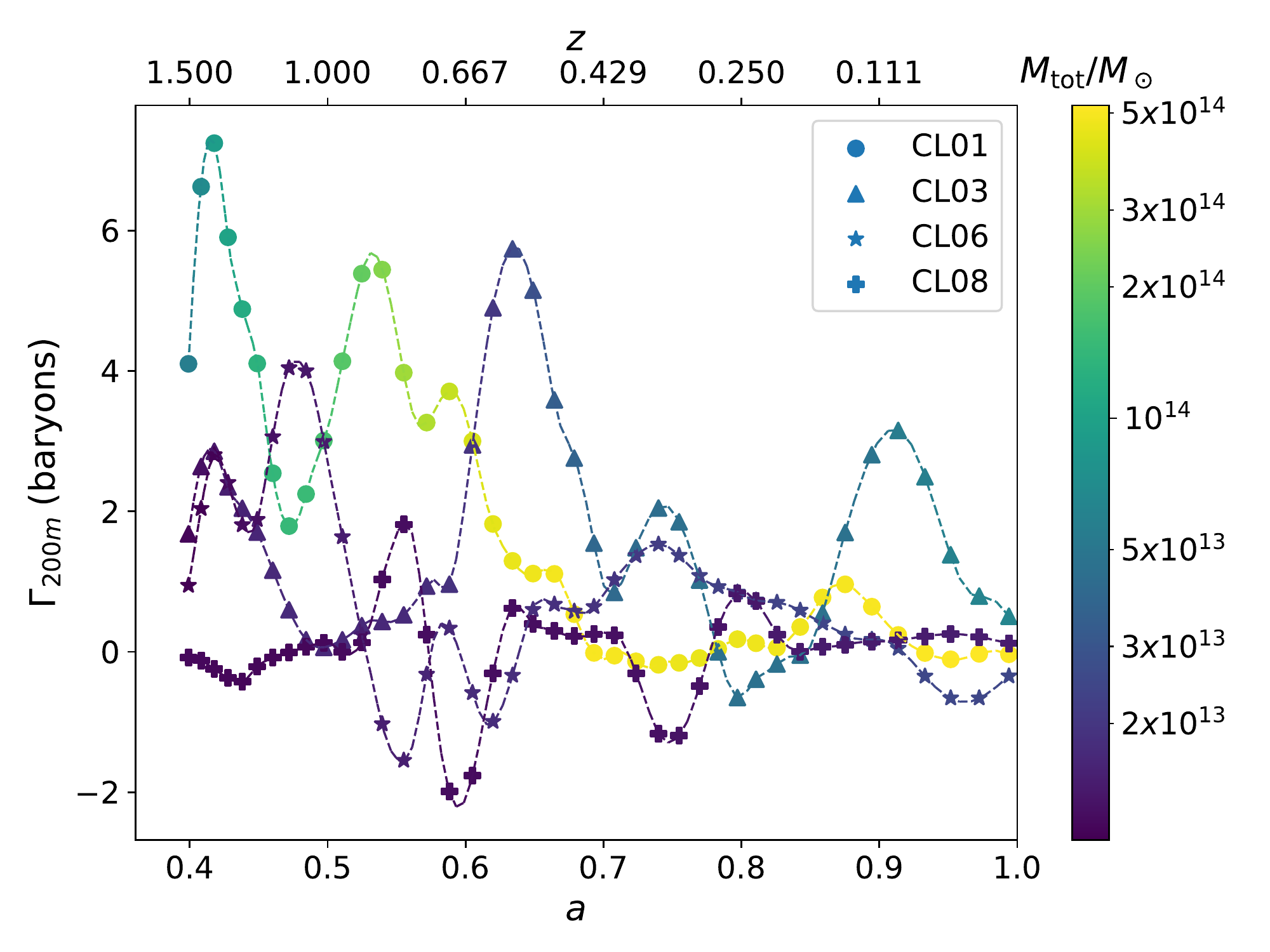}}
	\caption{MARs of four of the clusters in the sample, for DM (left panel) and for baryons (right panel). The MARs have been computed using a fourth order Savitzky-Golay filter with window length of 17 points (the MAR curves have been previously resampled with 100 points by linear interpolation, in order to have uniform spacing in the independent variable, $\log a$). Colours encode the total mass (baryons + DM). The shape of the dots refers to each of the clusters according to the legend. The colourbar and the legend in the right panel apply to both plots.}
	\label{fig:mar_all}
\end{figure*}

We have computed the baryonic and DM MARs of the clusters in our sample from $z=1.5$ to $z=0$, with respect to the masses measured within $R_{200m}$. A selection of them is presented in Figure \ref{fig:mar_all}, where the colour scale keeps track of the total mass. In these plots, the derivatives have been computed using fourth-order Savitzky-Golay filters with window length of 17 points.

The graphs show similar qualitative behaviour for the DM and baryonic MARs, reflecting the fact that gas traces DM to a first approximation. However, the most prominent peaks are typically more pronounced for the dark component than for their baryonic counterparts, implying that gas is generally accreting at a slower pace when compared to DM. This trend has already been pointed out by other studies (see, e.g., \citealp{Lau_2015}, where a similar conclusion is drawn from studying the radial velocity profiles of both components). As opposed to collisionless DM, collisional gas is supported by pressure and experiences ram pressure from the ICM \citep{Tormen_2004, Cen_2014, Quilis_2017}, shocks, etc. which contribute to slow down the infall.

A clear distinction is displayed between massive (CL01 and CL03) and low-mass (CL06 and CL08) objects. Massive clusters often present pronounced peaks in their MAR curves, typically associated to major mergers, which are still frequent as clusters continue growing and collapsing by $z \sim 0$. We analyse in further depth the relation between mergers and accretion rates in Sec. \ref{s:results_mar.mergers_surroundings}. Less massive clusters show flatter curves, pointing out that either they do not undergo merger events as significant as their massive homologues, or they also experience important mass losses during these events, as a consequence of their shallower potential wells. The latter idea is supported by the fact that the differences between baryonic and dark components are more remarkable in these systems. Clusters and groups with total mass $\lesssim 5 \times 10^{13} M_\odot$ do not seem to dominate as efficiently their neighbourhoods, and are therefore harassed by other systems.

\subsection{Effects of mergers and surrounding densities on the MARs}
\label{s:results_mar.mergers_surroundings}

\subsubsection{Identification and classification of mergers}

We define mergers as events where two cluster-sized haloes (and their respective baryonic counterparts) encounter and share a significant amount of mass (e.g., \citealp{Planelles_2009}). The merger tree of a massive halo can contain many progenitor haloes, most of which either are low-mass infalling substructures or contribute very little to the mass of the descendant halo. In order to identify halo mergers, we have imposed the following conditions:

\begin{enumerate}
	\item The distance between the centers of mass of the two progenitor candidates, $i$ and $j$, is less than the sum of their virial radii, i.e., their spheres of radius $R_\mathrm{vir}$ intersect:
	\begin{equation}
		d_{ij} \leq R_\mathrm{vir,i} + R_\mathrm{vir,j}
		\label{eq:mergers_conditiondistance}
	\end{equation}
	\item Each of the progenitor haloes gives, at least, 1$\%$ of its (DM) mass to the descendant halo.
	\item Each of the progenitor halo masses is greater than $1/10$ of the descendant mass. Mergers with haloes of smaller mass are regarded as smooth accretion.
\end{enumerate}

These conditions are conceptually similar to those of \cite{Chen_2019}, who nevertheless use more stringent values ($R_{500c}$ instead of $R_\mathrm{vir}$ and $10\%$ of shared mass between progenitor and descendant halo) in order to assess the merging times. We use the the presence/absence of mergers and the maximum mass ratio between the progenitors to distinguish three accretion regimes in the assembly history of a cluster, according to the following classification:

\begin{itemize}
	\item Major mergers: involve two haloes of comparable mass and are relatively unfrequent. They typically have an important impact on the structure of haloes. We take a mass ratio of $1:3$ as the threshold for these events \citep{Planelles_2009, Chen_2019}.
	\item Minor mergers: produce less significant disturbance on the objects, but are generally more frequent. Their mass lower threshold is slightly more arbitrary and varies through the literature. We use a ratio of $1:10$ as the threshold for minor mergers, as in \cite{Planelles_2009}.
	\item Smooth accretion: systems which experience no mergers above the $1:10$ mass ratio threshold are considered to undergo smooth accretion.
\end{itemize}

In our sample, clusters CL01, CL02 and CL03 exhibit periods of major and minor merging activity. CL06 does not experience any major mergers, but only minor ones. Last, no mergers have been identified in CL04 and CL08 and they are therefore smoothly accreting clusters throughout the considered redshift interval, $1.5 \geq z \geq 0$.

\subsubsection{Surrounding densities}
A significant part of the accreted mass in major mergers can end up lying outside the $R_\Delta$ boundary of the final halo and, hence, the corresponding spherical overdensity masses are not additive in such events \citep{Kravtsov_2012, More_2015}. In order to study how matter is deposited in the outskirts of galaxy clusters, and how this effect shapes the MAR curves, we quantify the densities in the surroundings of each cluster and its evolution with cosmic time in four non-overlapping, equally spaced radial bins, covering the region $1 \leq R/R_{200m} \leq 3$.

\begin{figure*}
	\centering
	{\includegraphics[width=0.5 \textwidth]{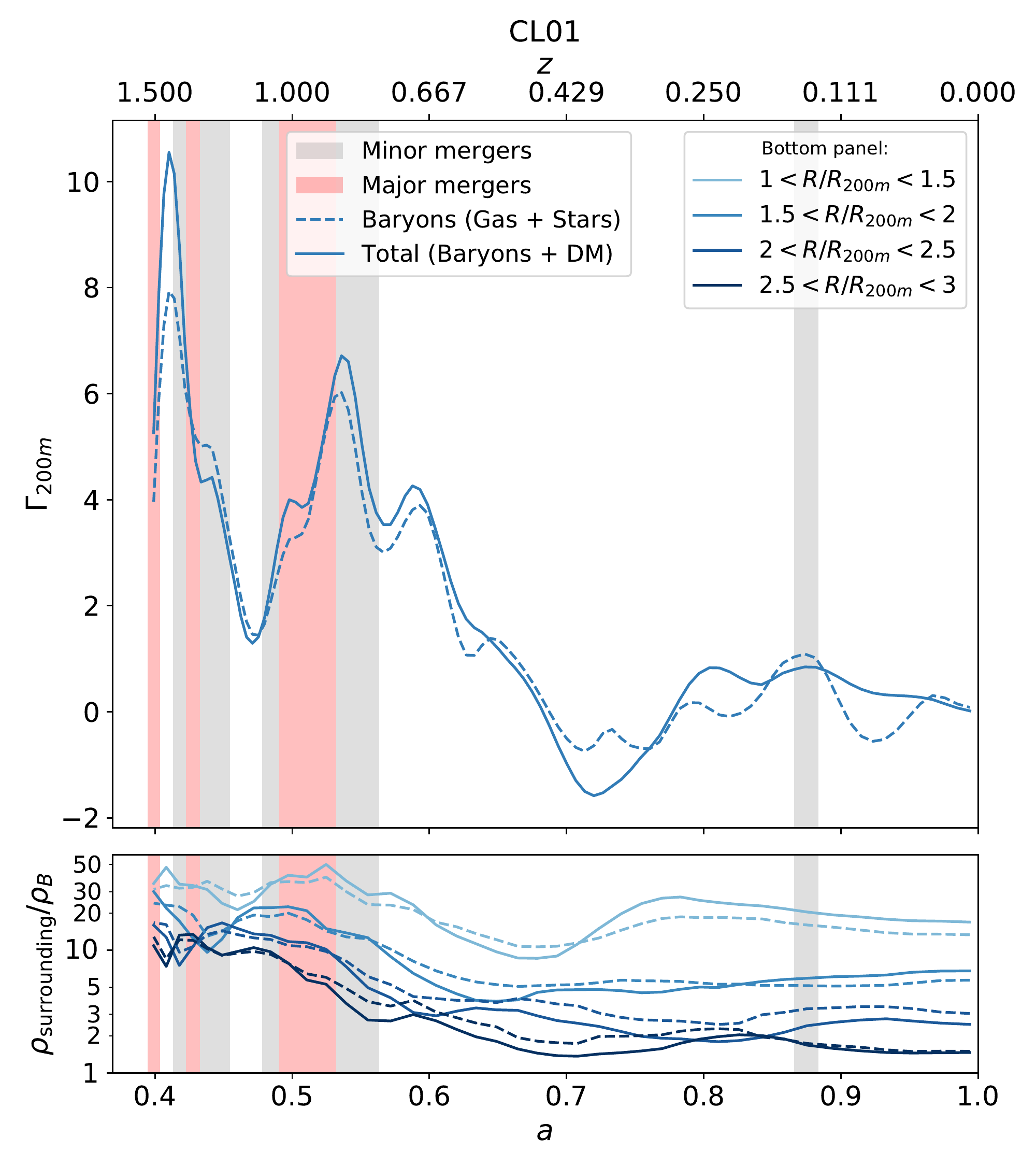}}~
	{\includegraphics[width = 0.5 \textwidth]{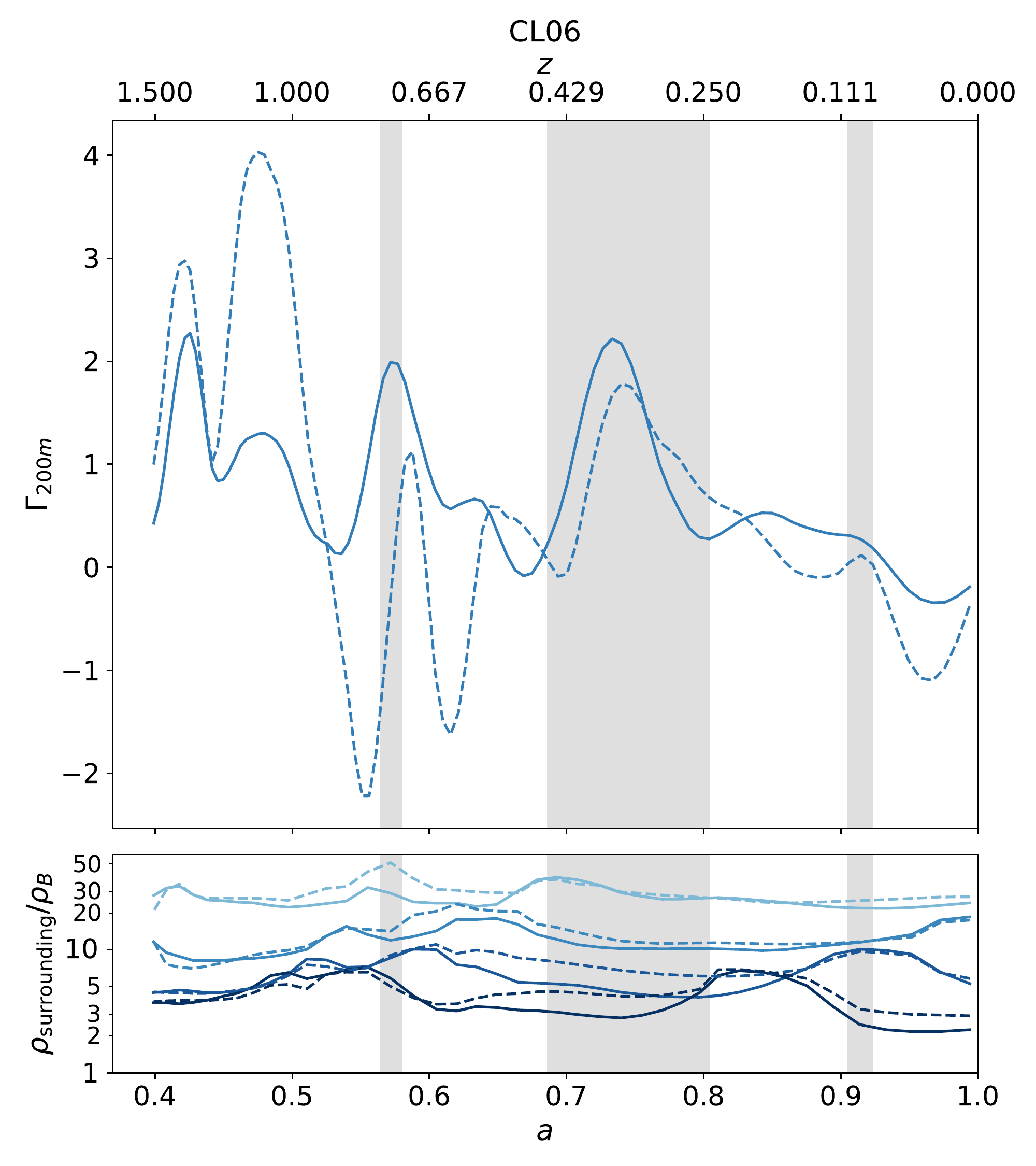}}
	\caption{Relation between the MARs, accretion regimes (ongoing merger events) and densities in the surroundings of the cluster, for clusters CL01 (left) and CL06 (right). In each panel, the upper plot shows the MARs for the total mass (solid line) and baryonic mass (dashed line). The lower panel displays the surrounding density in several clustercentric radial bins, in units of the background density of the Universe. The dashed line, corresponding to the baryonic density, has been normalised to the cosmic baryon fraction (i.e., it has been multiplied by $\Omega_m / \Omega_b$). The legend in the left panel applies to both plots.}
	\label{fig:mergers}
\end{figure*}

The results of the joint analyses of MARs, accretion regimes and surrounding densities are shown in Figure \ref{fig:mergers}, for clusters CL01 and CL06, as paradigmatic cases of a massive cluster which undergoes numerous major and minor mergers and a low-mass cluster which only suffers minor mergers, respectively. Both panels show the total (solid lines) and baryonic (dashed lines) MARs, the accretion regimes (background colour of the plot) and the surrounding densities (lower panels).

In the case of CL01, the differences in the MAR of DM and baryons are small in magnitude. Peaks in the MAR are undoubtedly associated to (major) merger events, as in the case of the displayed peaks around $z \sim 1.4$ and $z \sim 0.8$. It is also interesting to note how high MARs are mantained for a long time after the merger has taken place (particularly salient is the case of the merger at $z \sim 0.8$), as matter deposited outside $R_{200m}$ continues feeding the cluster --in a more quiescent way-- for several Gyr. In this respect, the lower panel shows how densities in the $1 \leq R/R_{200m} \leq 1.5$ region keep above $10 \rho_B$ until $z \sim 0.5$. 

Comparing the surrounding total and baryonic densities, the latter appears to evolve in a much smoother way than the former. At outer radii, $R \gtrsim 2 R_{200m}$, baryon surrounding densities (when normalised to the cosmic baryon fraction) tend to be higher than total densities. This reinforces the idea that gas, due to its pressure support, is deposited at larger radii than DM, which can more easily penetrate to inner regions. We find these general trends to be common for all the massive clusters in our sample which suffer major mergers.

On the other hand, CL06 exhibits significant differences in the behaviour of baryonic and total masses. The total MAR experiences peaks in correlation to the minor merger events. The baryonic component roughly follows these peaks, although their magnitude can differ significantly. There are also severe declines in the baryonic mass ($\Gamma_{200m} < 0$), which reflect the inability of low-mass systems and groups to keep their gas inside $R_{200m}$. This gas is typically expelled to larger clustercentric radii (as seen in the lower panel) and only slowly reaccreted afterwards.

\subsubsection{Correlation and time shift between surrounding densities and accretion rates}
\label{s:results_mar.mergers_surroundings.correlation}
As seen in the left panel of Fig. \ref{fig:mergers}, the density in the immediate neighbourhood of the cluster shows remarkable resemblance to the behaviour of the MAR, although a time shift between both curves is evident. In order to have an estimation of this shift, we compute the Spearman's rank correlation coefficient\footnote{The Spearman's rank correlation coefficient, $\rho_\mathrm{sp}$, assesses the monotonicity of the relation between two variables (without the need to assume linearity). $\rho_\mathrm{sp}$ is valued in $[-1,1]$, with higher absolute values implying a more monotonic relation and positive (negative) values corresponding to increasing (decreasing) relations.}, $\rho_\mathrm{sp}$, of $\rho_\mathrm{surrounding}(t)$ in the first radial bin with $\Gamma(t + \tau)$, and find the $\tau$ which maximises $\rho_\mathrm{sp}$. Below, we summarise the results of such analysis:

\begin{itemize}
	\item For cluster CL01, when total masses are considered, an optimal shift of $900 \, \mathrm{Myr}$ provides a rank correlation of $\rho_\mathrm{sp} = 0.803$. Restricting to the baryonic component enhances this correlation to $\rho_\mathrm{sp} = 0.895$, while increasing the time shift to $1.1 \, \mathrm{Gyr}$, as pressure effects slow down gas infall. These times are consistent with the shell crossing time of the fastest DM particles. A similar trend is shown by cluster CL03, although in this case the correlations are slightly weaker ($\rho_\mathrm{sp} = 0.744$ and $\rho_\mathrm{sp} = 0.691$, respectively).
	\item CL02, a massive cluster which only experiences a major merger at high redshift and undergoes a quiescent evolution therein, shows weaker correlations: $\rho_\mathrm{sp} = 0.623$ and $\rho_\mathrm{sp} = 0.572$, respectively.
	\item Less massive objects do not display significant correlations between these variables, as a result of their limited ability to capture matter (especially, gas). Indeed, in the lower panel of Fig. \ref{fig:mergers} a much flatter evolution of the surrounding densities is clearly noticeable.
\end{itemize}

\subsection{Impact of accretion on radial density profiles}
\label{s:results_mar.density_profiles}
\begin{figure*}
	\centering
	{\includegraphics[height=0.19\textheight]{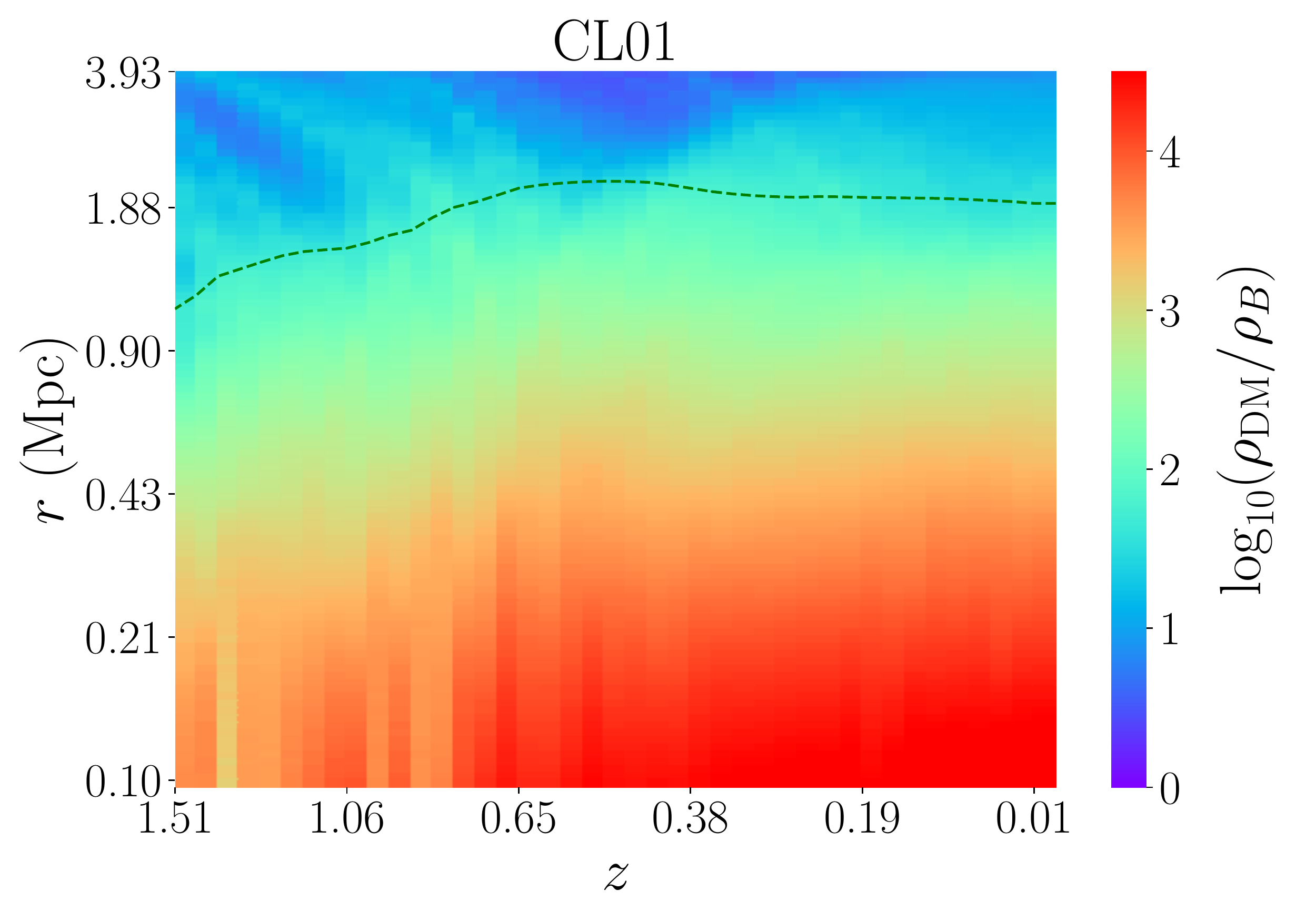}}~
	{\includegraphics[height=0.19\textheight]{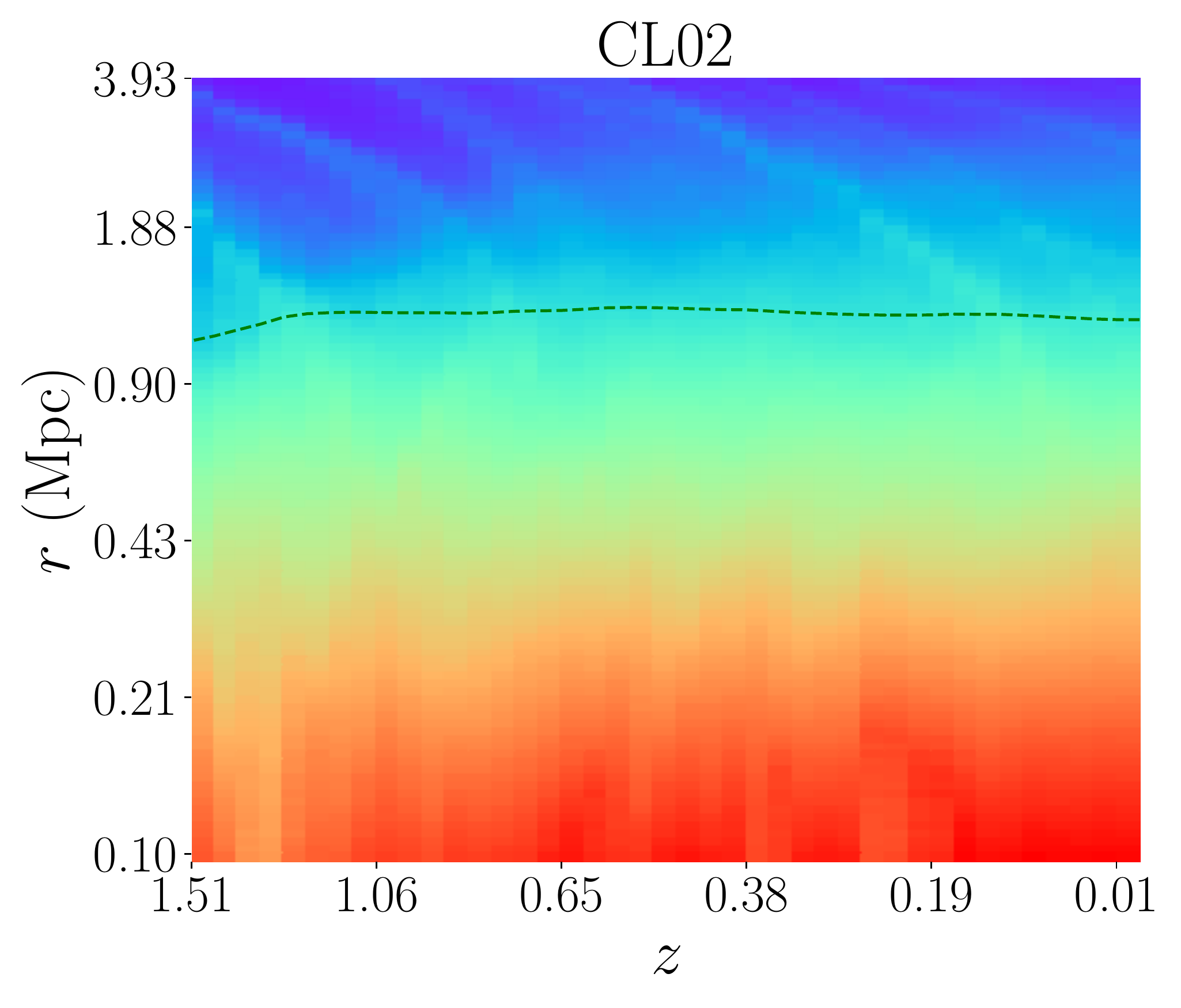}}~
	{\includegraphics[height=0.19\textheight]{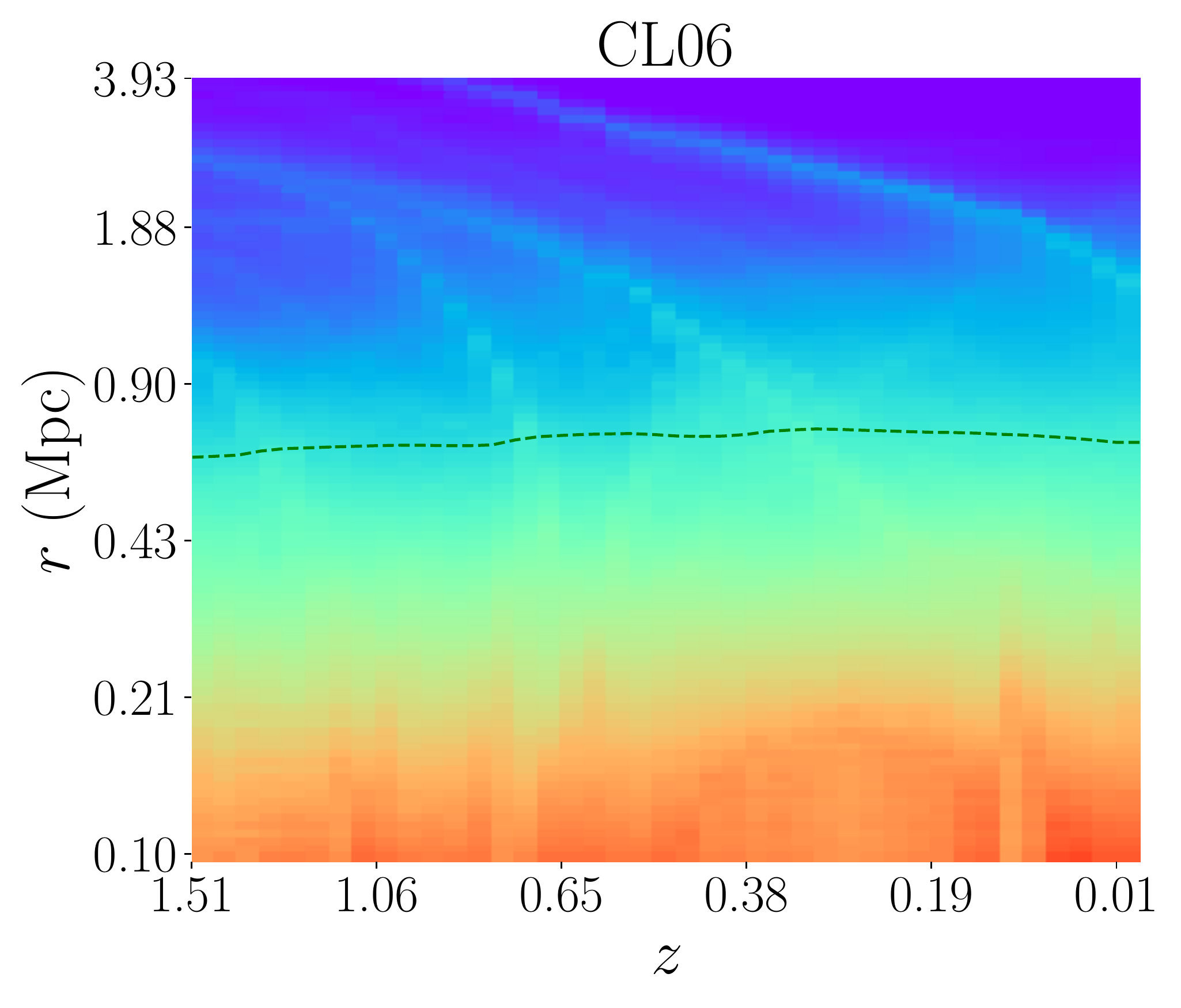}}
	{\includegraphics[height=0.19\textheight]{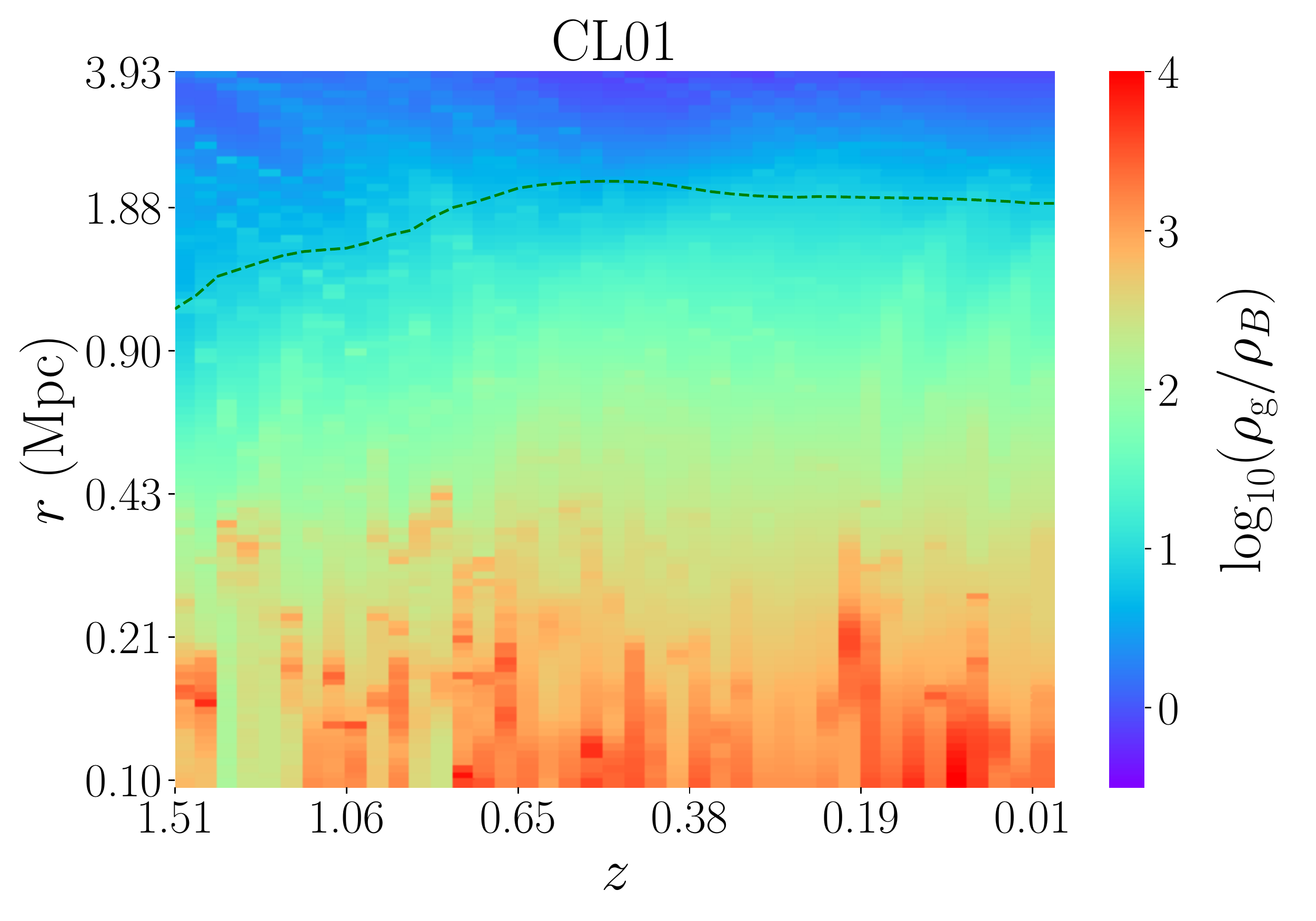}}~
	{\includegraphics[height=0.19\textheight]{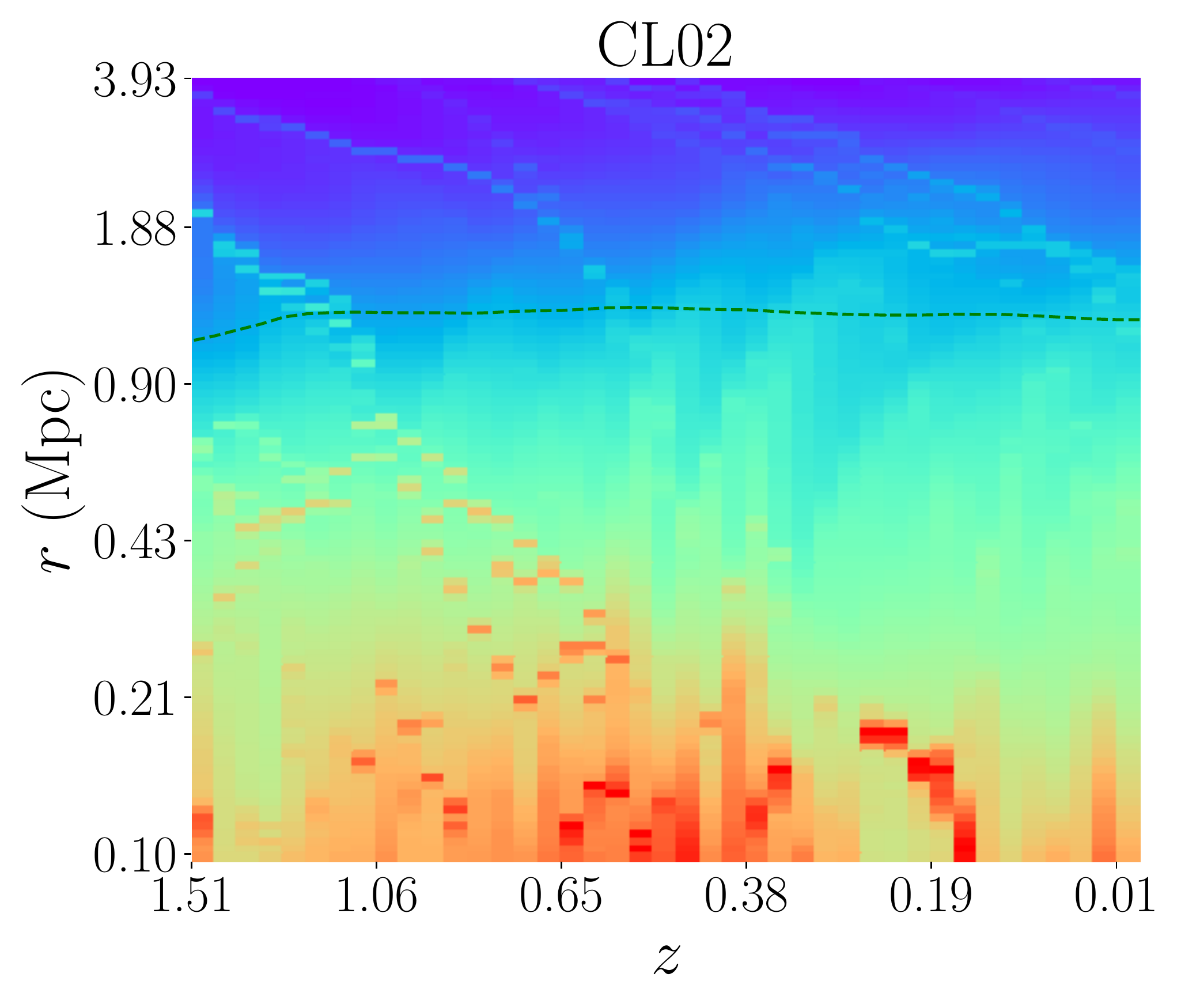}}~
	{\includegraphics[height=0.19\textheight]{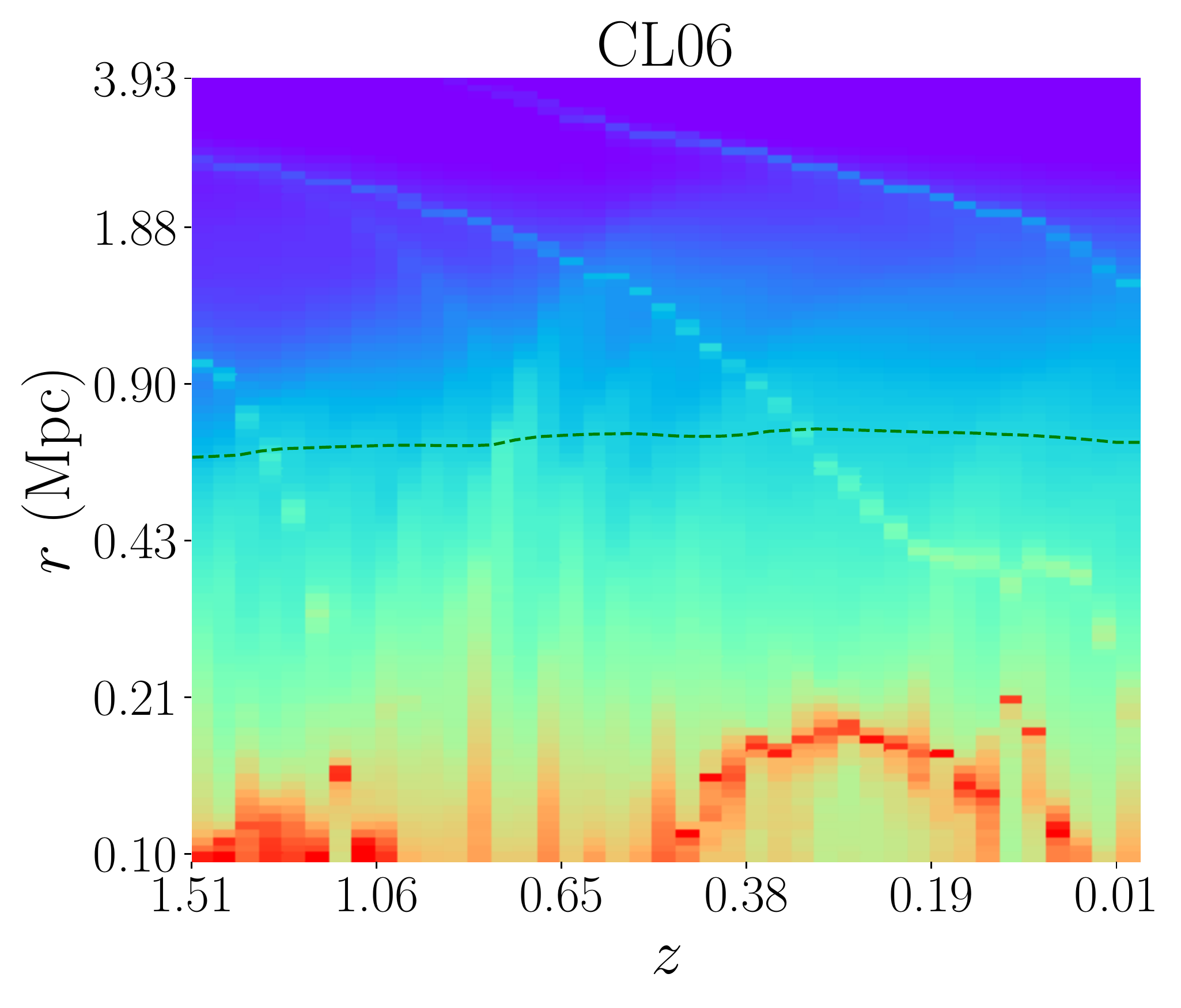}}
	{\includegraphics[height=0.192\textheight]{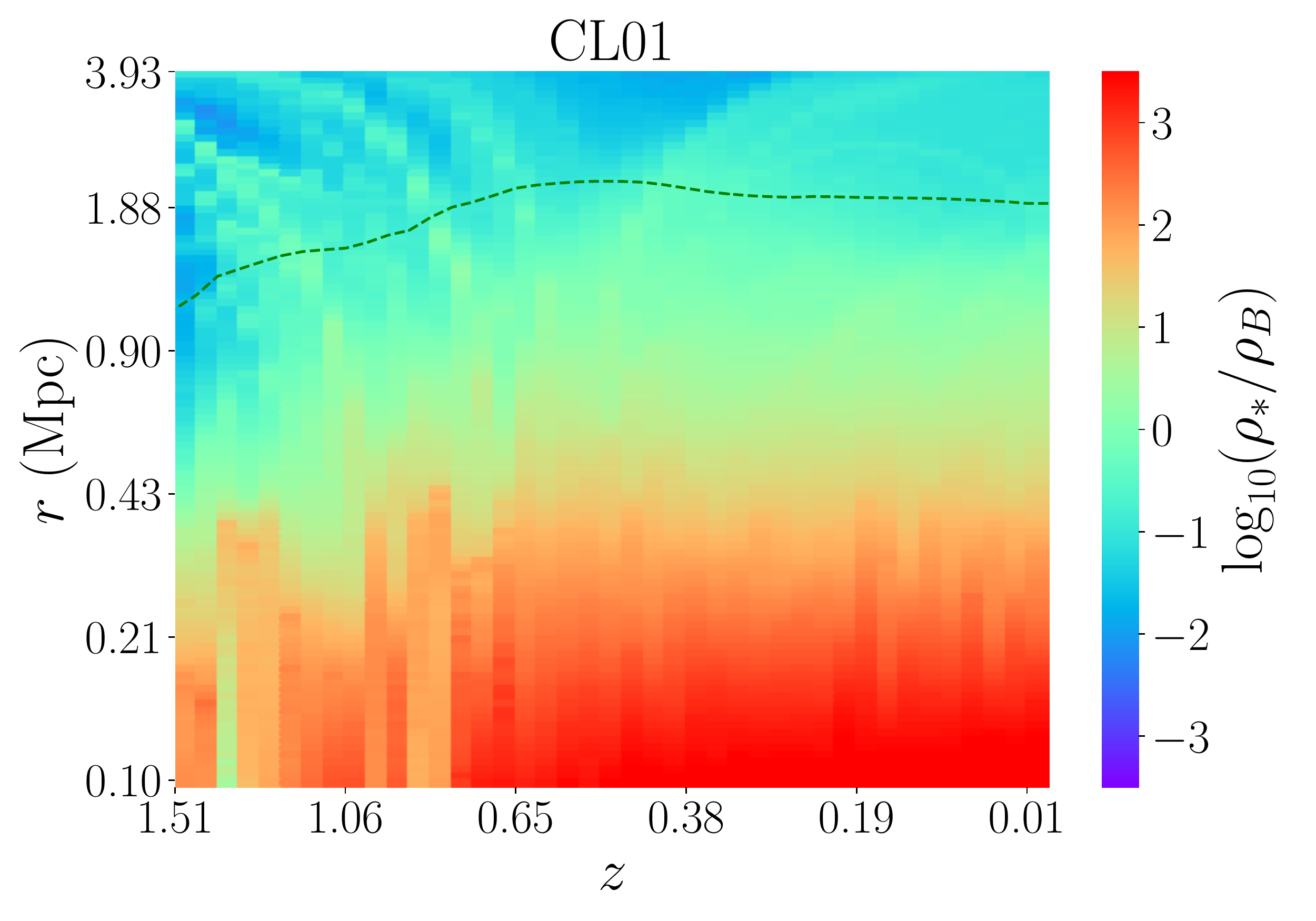}}~
	{\includegraphics[height=0.192\textheight]{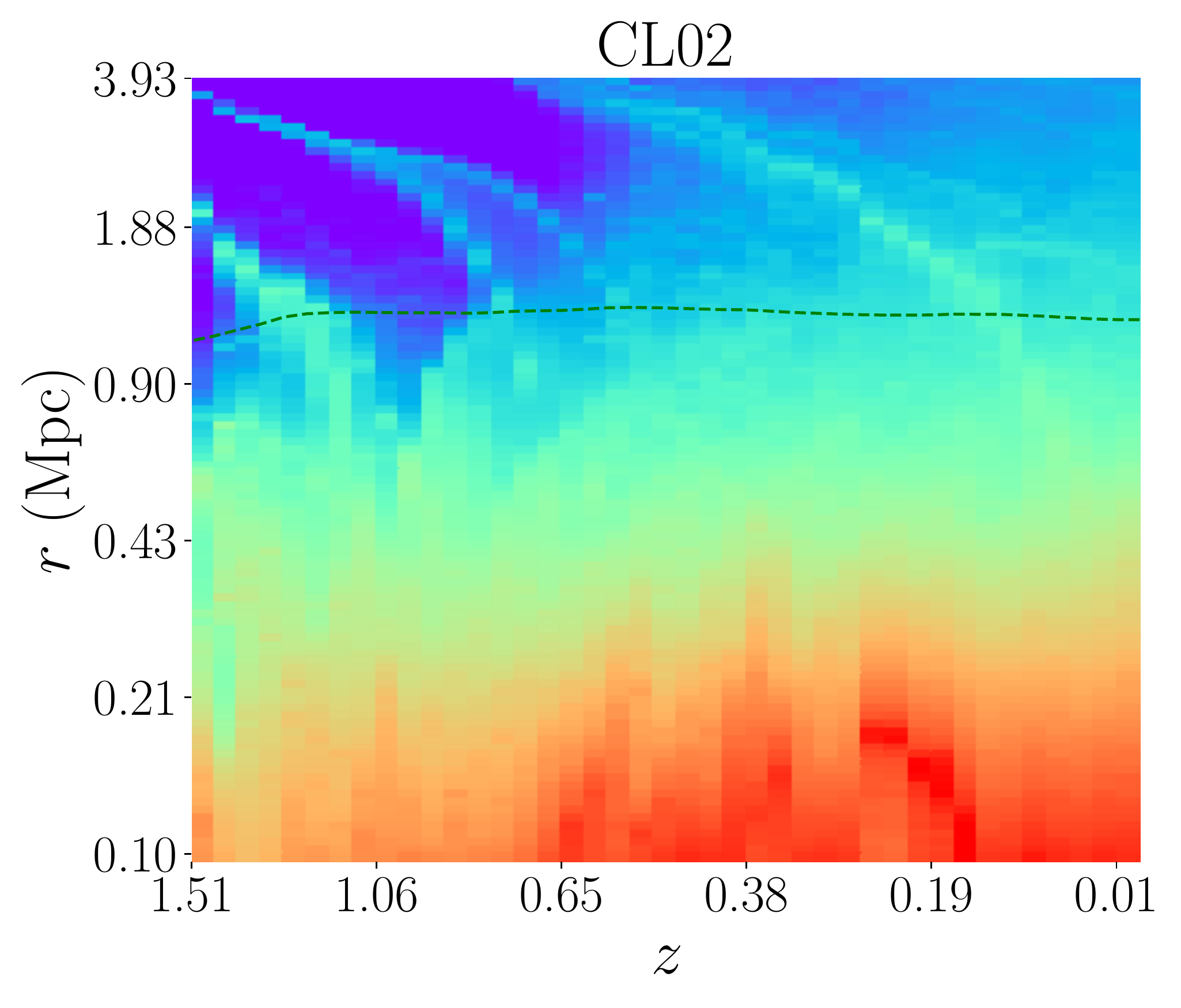}}~
	{\includegraphics[height=0.192\textheight]{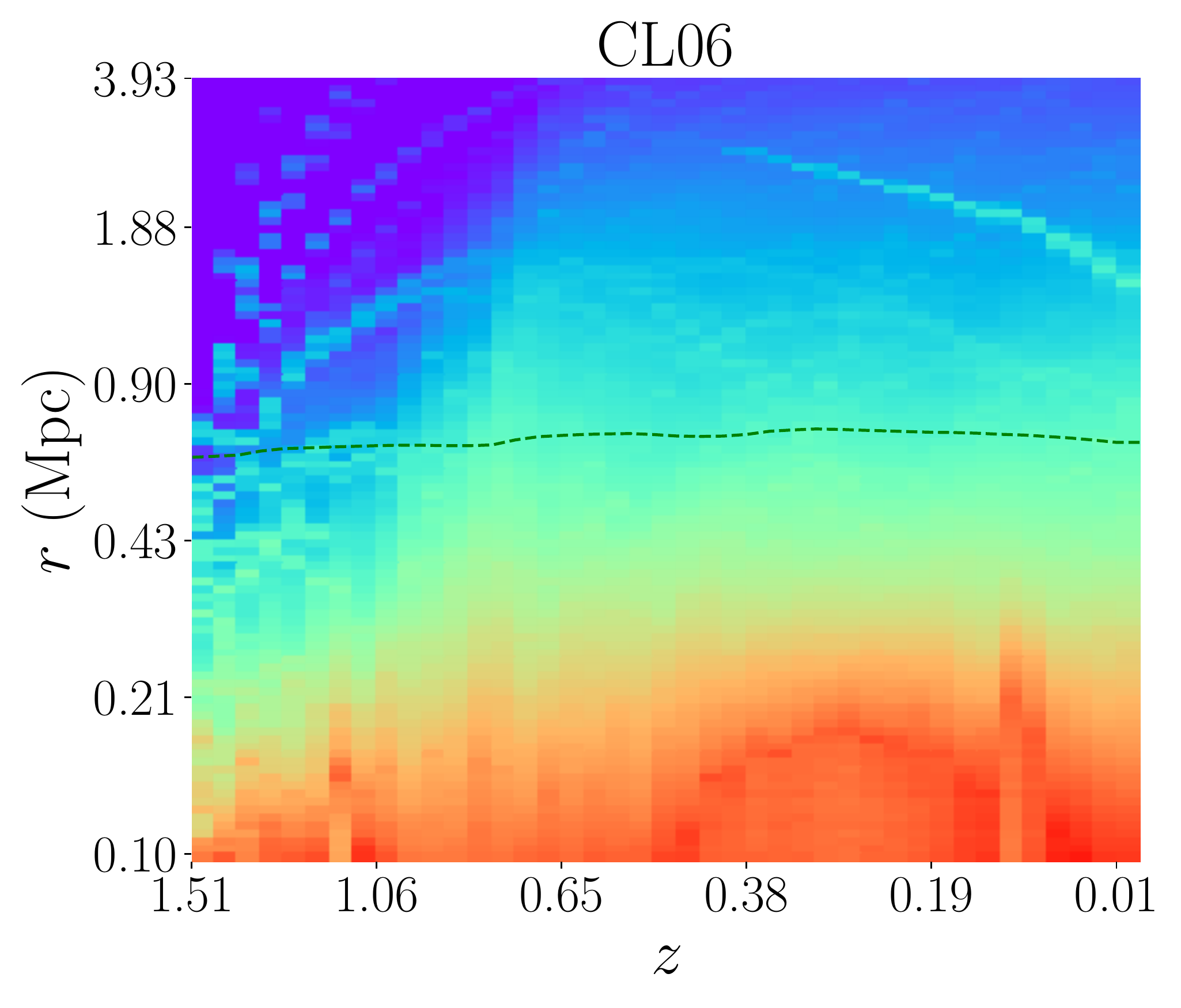}}
	\caption{The panels show the evolution of the comoving density profiles of the different material components for three clusters, from $z \simeq 1.5$ to $z=0$. The profiles have been taken with $100$ logarithmically spaced bins, from $100 \, \mathrm{kpc}$ to $4 \, \mathrm{Mpc}$, and taking center at the potential minima. Densities are always normalised to the background matter density of the Universe and the colour scale is logarithmic. Radial coordinates are comoving. From top to bottom, panels show DM, gas and stellar densities. From left to right, these quantities are presented for CL01, CL02 and CL06. The colour scale in the left plot of each row applies for all plots in the row. The green line in each panel marks the virial radius.}
	\label{fig:densityprofiles}
\end{figure*}

In this section, we analyse how mergers and strong accretion rates impact the inner distribution of the different material components in clusters by computing the comoving density profiles of DM, gas and stars. We graphically present these profiles for objects CL01, CL02 and CL06 as a function of redshift in Figure \ref{fig:densityprofiles}.

Note how the virial radii of CL02 (which only undergoes a major merger at $z \sim 1.4$) and CL06 (which only suffers minor mergers) are roughly constant, while CL01 (undergoing several major mergers) displays an important growth of this boundary.

Relating to this, the inner regions of dark matter comoving density profiles are mostly constant in time for CL02 and CL06, suggesting that these structures are already collapsed by $z \sim 1$ and the innermost radii ($r \lesssim R_{2500m}$) do not get disturbed by minor mergers and smooth accretion. This result is consistent with \cite{More_2015}, who find that the mass inside $4 r_s$ (being $r_s$ the scale radius of the NFW, \citealp{NFW_1997}, profile) evolves relatively slowly for $z \lesssim 1 - 2$. Conversely, CL01 does experience important disturbances in its DM profile, especially around $z \sim 0.8$. The enhanced MAR during $1.1 \gtrsim z \gtrsim 0.7$, associated to a major merger event (see Fig. \ref{fig:mergers}), appears to substantially increase the central density.

Gas density profiles are clumpier than DM ones and are affected by miscentering issues (specially in the case of CL06). The centers of the gaseous and dark components do not necessarily coincide (see, e.g., \citealp{Forero-Romero_2010, Cui_2016}), especially when there is ongoing merging activity (i.e. departures from dynamical equilibrium). Lines of decreasing radii with decreasing redshift reflect the infalling orbits of massive structures being accreted, mainly in galaxy cluster mergers. These streams of matter are better recognised in the stellar component, as the stellar mass is more concentrated towards the center of the infalling cluster and leaves a sharper imprint on the density profile. The redshifts at which these streams cross the halo $R_\mathrm{vir}$ boundary appear to be consistent with the periods of mergers according to the classification of Sec. \ref{s:results_mar.mergers_surroundings}.
	
The profiles of the gaseous component also suggest the presence of gas being deccreted or expelled outside the virial radius. This is particularly notorious for CL02 and CL06, and hints that dynamical interactions between clusters can extract gaseous matter to outer radii (e.g., through gas sloshing; see for example \citealp{Markevitch_2001, Roediger_2011}).

\subsection{Velocity profiles. Alternative MAR definitions}
\label{s:results_mar.velocity_profiles}
\begin{figure}
	\centering
	{\includegraphics[width=0.5\textwidth]{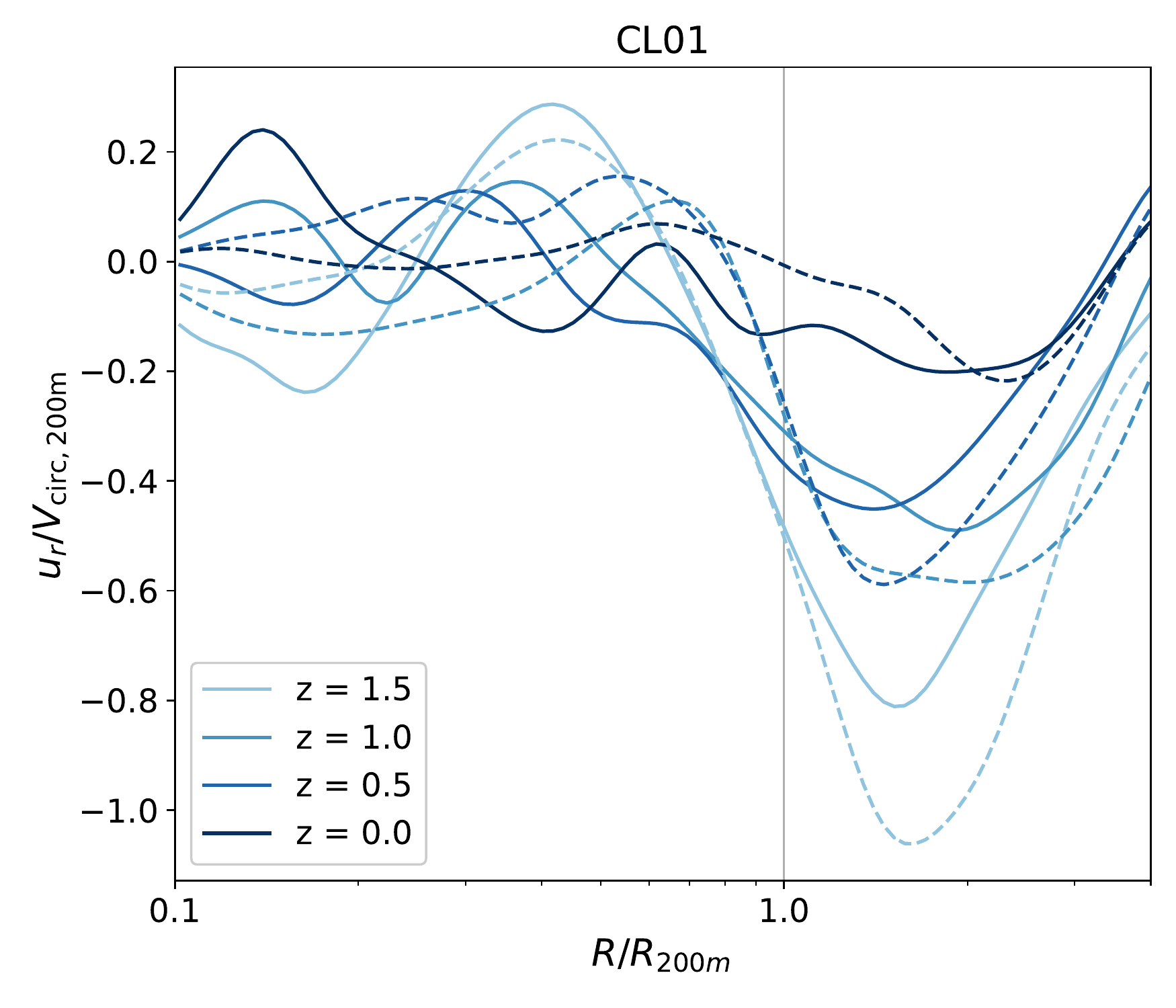}}
	\caption{Radially-averaged radial physical velocity profiles for the cluster CL01 at 4 different redshifts. Solid lines represent the radial velocities of gas, while dashed lines correspond to DM. The profiles have been taken with $100$ bins logarithmically spaced between $0.1 R_\mathrm{200m}$ and $4 R_\mathrm{200m}$ and smoothed with a gaussian filter with window length of 4 points.}
	\label{fig:velocity_profile_CL01}
\end{figure}

The dynamics of accretion onto galaxy clusters and the differential behaviour of DM and gas can be explored from the radially-averaged profiles of radial velocity in the cluster outskirts. In Figure \ref{fig:velocity_profile_CL01}, we present such profiles for the massive central cluster CL01, for four evenly spaced redshifts from $z=1.5$ to $z=0$. The magnitude represented in the figure corresponds to the physical velocity, $u_r \equiv v_r + H(z) r / (1+z)$, where $r$ and $v_r$ are, respectively, the comoving clustercentric radial coordinate and the peculiar velocity. This quantity has been normalised to the circular velocity at $R_{200m}$, $V_\mathrm{circ,200m} = \sqrt{G M_{\mathrm{200m}} / R_\mathrm{200m}}$.

The interior of the cluster ($r \lesssim 0.8 R_\mathrm{200m}$) does not present particularly strong inflows nor outflows, as these regions are already collapsed and relatively stable. Accretion flows are dominant in the cluster outskirts ($r \gtrsim R_\mathrm{200m}$), where radial velocities drop sharply and reach a minimum located around $1.5 \lesssim r/R_{200m} \lesssim 2$. The position of the velocity minimum with respect to $R_\mathrm{200m}$ does not show a clear redshift evolution in the case of CL01, in consistency with the general behaviour pointed out by \cite{Lau_2015}. At large radii, radial velocities increase, as the Hubble flow term begins to dominate the dynamics of both material components.

The magnitude of the velocity at the minima shows a remarkable redshift evolution. Both, gas and DM, have larger infall velocities --when compared to the circular velocity-- at earlier times and, consequently, their MARs shown in Fig. \ref{fig:mergers} are larger. Even though gas and DM radial velocity curves exhibit a similar pattern in the cluster outskirts, their different magnitudes highlight that DM is being accreted slightly faster, as it is not pressure supported and does not feel hydrodynamical effects, such as ram pressure from the ICM, shock heating, etc. as it falls into the cluster.

\subsubsection{Comparing $\Gamma_\Delta$ to $\alpha_\Delta$}
\label{s:results_mar.velocity_profiles.compare}
In \cite{Lau_2015}, the authors suggest a different proxy for the instantaneous MAR, defined as the radial infall velocity of DM ($v_r^\mathrm{DM}$) at some radius $r = R_\alpha$, in units of the circular velocity at $R_\Delta$:

\begin{equation}
	\alpha_\Delta = \frac{v_r^\mathrm{DM}(r = R_\alpha)}{V_\mathrm{circ,\Delta}}
	\label{eq:mar_Lau15}
\end{equation}

More negative $\alpha_\Delta$ indicates more rapid infall of matter, and thus $\alpha_\Delta$ should be anticorrelated to $\Gamma_\Delta$. \cite{Lau_2015} report that taking $R_\alpha = 1.25 R_{200m}$ maximises the anticorrelation of $\alpha_{200m}$ and $\Gamma_\Delta^{[a_1, a_0]}$, with $a_1 = 0.67$. Note, however, that $\Gamma_\Delta^{[a_1, a_0]}$ is the averaged MAR over the last $\sim 5 \, \mathrm{Gyr}$.

In order to compare $\alpha_\Delta$ to the instantaneous $\Gamma_\Delta$ MAR proxy defined in Eq. (\ref{eq:mar_derivative}), we have computed $\alpha_\Delta$ for all the clusters in our sample and all the simulation outputs available, from $z=1.5$ to $z=0$. We find that, taking $R_\alpha = R_{200m}$, the Spearman rank correlation between $\alpha_{200m}$ and $\Gamma_{200m}$ is $\rho_\mathrm{sp} = -0.832$. Larger values of $R_\alpha / R_{200m} = 1.25, \, 1.5$ weaken this anticorrelation down to $\rho_\mathrm{sp} = -0.719$ and $\rho_\mathrm{sp} = -0.556$, respectively.

\begin{figure}
	\centering
	{\includegraphics[width=0.5\textwidth]{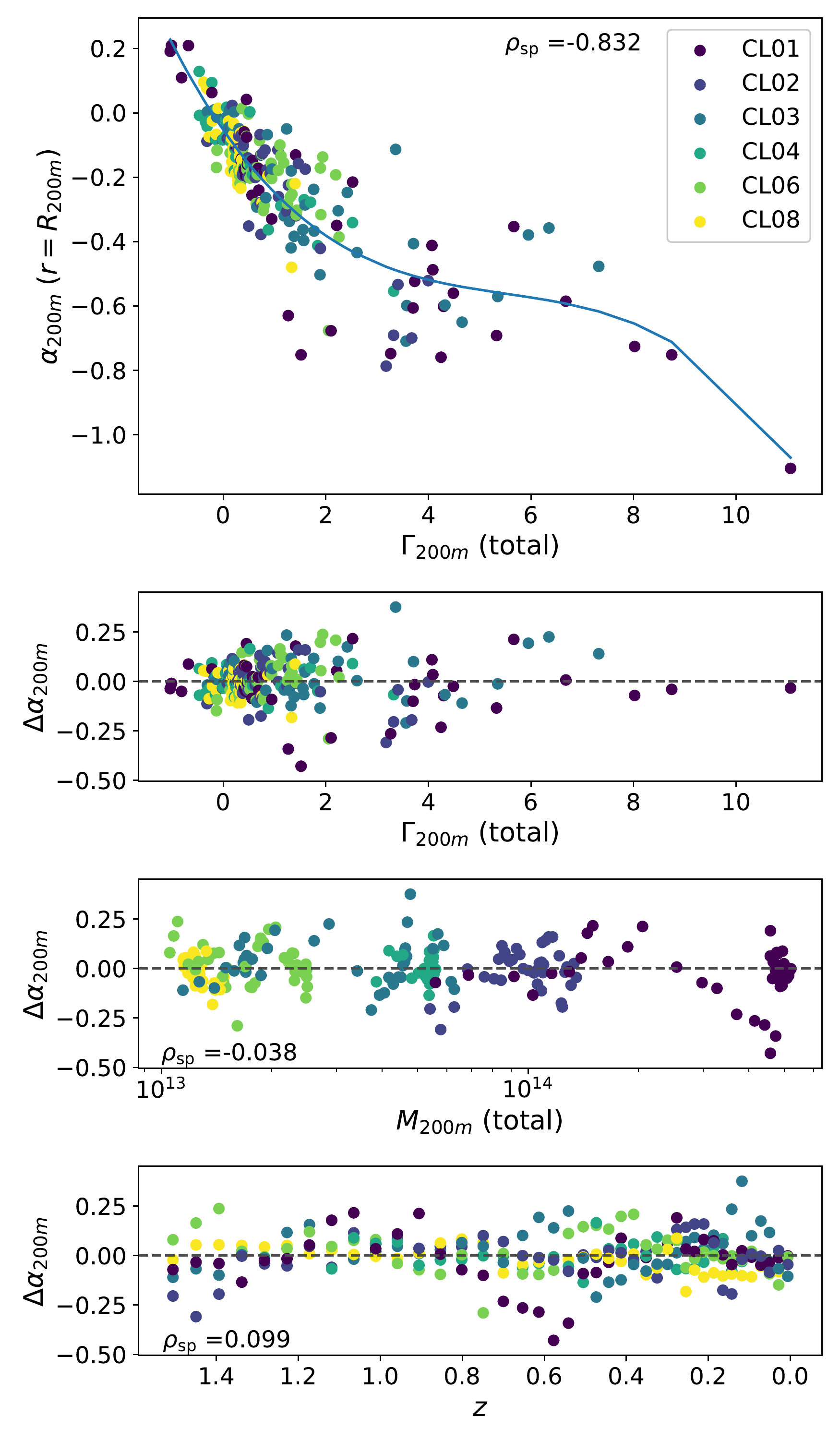}}
	\caption{Relation between the instantaneous MAR proxies $\Gamma_{200m}$ (used in this work) and $\alpha_{200m}$ \citep{Lau_2015}, combining all the snapshots in $1.5 \geq z \geq 0$. The upper panel contains the scatter plot of both variables and a nonparametric fit using smoothing splines. The residuals of this fit, $\Delta \alpha_{200m}$ are used in the lower panels to rule out redshift or mass dependences in this relation, as they appear to be uncorrelated to $M_{200m}$ and $z$.}
	\label{fig:alpha_vs_gamma}
\end{figure}

In the upper panel of Figure \ref{fig:alpha_vs_gamma}, we present a scatter plot of the two instantaneous MAR proxies, where the dots have been coloured to identify each cluster. All snapshots, from $z=1.5$ to $z=0$, are included in this comparison. A nonparametric fit with smoothing splines is shown as the blue line. To rule out any possible dependences of the scatter in this relation with other variables, the panels below show the distribution of residuals with respect to the fit, $\Delta \alpha_{200m}$, as a function of $\Gamma_{200m}$, $M_\mathrm{200m}$ and redshift $z$. No significant residual dependences on $M_\mathrm{200m}$ and $z$ are found. Consistently, the residuals are uncorrelated to these variables, with rank correlation coefficients $\rho_\mathrm{sp} = -0.038$ and $\rho_\mathrm{sp} = 0.099$, respectively. Hence, $\alpha_{200m}$ and the instantaneous $\Gamma_{200m}(a)$ are consistent probes of the MAR, and their relation is mostly independent of redshift or cluster mass.

\section{Angular distribution of the mass flows and thermodynamical properties}
\label{s:results_angular}

This section focuses on the characterisation of the angular distribution of accreting gas. This topic has not been extensively covered in the literature, but it is of utmost interest in order to assess the complex interplay of clusters with their environments, which shapes the accretion patterns. In Sec. \ref{s:results_angular.method} we present a simple method for estimating and presenting the accretion flows. Its results are discussed in Sec. \ref{s:results_angular.results}.

\subsection{Estimation of the mass fluxes through the cluster boundary}
\label{s:results_angular.method}

Let us consider a cluster at redshift $z$, delimited by a spherical boundary $R_\mathrm{bdry}$ (e.g., $R_\mathrm{vir}$) around it. We quantify the mass flows by computing an estimated flux from the peculiar velocity and density fields as described in the following paragraphs.

We assume each gas cell, with density contrast $\delta_\mathrm{cell}$ and physical volume $\Delta V_\mathrm{cell}$, as a particle located at its geometrical center, with mass $m_\mathrm{cell} = \rho_B(z) (1 + \delta_\mathrm{cell}) \Delta V_\mathrm{cell}$ and the peculiar velocity $\vb{v} = a \dv{\vb{x}}{t}$ given by the corresponding cell-averaged velocity. All cells, regardless of the refinement level they belong to, are considered on equal footing. For each cell, let $r$ be the radial comoving distance to the cluster center and $v_r$ its clustercentric radial peculiar velocity. With these definitions, we estimate the fluxes accross $r = R_\mathrm{bdry}$ according to the following rules:

	\begin{itemize}
		\item Given a cell within the spherical boundary, $r < R_\mathrm{bdry}$, we mark it as an escaping cell if $r + \frac{v_r}{a} \Delta t > R_\mathrm{bdry}$.
		\item A cell outside the radial boundary $r > R_\mathrm{bdry}$ is marked as an entering cell if $r + \frac{v_r}{a} \Delta t < R_\mathrm{bdry}$.
	\end{itemize}
	
The time interval, $\Delta t$, used for the estimation of the fluxes has to be chosen as a compromise between angular resolution (as higher $\Delta t$ increases the number of entering and escaping cells) and accuracy (radial flows are not necessarily maintained for arbitrarily large $\Delta t$). We have set $\Delta t$ to the time difference between consecutive snapshots (ranging from $\sim 60 \, \mathrm{Myr}$ to $\sim 300 \, \mathrm{Myr}$ in this particular simulation).

Once the entering and escaping cells have been found, the angular distribution of accreted gas is computed by binning the complete solid angle around the cluster in the clustercentric spherical angles, $\phi$ and $\cos \theta$ (so that angular sectors at all latitudes subtend the same solid angle). We define the spherical coordinate system with respect to the major axis of the cluster total mass distribution. This is motivated by $N$-Body simulations having shown that most of the mergers and accretion of DM occur through the filaments connecting a cluster to its nearest massive neighbour \citep{Lee_2007} and, consequently, major axes tend to be aligned with such filaments \citep{Lee_2008}. In order to do so, the characterisation of the shape of matter distributions is discussed in Sec. \ref{s:results_angular.method.shape}. In our analyses for cluster CL01, we split the solid angle in $n_\phi \times n_\theta = 80 \times 80$ bins. Increasing the number of bins beyond this quantity does not result in any significant enhancement in the description of the mass fluxes, as the resolution gets constrained by the cell sizes at the $r=R_\mathrm{bdry}$ boundary.

For each entering (escaping) cell, we assign all its mass to the bin corresponding to its angular position, yielding the distribution of accreting (deccreting) matter. Their subtraction is the net mass flowing across the $r=R_\mathrm{bdry}$ boundary. Finally, we compute the mass radial flux by normalising this quantity as follows:

\begin{equation}
	j_M = \frac{\Delta M}{R_\mathrm{bdry}^2 \Delta \Omega \Delta t}
	\label{eq:massflux}
\end{equation} 

\noindent where $\Delta M = \Delta M_\mathrm{enters} - \Delta M_\mathrm{escapes}$ and $\Delta \Omega = \frac{4 \pi}{n_\phi n_\theta}$. Note that we take, as sign convention, that $j_M > 0$ when matter is infalling (being accreted).

\subsubsection{Validity of the method. Spatial and temporal coherence of the radial flows.}

The procedure proposed above relies on the implicit assumption that gas velocities measured at one code output are persistent during the time interval, $\Delta t$, used to estimate which gas cells cross the cluster boundary. In the following lines, we briefly argue that this approximation is, indeed, applicable to our system, thus justifying the validity of the method.

The basic scheme of the performed test consists on the comparison of the radial velocity dispersion, $\sigma_r$, with the radial velocity, $v_r$. For each entering cell, we compute its total and radial velocity dispersions as the standard deviation of such quantities in the neighbouring $5 \times 5 \times 5$ cells. Such analysis yields the following conclusions: 

	\begin{itemize}
		\item The gas flows are eminently radial. The mean radial projection of the velocity of the entering cells has magnitudes $0.8 \lesssim |v_r| / v \lesssim 0.9$ for all code outputs.
		\item Radial velocity dispersions are consistent, only slightly above the isotropic value, $\sigma_r^\mathrm{iso} = \frac{\sigma_v}{\sqrt{3}}$. For most of the snapshots, $\frac{\sigma_r}{\sigma_r^\mathrm{iso}} \sim 1 - 1.1$. Thus, the relative velocity dispersion in the radial direction $\sigma_r / |v_r|$ is much smaller than in the directions tangential to the flow.
		\item Indeed, $|v_r| / \sigma_r$ takes mean values between 10 and 40, indicating that, in the neighbourhood of a cell, radial gas flows are spatially coherent.
	\end{itemize}

As a result of the radial flows being spatially coherent, one should expect the shear forces between neighbouring cells to be small along the radial direction. Therefore, turbulence is not expected to have a severe impact on the overall distribution of radial flows. However, this analysis does not guarantee the persistence of the flows during an arbitrarily large $\Delta t$. Once the spatial coherence of the flows has been confirmed, these can be assumed to be persistent between consecutive snapshots, provided that the angular distributions for consecutive snapshots are temporally coherent, i.e., they show similar structures and temporal changes are gradual. 

Note, however, that this analysis has considered the mean values of $\sigma_r$ and $v_r$ across the whole boundary. It is still posible that turbulence is relevant on the radial flows in small angular regions, thus introducing spurious noise into our flux maps. 

\subsubsection{Characterisation of clusters' shapes}
\label{s:results_angular.method.shape}
The procedure for studying the mass flows presented above requieres finding the major axis of the cluster, as we use it to define the spherical coordinate system tied to the cluster.

A widely extended method in the literature to estimate the shape of DM haloes (e.g., \citealp{Cole_1996, Planelles_2010, Knebe_2010}, among others) relies on finding the eigenvalues and eigenvectors of the shape tensor,
	
\begin{equation}
	\vb{S} = \frac{1}{M_\mathrm{tot}}\int_{\mathcal V} \rho(\vb{r}) \vb{r} \otimes \vb{r} \dd V,
	\label{eq:shapetensor}
\end{equation}

\noindent where the integration extends to the spherical overdensity boundary definition and $M_\mathrm{tot} = \int_{\mathcal V} \rho(\vb{r}) \dd V$ is the enclosed mass. Even though this approach produces sensible estimates, it might be contradictory that a spherical integration volume is used to characterise the shape of a triaxial object. A more robust method uses an iterative procedure aimed to adapt the integration volume to the --initially unknown-- shape of the mass distribution \citep{Zemp_2011}. We have implemented a method partially based on the one described in the aforementioned reference, which better suites our purposes. The main steps can be summarised as follows:

	\begin{enumerate}
		\item The first iteration uses an sphere of radius $R_\mathrm{bdry}$ as the integration volume. Let us call $\{\vb{\hat v_1}, \vb{\hat v_2}, \vb{\hat v_3} \}$ the orthonormal set of eigenvectors of $\vb{S}$, where their corresponding eigenvalues are such that $\lambda_1 < \lambda_2 < \lambda_3$. That is to say, the third vector points in the direction of the tentative major axis.
		\item These values are used to define the new integration volume, which is given by the ellipsoid
		\begin{equation}
			\frac{(\vb{\hat v_1} \cdot \vb{r})^2}{a_1^2} + \frac{(\vb{\hat v_2} \cdot \vb{r})^2}{a_2^2} + \frac{(\vb{\hat v_3} \cdot \vb{r})^2}{a_3^2} \leq 1,
			\label{eq:ellipsoid}
		\end{equation}
		\noindent where $a_i$ is the semiaxis length corresponding to $\vb{\hat v_i}$. The semiaxes are rescaled so that their squares are proportional to the corresponding shape tensor eigenvalues. However, this premise does not fix the magnitude of the semiaxes (only their quotients) and, therefore, a choice regarding the normalisation has to be made.
		
		In \citet{Zemp_2011}, the new semiaxes lengths are defined so that the major semiaxis is preserved. This election is motivated by their aim of measuring the shape at different distances, which they label by the major semiaxis length. However, this method shrinks the volume along the direction of the minor and intermediate axis and, therefore, dramatically increases the enclosed overdensity.
		
		With the aim of describing the overall shape of the cluster without changing significantly the enclosed overdensity, in this work we have chosen to preserve the enclosed volume. Let $r_i \equiv \sqrt{\frac{\lambda_i}{\lambda_3}}$, for $i=1,2$. Then, it is easy to check that volume is preserved by rescaling the axes so that:
			\begin{equation}
				a_1 = \sqrt[3]{\frac{r_1^2}{r_2}} R_\mathrm{bdry}, \quad 
				a_2 = \sqrt[3]{\frac{r_2^2}{r_1}} R_\mathrm{bdry}, \quad 
				a_3 = \frac{1}{\sqrt[3]{r_1 r_2}} R_\mathrm{bdry}
				\label{eq:rescale_axes}
			\end{equation}
		\item $\vb{S}$ is estimated in the new volume, yielding a new set of eigenvectors and eigenvalues. Then, step (ii) can be repeated with these new values. This iterating scheme is repeated until convergence.
	\end{enumerate}

As in \citet{Zemp_2011}, convergence is assessed by the change in the semiaxes ratios, $r_1$ and $r_2$, in two consecutive iterations. We stop the iterative scheme when the relative change in these magnitudes is smaller than $10^{-3}$. 

This process is first done for the latest code output, thus characterising the shape of mass distributions at $z=0$. In order to trace the principal axes to higher redshifts, the same procedure is repeated on each code output. However, after the principal axes have been determined, it could happen that the eigenvalues have changed their order. In order to prevent sudden rotations of the axes, the principal axes on a given snapshot, $\vb{\hat{v}_i}$, are matched to the ones in the previous snapshot, $\vb{\hat{v}'_{j(i)}}$, by pairing up the vectors so that $\sum_i |\vb{\hat{v}_i} \cdot \vb{\hat{v}'_{j(i)}}|$ is maximised, i.e., by performing the smallest possible rotation.

\subsection{Angular distribution of the mass flows in the principal axes system}
\label{s:results_angular.results}
\begin{figure*}
	\centering
	{\includegraphics[width=0.5\textwidth]{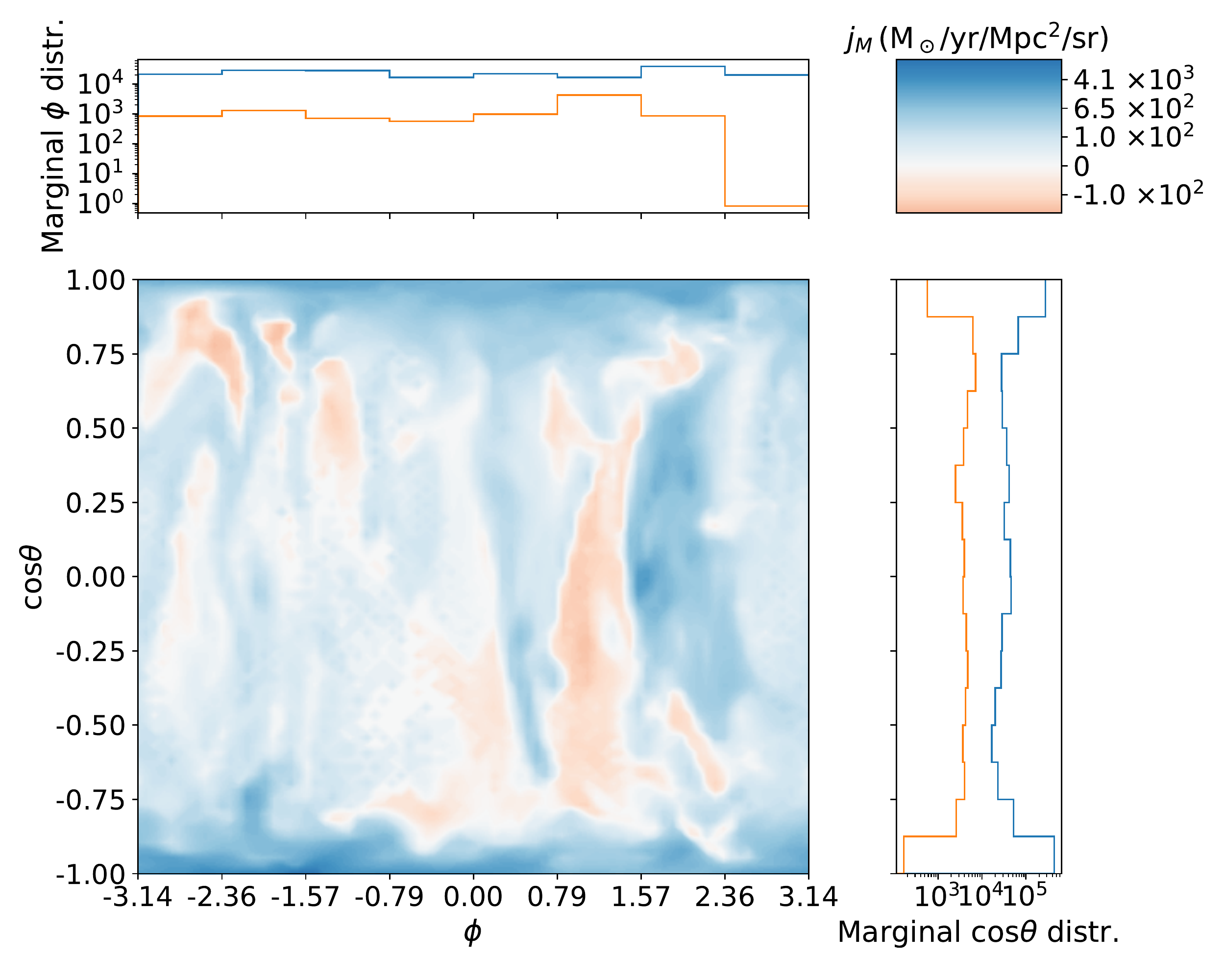}}~
	{\includegraphics[width=0.5\textwidth]{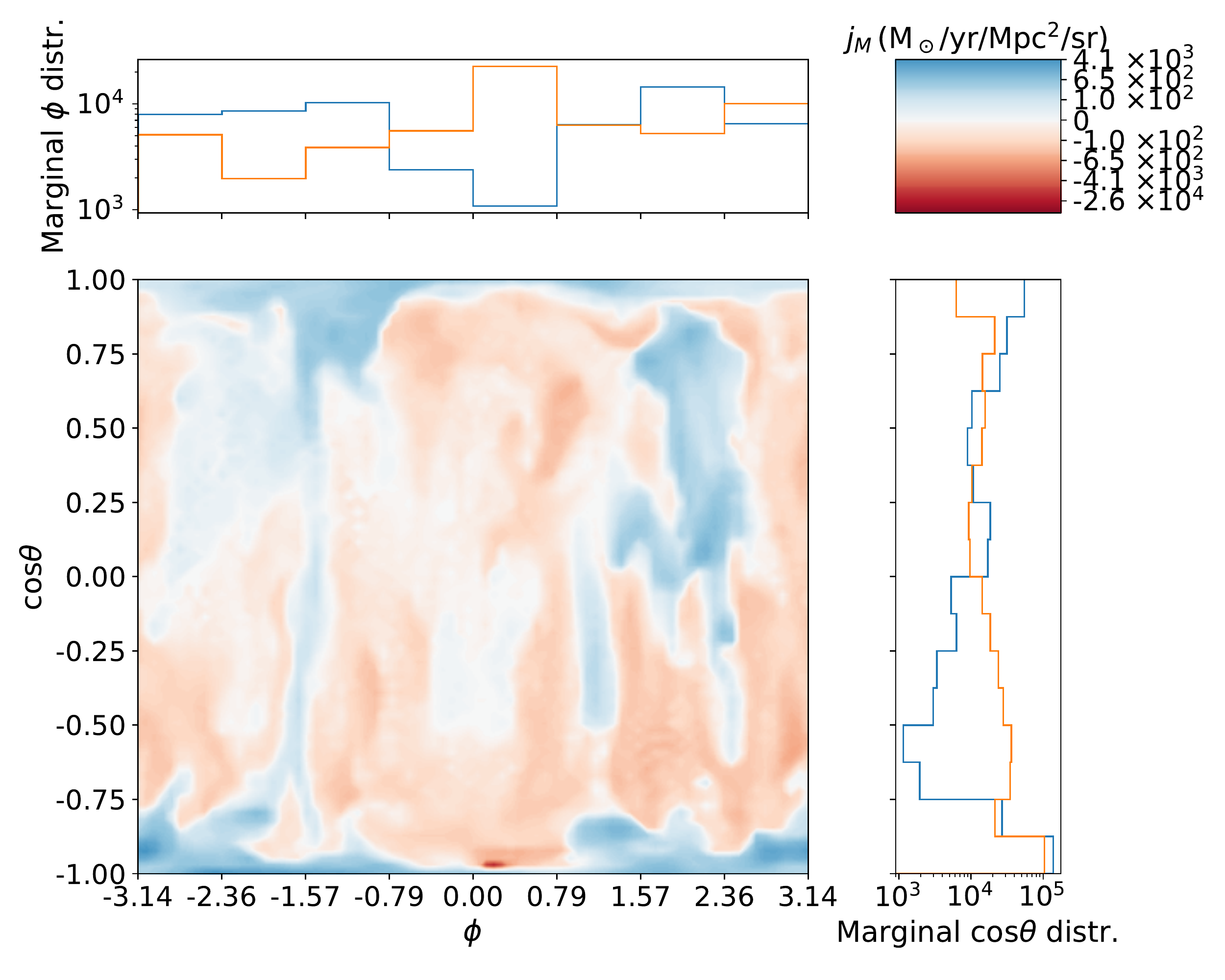}}
	\caption{Angular distribution of mass flows through the $r = R_\mathrm{vir}$ surface of cluster CL01 at $z \simeq 0.81$ (during a major merger; left panel) and at $z \simeq 0.35$ (in the smooth accretion regime; right panel). The colour scale encodes the mass flux density in symmetric logarithmic scale (orange tones indicate escaping gas, while blue tones indicate gas being accreted). The upper histogram shows the marginal azimuthal distribution of the entering (blue) and escaping (orange) fluxes, $\int_{-1}^{1} j_M(\theta, \phi) \dd (\cos \theta)$. Likewise, the histogram on the right shows the marginal polar distribution of the fluxes, $\int_{-\pi}^{\pi} j_M(\theta, \phi) \dd \phi$.}
	\label{fig:angular_massflows}
\end{figure*}

As a result of applying the method described in Sec. \ref{s:results_angular.method} to cluster CL01, Figure \ref{fig:angular_massflows} presents the angular distribution of the mass flows at two characteristic moments in the evolution of this object. The left panel shows the angular distribution of accreting and deccreting gas at $z \simeq 0.81$, right after a major merger (see Fig. \ref{fig:mergers}), while the baryon accretion rate is still high. The right panel presents the same information for a quiescent stage in the evolution of the cluster, while $\Gamma_\mathrm{vir}^\mathrm{baryons} \approx 0$. Besides the flux maps, the adjacent plots show the polar and azimuthal marginal distributions of accreting (blue line) and deccreting (orange line) gas. In order to produce these plots, one needs to represent positive (entering matter) and negative (escaping matter) values which span a broad range of orders of magnitude. To do so, we have implemented a symmetric logarithmic scale (see Appendix \ref{s:appendix.symlog}).

The marginal $\cos \theta$ distribution of both panels reveals that gas accretion fluxes are maximum in the polar regions (defined with respect to the major axis), corresponding to the gas infalling along the cosmic filaments \citep{Lee_2007, Lee_2008}. Indeed, the regions limited by $\left|\cos \theta \right| > 0.875$ (which correspond to $1/8$ of the total solid angle) account for $60\%$ and $54\%$ of the total gas mass inflow, for the snapshots shown in the left and right panels of Fig. \ref{fig:angular_massflows}, respectively.

Outside the polar region, gas accretes in a much smoother manner. However, far from being isotropic, the flows present intricate structures. At $z \simeq 0.81$, during the high-accretion regime, mass inflows dominate almost everywhere, but some structures stretched along the $\theta$ direction stand out. At $z \simeq 0.35$, the baryonic component has a net MAR of $\Gamma_\mathrm{vir}^\mathrm{baryons} \simeq -0.5$, which could be tempting to interpret as the absence of strong mass flows. However, the right panel in Fig. \ref{fig:angular_massflows} shows that this interpretation is not a good descriptor of the physical scenario. The low absolute value of $\Gamma$ emerges as a result of a rather complex counterbalance between strong inflows and outflows. Once again, stretched structures of accretion and deccretion at nearly constant $\phi$ are present. Accretion continues dominating in the polar region, while loss of gas is prevalent elsewhere.

\subsubsection{Multipolar expansion of the mass fluxes}
The plots in Fig. \ref{fig:angular_massflows} depict a complex pattern of gas inflows and outflows. In order to extract quantitative information from these accretion patterns, we develop the mass flux through the $r = R_\mathrm{bdry}$ boundary in the basis of real spherical harmonics,

\begin{equation}
	j_M(\theta, \phi) = \sum_{l=0}^{\infty} \sum_{m=-l}^{l} c_{lm} \mathcal{Y}_\mathrm{lm}(\theta, \phi),
	\label{eq:multipolar_expansion}
\end{equation}

\noindent where $c_\mathrm{lm} = \oiint \dd \Omega \mathcal{Y}_\mathrm{lm}(\theta, \phi) j_M(\theta, \phi) \in \mathbb{R}$ (see Appendix \ref{s:appendix.realspharm}). However, the discrete sampling of the mass flux in $n_\phi \times n_\theta$ bins constrains the maximum degree, $l$, that can be faithfully reconstructed. As the spherical harmonics of degree $l$ represent variations on angular scales $\pi / l \, \mathrm{rad}$, the sum over $l$ shall be limited to $l_\mathrm{max} \equiv \min(n_\phi/2, n_\theta)$, i.e., the Nyquist frequency of the grid\footnote{This condition simply corresponds to requiring that the function is sampled twice per period.}. In our case, having $n_\phi \times n_\theta = 80 \times 80$, the multipolar expansion ought to be cut at $l_\mathrm{max} = 40$.

\begin{figure*}
	\centering
	{\includegraphics[width=0.24\textwidth]{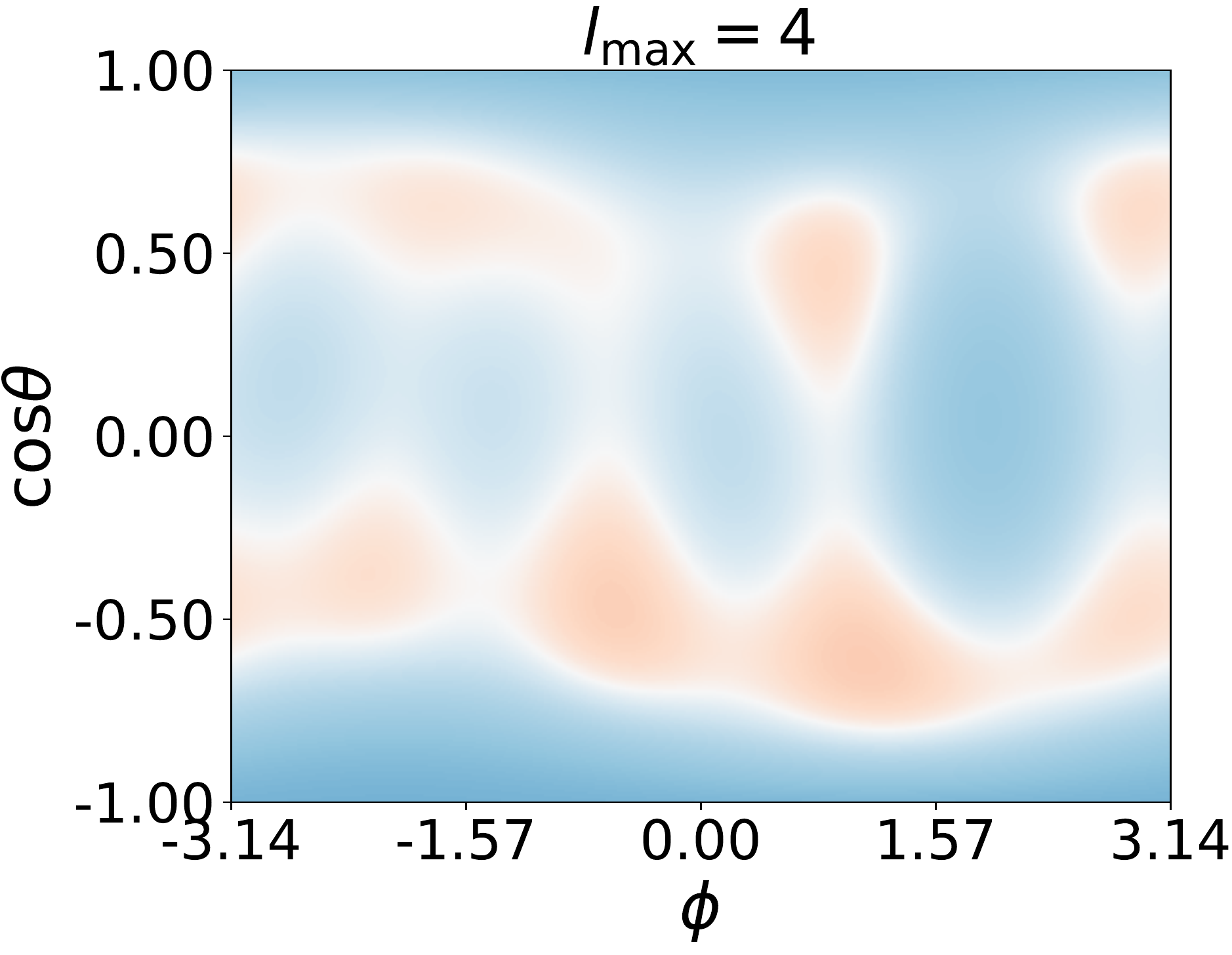}}~
	{\includegraphics[width=0.24\textwidth]{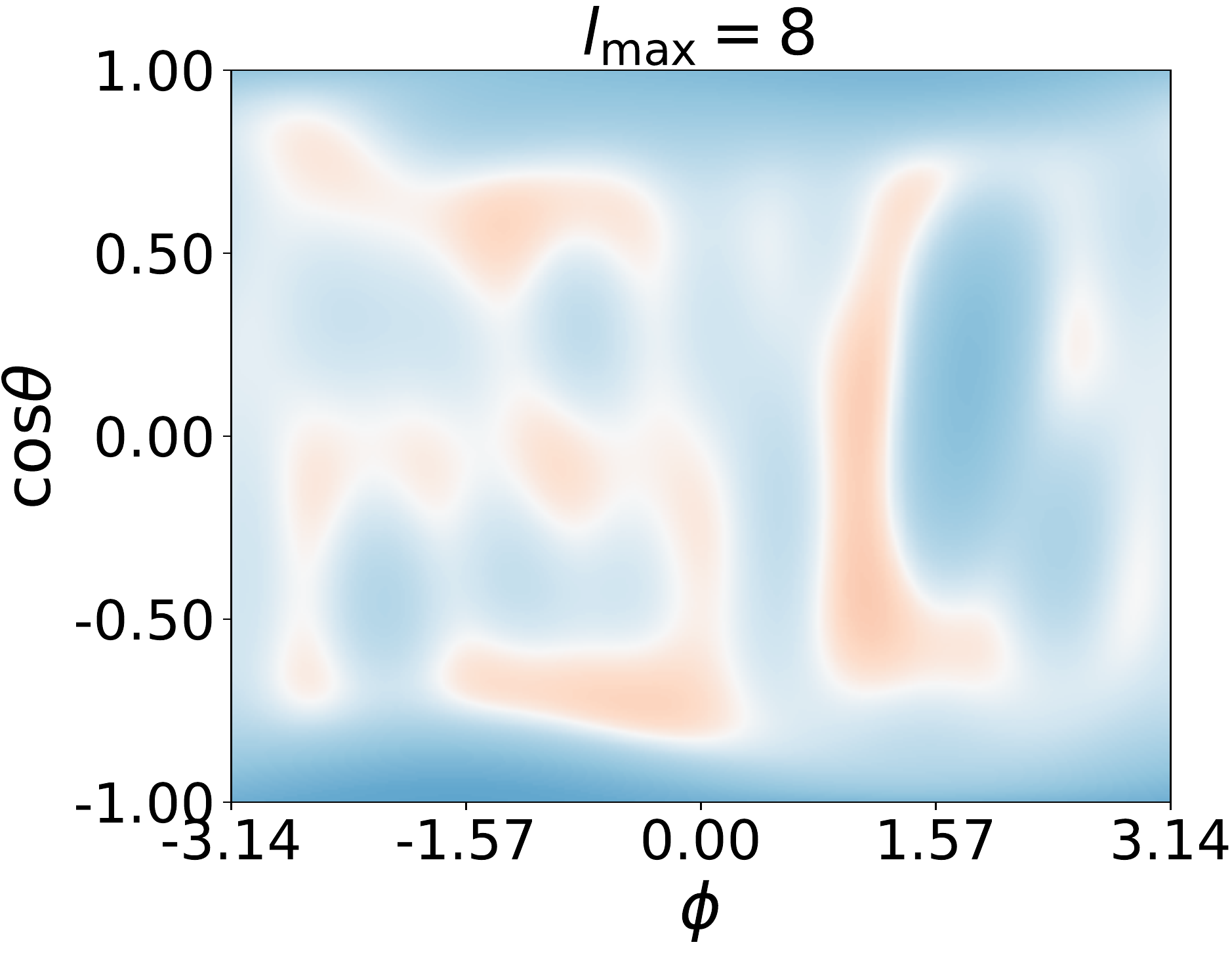}}~
	{\includegraphics[width=0.24\textwidth]{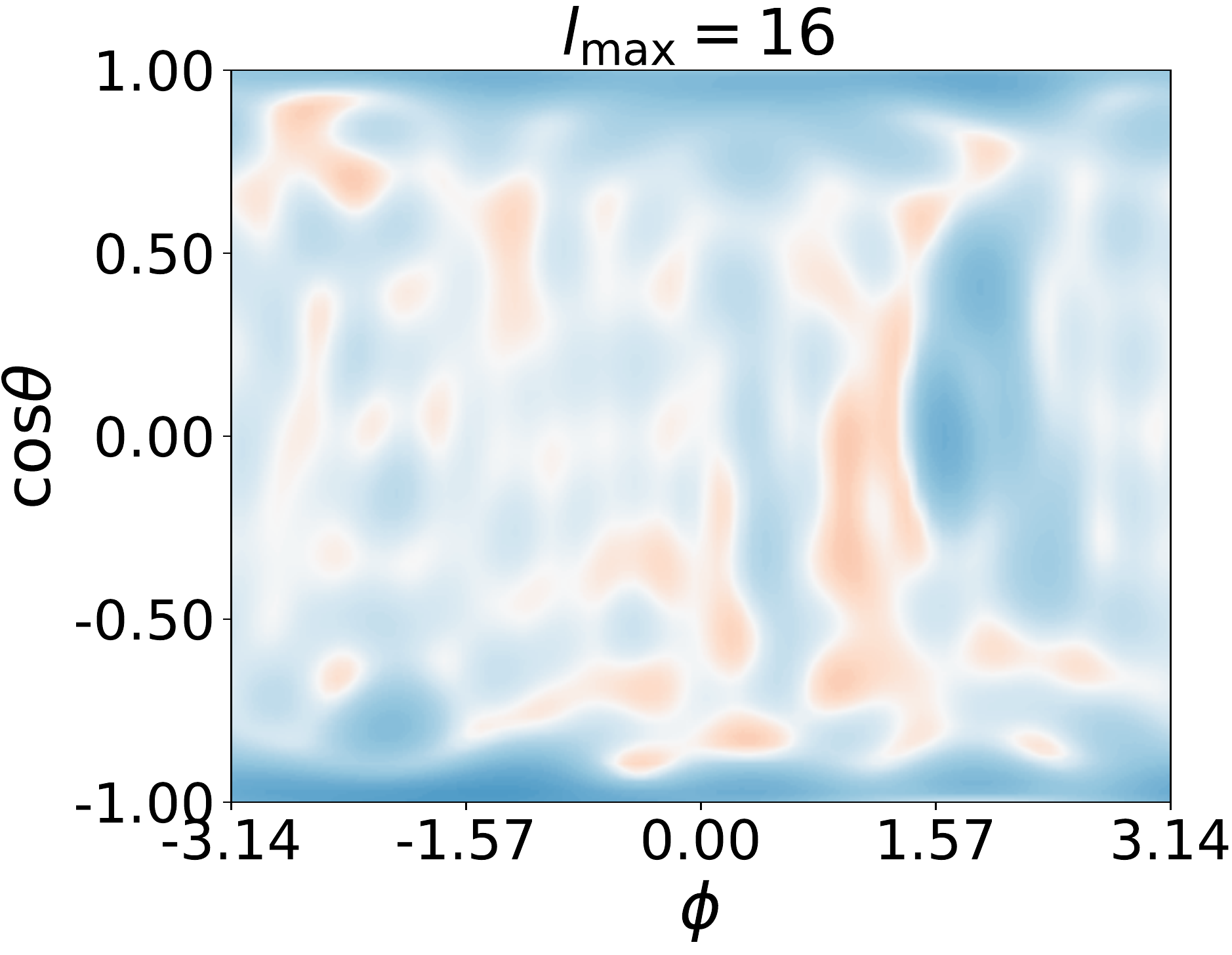}}~
	{\includegraphics[width=0.24\textwidth]{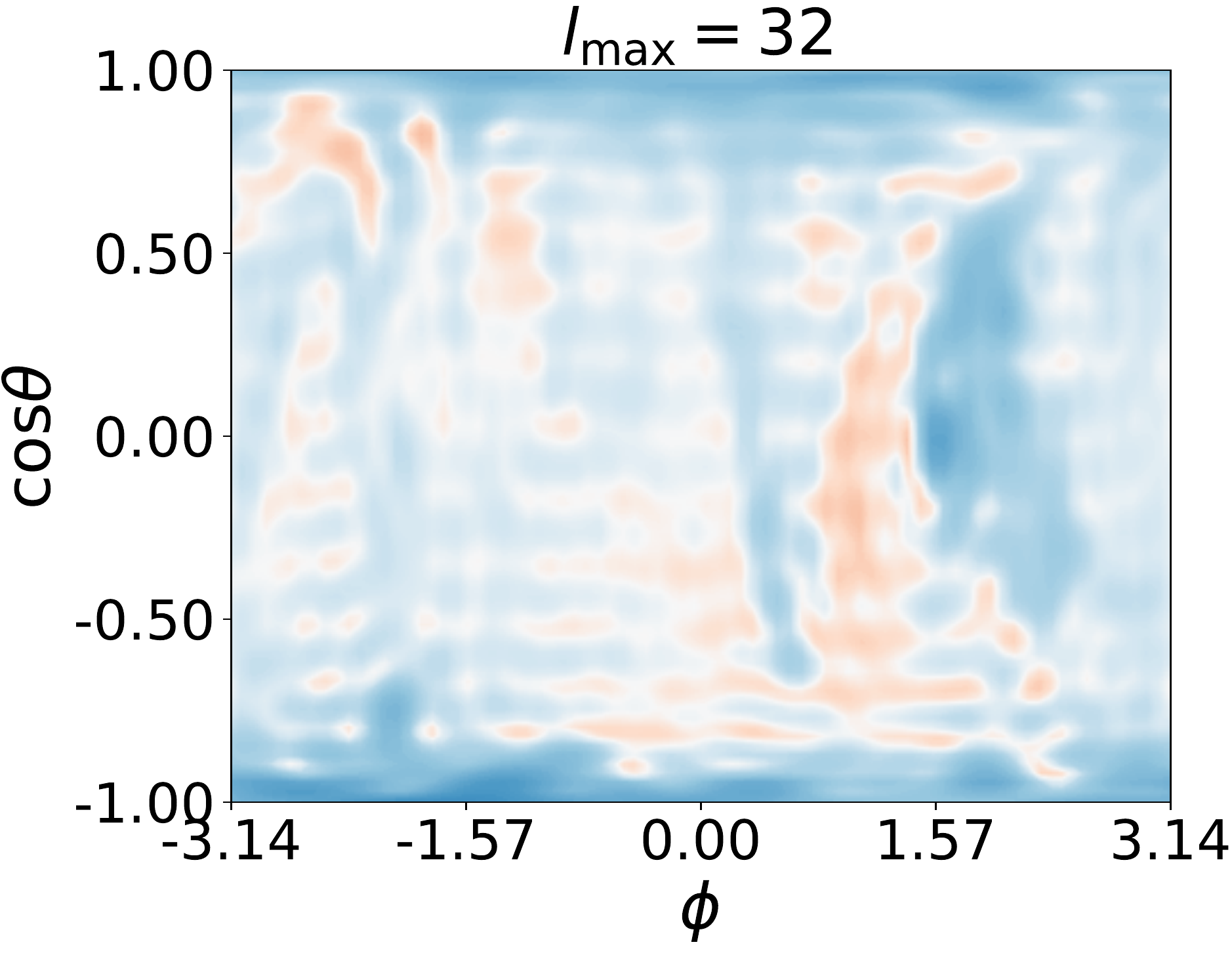}}
	{\includegraphics[width=0.24\textwidth]{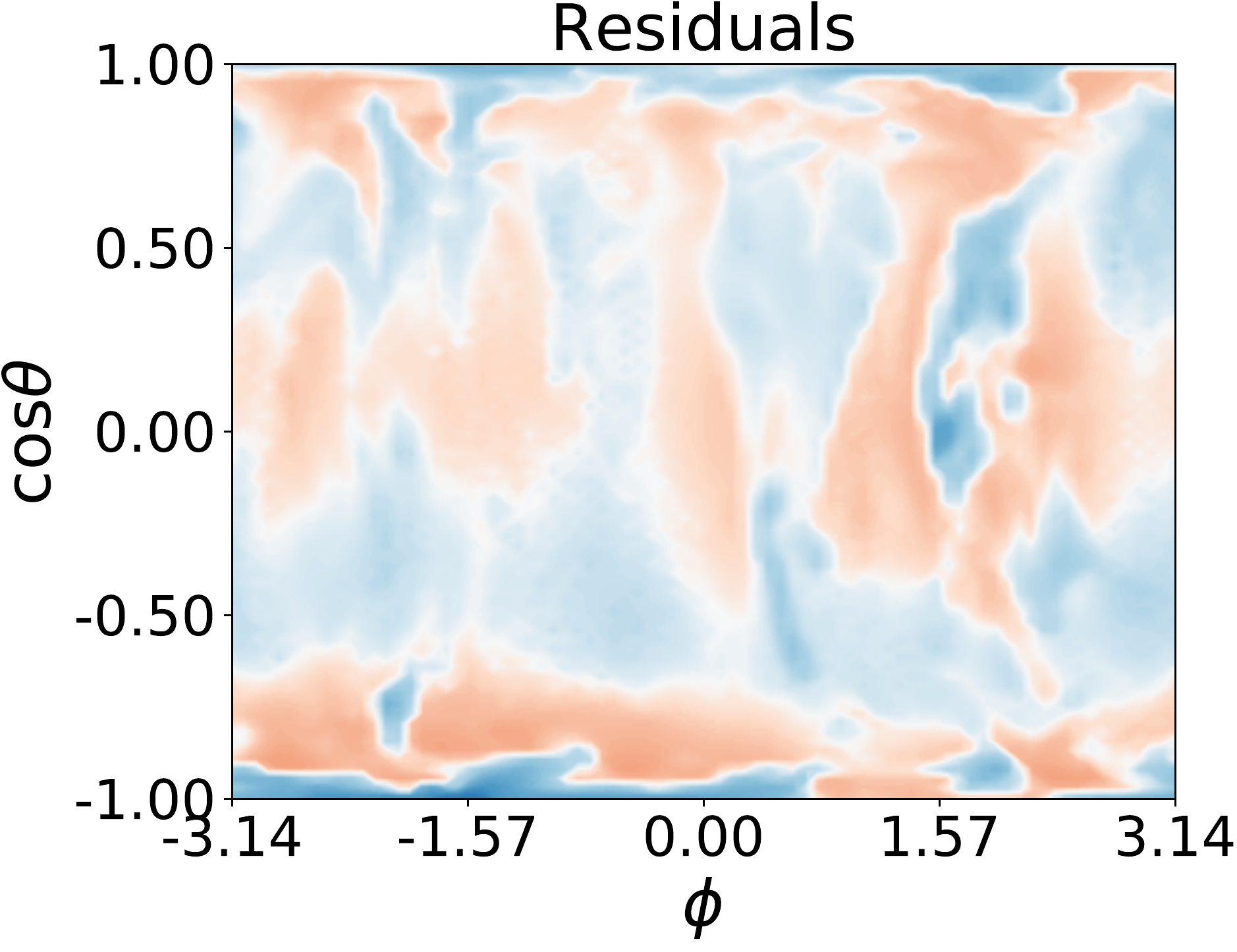}}~
	{\includegraphics[width=0.24\textwidth]{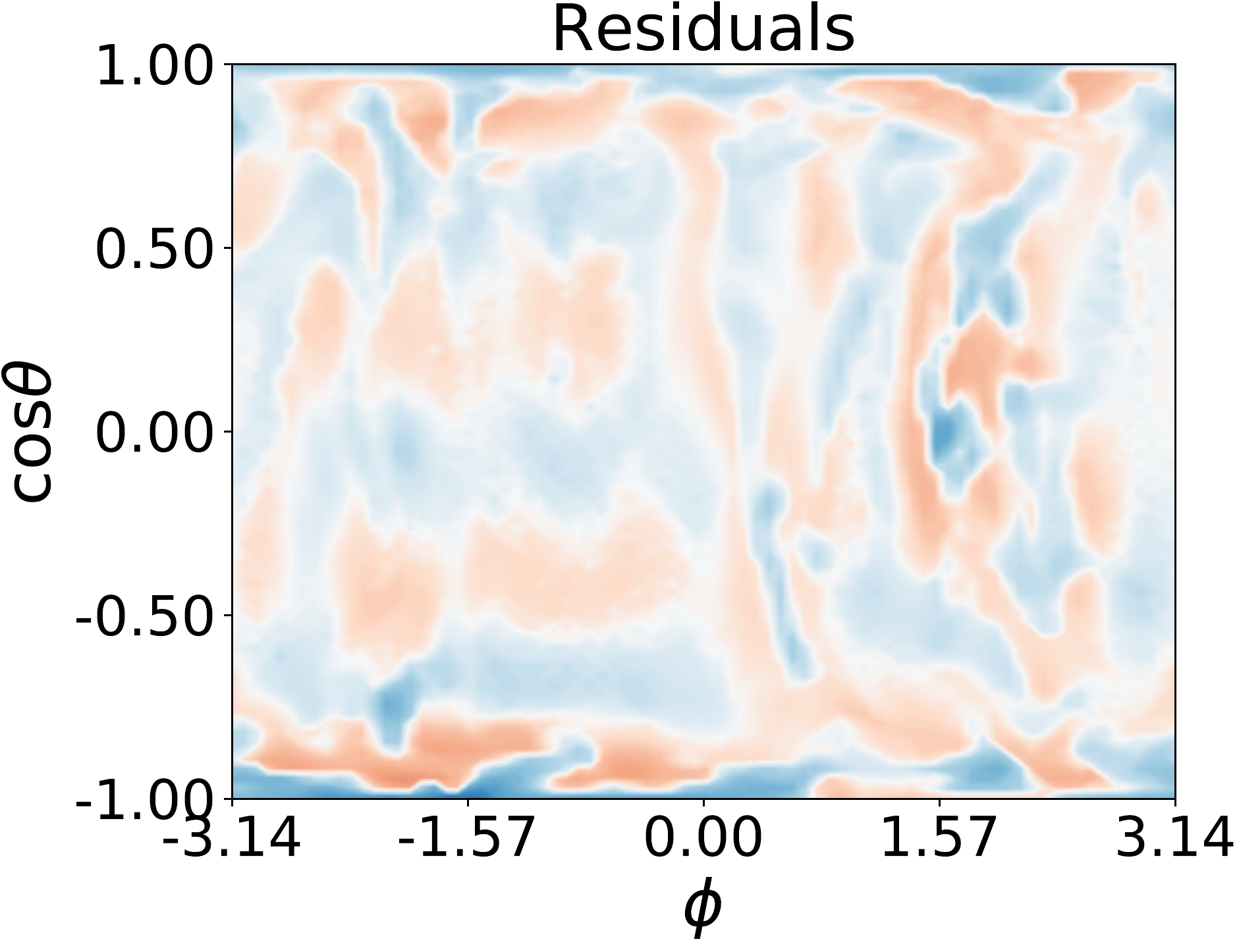}}~
	{\includegraphics[width=0.24\textwidth]{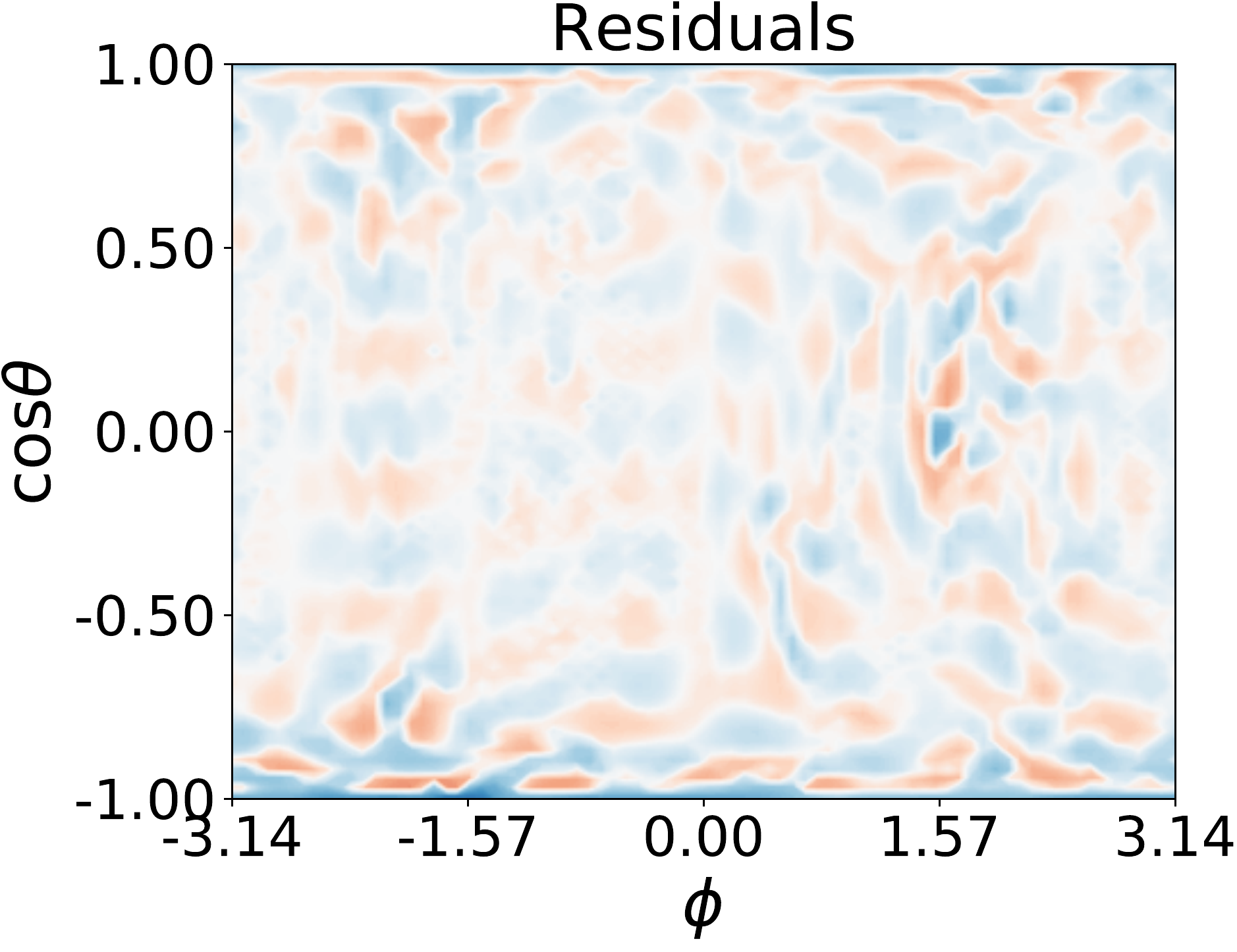}}~
	{\includegraphics[width=0.24\textwidth]{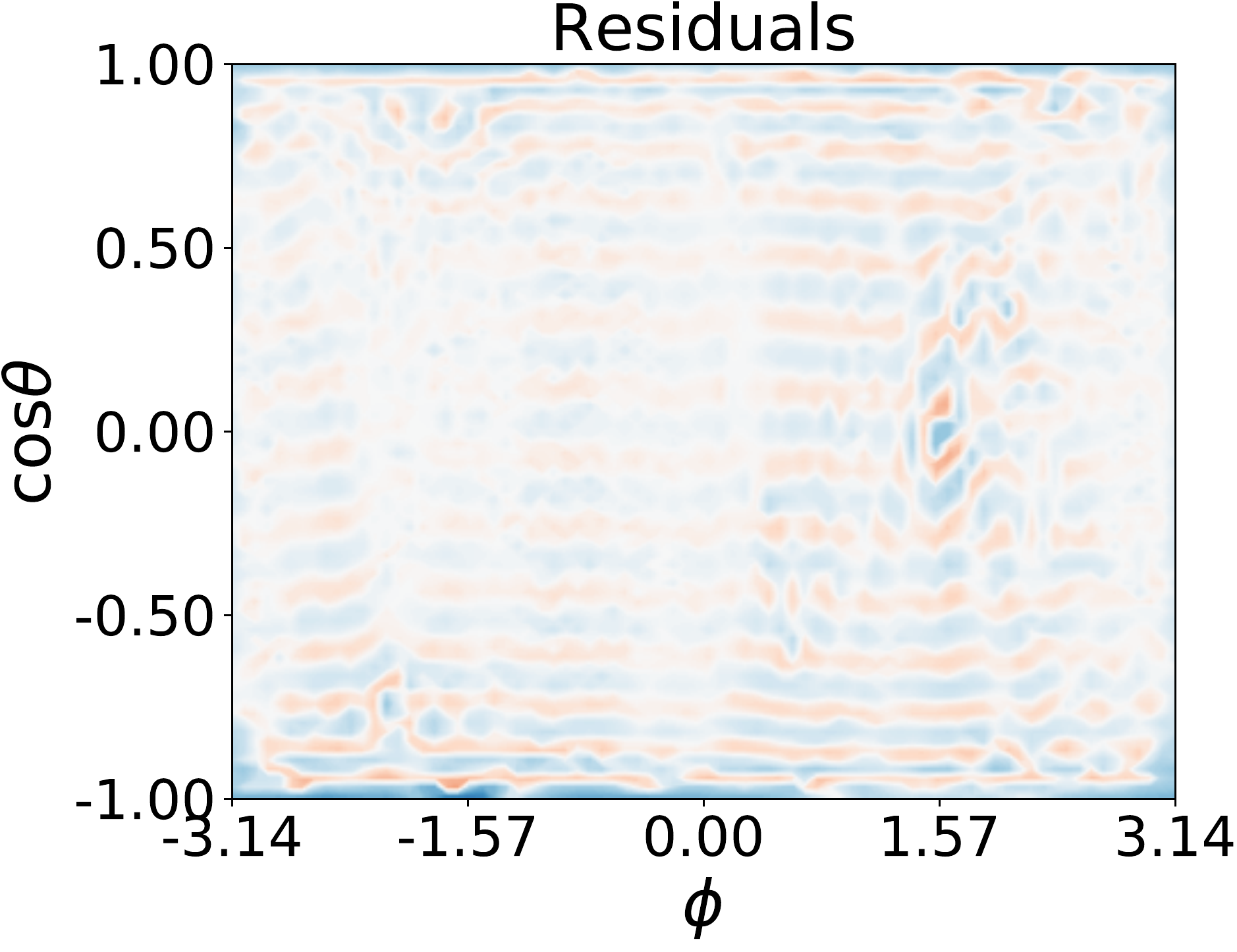}}
	\caption{Reconstruction of the mass flux (corresponding to the left panel of Fig. \ref{fig:angular_massflows}) with increasingly higher degree spherical harmonic components: from left to right, $l_\mathrm{max} = 4, \, 8, \, 16, \, 32$. The upper panels present the flux reconstructed by adding all the terms in the multipolar expansion up to $l=l_\mathrm{max}$. The lower panels show the residuals of this expansion, i.e., the subtraction of the reconstructed flux from the real flux. The colour scale of all panels has been kept the same as in the left panel of Fig. \ref{fig:angular_massflows}.}
	\label{fig:angular_massflows_spharm}
\end{figure*}

Figure \ref{fig:angular_massflows_spharm} shows how higher degree components capture the details of the accretion pattern on increasingly smaller angular scales, using the accretion fluxes in the left panel of Fig. \ref{fig:angular_massflows} as an example. From left to right, columns present the \textit{reconstructed} fluxes in their upper panel, computed by cutting the series at $l_\mathrm{max} = 4, \, 8, \, 16 \; \text{and} \; 32$. The lower panel corresponds to the residuals with respect to the real flux. Even though low order degree spherical harmonics represent the dominant contributions, as it will be seen below, higher degrees need to be reached in order for the reconstructed flux to resemble the real one. We quantify the goodness of the fit by computing the rms relative difference between the real and the reconstructed flux. This quantity steadily decreases with increasing $l_\mathrm{max}$ up to $l_\mathrm{max} \sim 25$, where it stalls as the high-order multipoles involved vary in angular scales comparable to our sampling of the flux. 

\subsubsection{Power spectrum of the angular distribution of accretion fluxes}

In order to extract the contribution of the different angular scales to the accretion flux, we define the \textit{power spectrum of the angular distribution of mass flows} as the average of the square of the $c_{lm}$ coefficients for a given $l$:

\begin{equation}
	P(l) = \frac{1}{2l + 1} \sum_{m=-l}^{l} c_{lm}^2
	\label{eq:power_spectrum}
\end{equation}

\begin{figure*}
	\centering
	{\includegraphics[height=0.25\textheight]{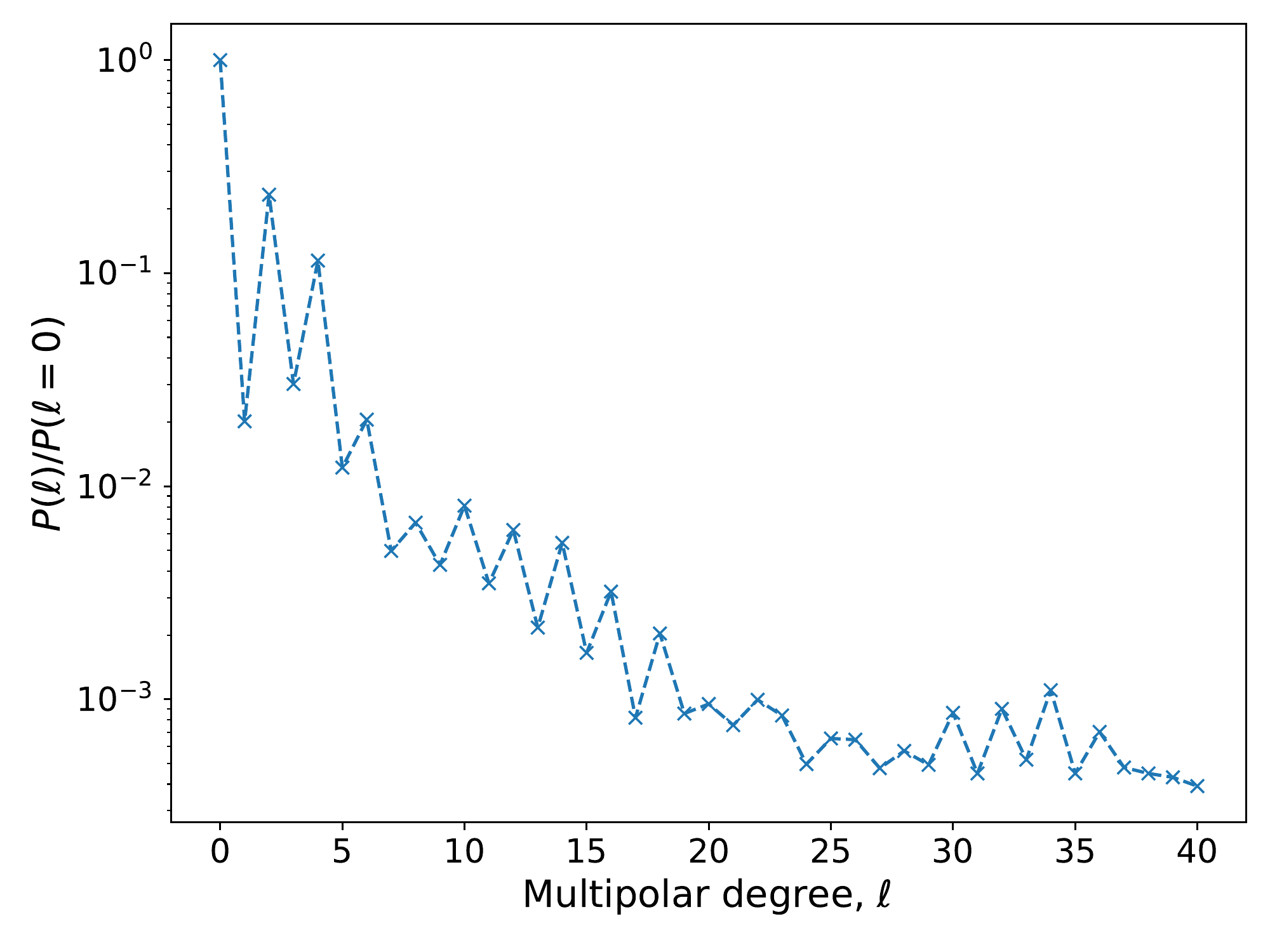}}~
	{\includegraphics[height=0.253\textheight]{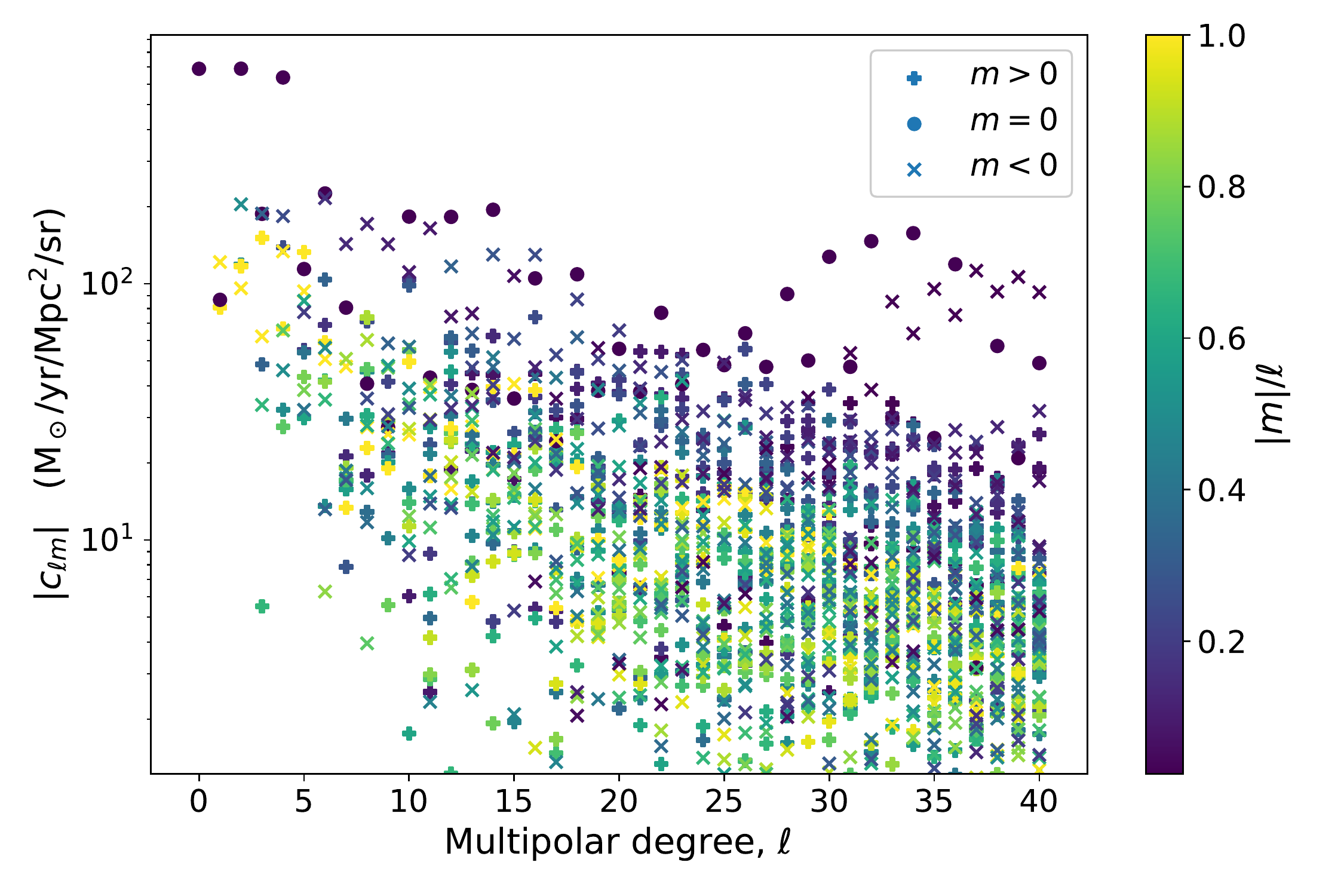}}
	\caption{The left panel presents the power spectrum of the mass flows, $P(l)$, normalised to the value corresponding to the monopolar order, $P(l=0)$. The right panel contains the absolute value of the individual $c_{lm}$ coefficients, as a function of the multipolar degree $l$. The spherical harmonic order, $m$, is encoded in the colour scale as $|m|/l$, with the sign given by the shape in the legend. Both representations have been computed for the accretion fluxes of CL01 at $z \simeq 0.81$ through $R_\mathrm{vir}$, corresponding to the accretion map on the left panel of Fig. \ref{fig:angular_massflows}.}
	\label{fig:angular_powerspectrum}
\end{figure*}

This quantity encodes the magnitude of the contribution of components varying on angular scales $\sim 180^\mathrm{o} / l$ to the flux. The left panel in Figure \ref{fig:angular_powerspectrum} exemplifies this by showing this quantity computed for the same mass flux map in the left panel of Fig. \ref{fig:angular_massflows}.

The overall behaviour of the power spectrum, decreasing with $l$, reveals that the components corresponding to fluctuations on large angular scales (i.e., low $l$; especially $l = 0, \, 2 \, \text{and} \, 4$) are dominant, while the small scale components only provide small corrections. Consequently, the overall accretion and deccretion patterns can be described, to a great extent, by looking at these simpler, main contributions. Interestingly, the power spectrum shows a clear odd-even effect: even $l$ components are systematically larger than odd $l$ components, implying that parity-even modes are responsible for the bulk of the gas flows through the $R_\mathrm{vir}$ boundary at this particular code output. Note, however, that although this effect is sustained in time and can be easily understood for the lower degree components ($l = 0 \, \text{and} \, 2$, which are further discussed below), the same is not true for higher degree components, where this effect is not so clear through all iterations.

On the right panel of Fig. \ref{fig:angular_powerspectrum}, the magnitudes of the $c_{lm}$ coefficients (for the same system and code output) are represented, offering complementary information to the power spectrum (which only gives the average over the orientations of the different multipoles, $m$, for a given $l$). In the figure, the horizontal axis encodes the degree, $l$, while the colour scale is used to indicate the order, $m$, of the real spherical harmonic. In this particular case, three $m=0$ components completely dominate the mass flux: namely, $l=0, \, 2, \, \text{and} \, 4$. Their relative weights undergo significant evolution during the redshift interval of this study, $1.5 \leq z \leq 0$. Let us further describe the lowest order components:

\subparagraph{The monopolar, isotropic, or smooth component ($l=0$).} This spherically symmetric contribution represents the average isotropic mass flux across the $r = R_\mathrm{bdry}$ boundary and, hence, can be interpreted as a MAR estimate. In Sec. \ref{s:results_angular.results.monopolar}, we explore the possibility of using it as a MAR proxy and compare it to $\Gamma_{200m}$.

\subparagraph{The dipolar components ($l=1$), or \textit{headwind}.} These multipoles, which account for an excess of accretion flows on a hemisphere and a defect in the antipodal one, can be visualised as the resulting mass flow pattern of an object moving through a dense medium. Although we do not explore it in this manuscript, its usage appears to be promising in the exploration of phenomena like ram pressure stripping \citep{Quilis_2000, Quilis_2017}.

\subparagraph{The aligned quadrupole or filamentary component ($l=2$, $m=0$).} This component represents a strong mass inflow through the poles (i.e., through the directions of the major axis) and an outflow near the equatorial regions. Thus, it is the natural candidate to account for the contribution of the mass flux through the cosmic filaments that connect a cluster with a near, massive neighbour.

\subsubsection{The monopolar component as a MAR estimate}
\label{s:results_angular.results.monopolar}

\begin{figure}
	\centering
	{\includegraphics[width=0.5\textwidth]{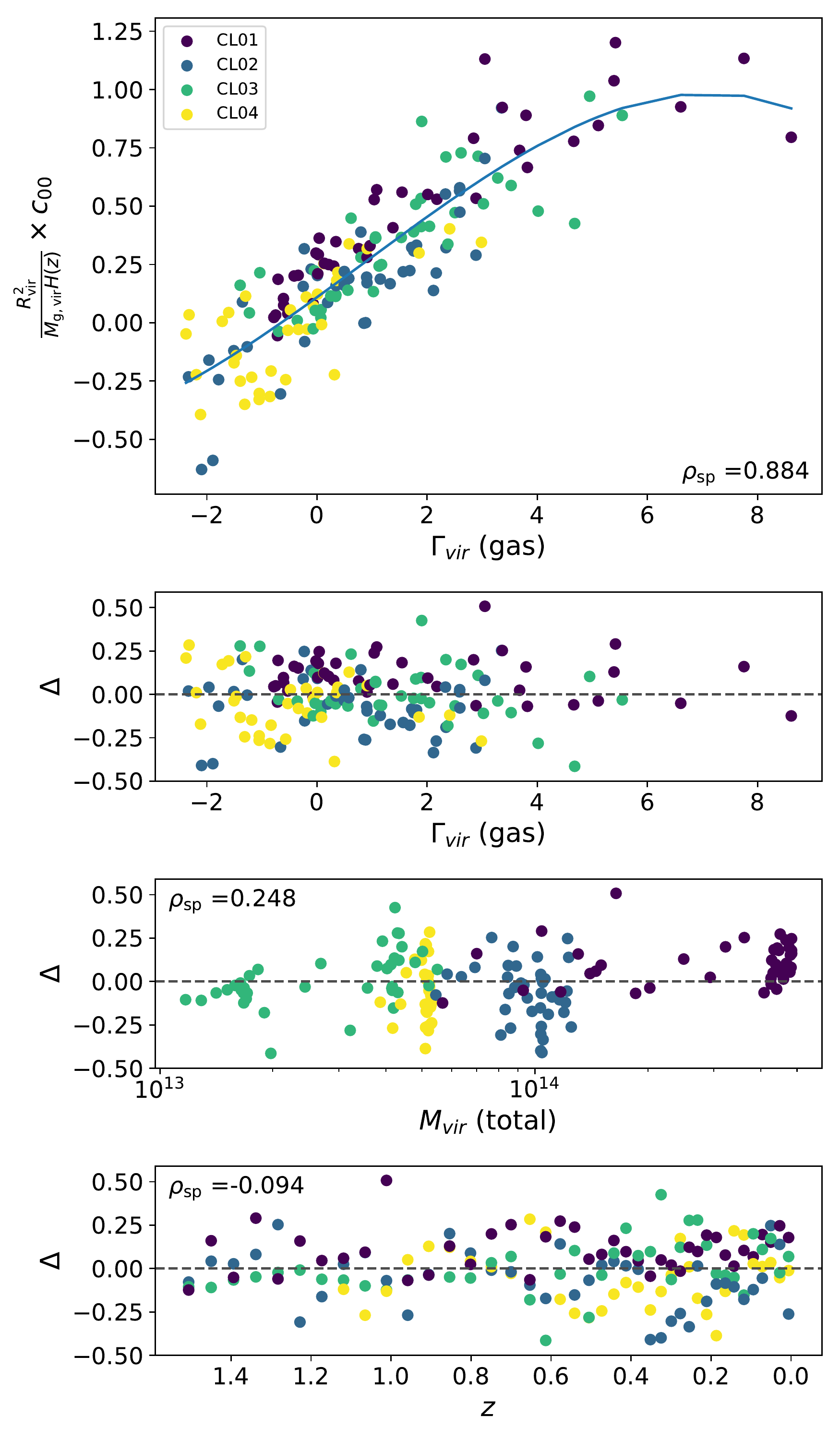}}
	\caption{Relation between the instantaneous MAR proxies $\Gamma_{200m}$ and $c_{00}$, combining all the snapshots in $1.5 \geq z \geq 0$. The latter has been normalised to cancel the expected dependences on redshift and mass. The upper panel contains the scatter plot of both variables and a nonparametric fit using smoothing splines (blue line). The residuals of this fit, $\Delta$, are used in the lower panels to assess whether there is still a redshift or mass dependence in this relation.}
	\label{fig:c00_vs_gamma}
\end{figure}

As already introduced, the monopolar coefficient, $c_{00}$, is a measurement of the angularly-averaged mass flux, and can thus be used in order to quantify the MAR of a cluster. In order to check to which extent does $c_{00}$ agree with other MAR estimates, Figure \ref{fig:c00_vs_gamma} presents the analyses of the correlation and the residual dependences between $c_{00}$ and $\Gamma_{200m}$, in the same way we have done for $\alpha_{200m}$ in Sec. \ref{s:results_mar.velocity_profiles}. Note, however, that $c_{00}$ is a measure of the isotropic mass flux, while $\Gamma_\mathrm{vir}$ quantifies the logarithmic increase in the enclosed mass. In order to provide a comparison where no residual dependences with the mass or redshift are expected, the vertical axes in Fig. \ref{fig:c00_vs_gamma} show $\frac{R_\mathrm{vir}^2}{M_\mathrm{g,vir} H(z)} \times c_{00}$, instead of just $c_{00}$. This normalisation can be understood from the following reasoning:

The mass accretion rate, $\dot M \equiv \dv{M}{t}$, can be got from the mass flux by integrating it over the surface of a sphere of radius $R_\mathrm{bdry}$. Since all the non-monopolar spherical harmonics average to zero, introducing the multipolar expansion of $j_M(\theta, \phi)$, Eq. (\ref{eq:multipolar_expansion}), into the surface integral allows to write the mass accretion rate as $\dot M = \sqrt{4\pi} R^2 c_{00}$. The same quantity can be obtained from $\Gamma_\Delta$, as $\dot M \equiv \dv{M}{t} = \dv{M}{\log M} \dv{\log M}{\log a} \dv{\log a}{t} = M \Gamma H(z)$. Therefore, $R^2 c_{00}$ scales as $M \Gamma H(z)$, justifying our choice to compare $\Gamma_\mathrm{vir}^\mathrm{gas}$ with $\frac{R_\mathrm{vir}^2}{M_\mathrm{g,vir} H(z)} \times c_{00}$. Note that, additionally, this normalisation cancels out the dimensions of $c_{00}$.

The upper panel in Fig. \ref{fig:c00_vs_gamma} shows that, even though there is considerable scatter, these two MAR proxies are tightly correlated ($\rho_\mathrm{sp} = 0.884$). With this sample, the relation appears to stall at high values of $\Gamma_\mathrm{vir}^\mathrm{gas}$. However, the reduced number of observations in this high-accretion regime prevents us to draw any robust conclusion to this respect. The lower panels present the distribution of residuals with respect to the independent variable, $\Gamma_\mathrm{vir}^\mathrm{gas}$, mass and redshift. The residuals appear to be uncorrelated to redshift, suggesting that the normalisation applied to $c_{00}$ cancels the redshift evolution of the $\Gamma - c_{00}$ relation. Likewise, the residual dependence with the mass, if any, is weak ($\rho_\mathrm{sp} = 0.248$).

\subsubsection{Relative weight of the smooth and the filamentary contributions}
Aiming to assess the relative weight of the filamentary (aligned quadrupolar) and isotropic (monopolar) contributions, we define the following parameter:

\begin{equation}
	\beta \equiv \frac{c_{20}}{|c_{00}| + |c_{20}|}
	\label{eq:defbeta}
\end{equation}

Note that $\beta$ is valued in the interval $[-1, \, 1]$. Positive values imply that the aligned quadrupole contributes to increase the gas mass flux in the regions close to the major semiaxis, corresponding to the gas infalling through the cosmic filaments, while the unlikely scenario in which $\beta < 0$ would correspond to a decreased gas mass flux through these regions. If these two components have comparable weights, $|c_{20}| \sim |c_{00}|$, then $|\beta| \sim 1/2$. Likewise, $|\beta| \sim 1$ ($\beta \sim 0$) indicates that the filamentary (smooth) component is the dominant contribution.

\begin{figure}
	\centering
	{\includegraphics[width=0.5\textwidth]{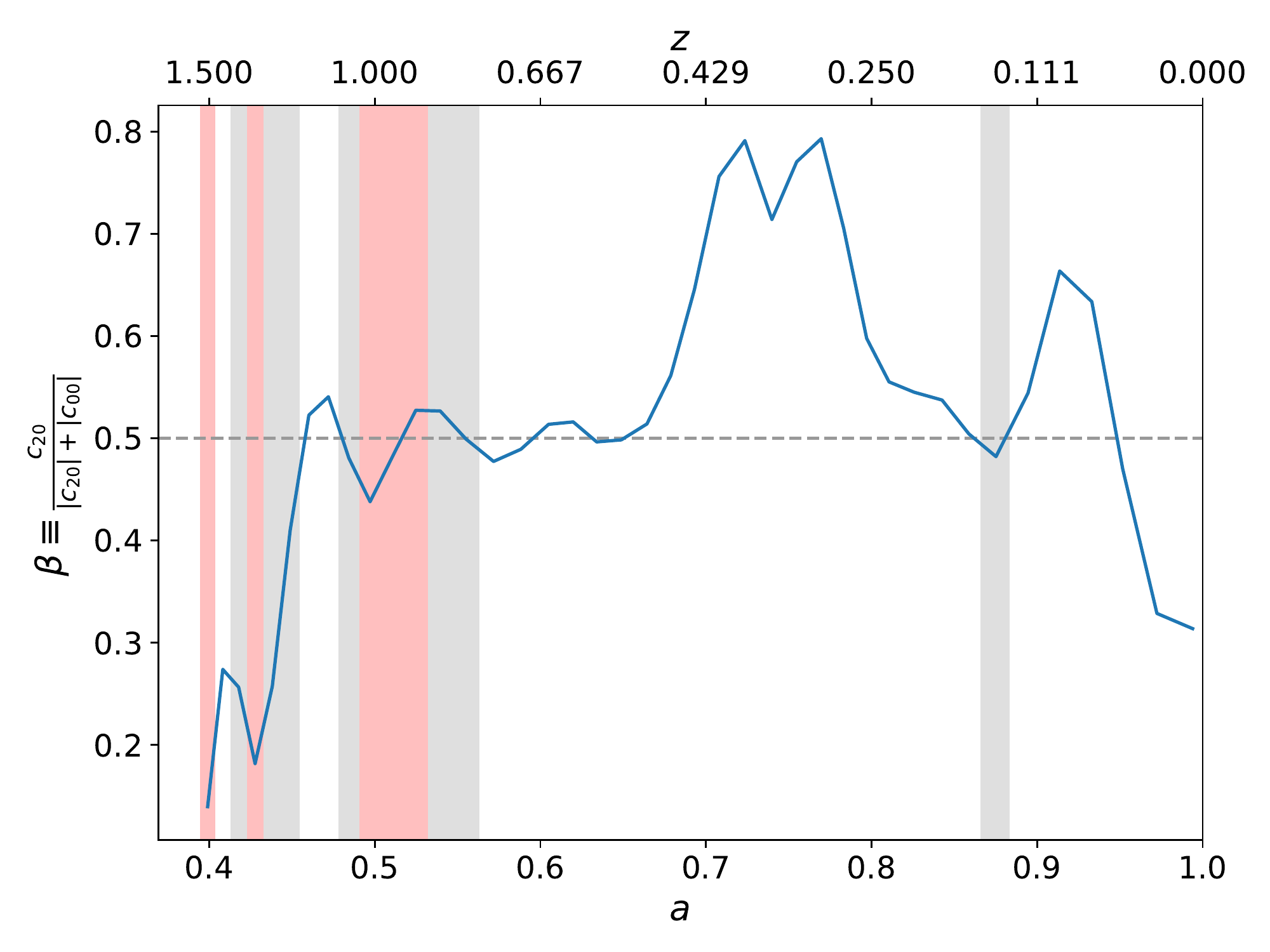}}
	\caption{Evolution of the parameter $\beta = \frac{c_{20}}{|c_{00}| + |c_{20}|}$, measuring the relative importance of the filamentary component of the gas mass inflows, for the cluster CL01. The gray dashed line marks the value $\beta = 1/2$. The gray and red shaded regions indicate the accretion regimes, as in Fig. \ref{fig:mergers}.}
	\label{fig:quadrupolar_to_monopolar}
\end{figure}

As an example, Figure \ref{fig:quadrupolar_to_monopolar} presents the evolution of this quantity for the cluster CL01. The value of $\beta$ fluctuates around the value of $\beta = 1/2$, which separates the filamentary-dominated or smooth-dominated accretion regimes. These values of $\beta$ do not seem to correlate with the merging regimes studied in Sec. \ref{s:results_mar.mergers_surroundings}. Note, however, that this result has been obtained for a particular cluster, and the method should be applied to a whole sample of massive clusters in order to draw any statistically significant conclusions.

\subsection{Thermodynamical properties of the accreted gas}
\label{s:results_angular.thermo}
The method described in Sec. \ref{s:results_angular.method} can also be applied to measure any property of the accreted and deccreted gas, such as its temperature and entropy. For each of the $n_\phi \times n_\theta$ solid angle bins, its temperature and entropy are computed as the mass-weighted average of such quantities over all the entering cells assigned to the bin. Note, however, that the angular regions where no gas is inflowing cannot be assigned a temperature nor entropy, resulting in `gaps' in the corresponding maps. 

With the purpose of exploring the thermodynamic differences between the gas accreted from the smooth component and from the filamentary component, we have assigned a temperature and an entropy to each multipolar component as 

\begin{equation}
	X_{lm} \equiv \frac{\int_{\tilde \Omega} \dd \Omega X(\theta, \phi) \mathcal{Y}_{lm}(\theta, \phi)}{\int_{\tilde \Omega} \dd \Omega \mathcal{Y}_{lm}(\theta, \phi)}
	\label{eq:multipolar_component_temperature}
\end{equation}

\noindent where $X$ represents either the temperature or the entropy, and the integration domain $\tilde \Omega$ is the region where $\mathcal{Y}_{lm}(\theta, \phi) > 0$ (and thus the correspondent component represents infall of matter) and $X(\theta, \phi)$ is defined (i.e., at least one gas cell has been marked as entering through the bin and therefore its temperature and entropy can be defined).

\begin{figure*}
	\centering
	{\includegraphics[width = 0.33 \textwidth]{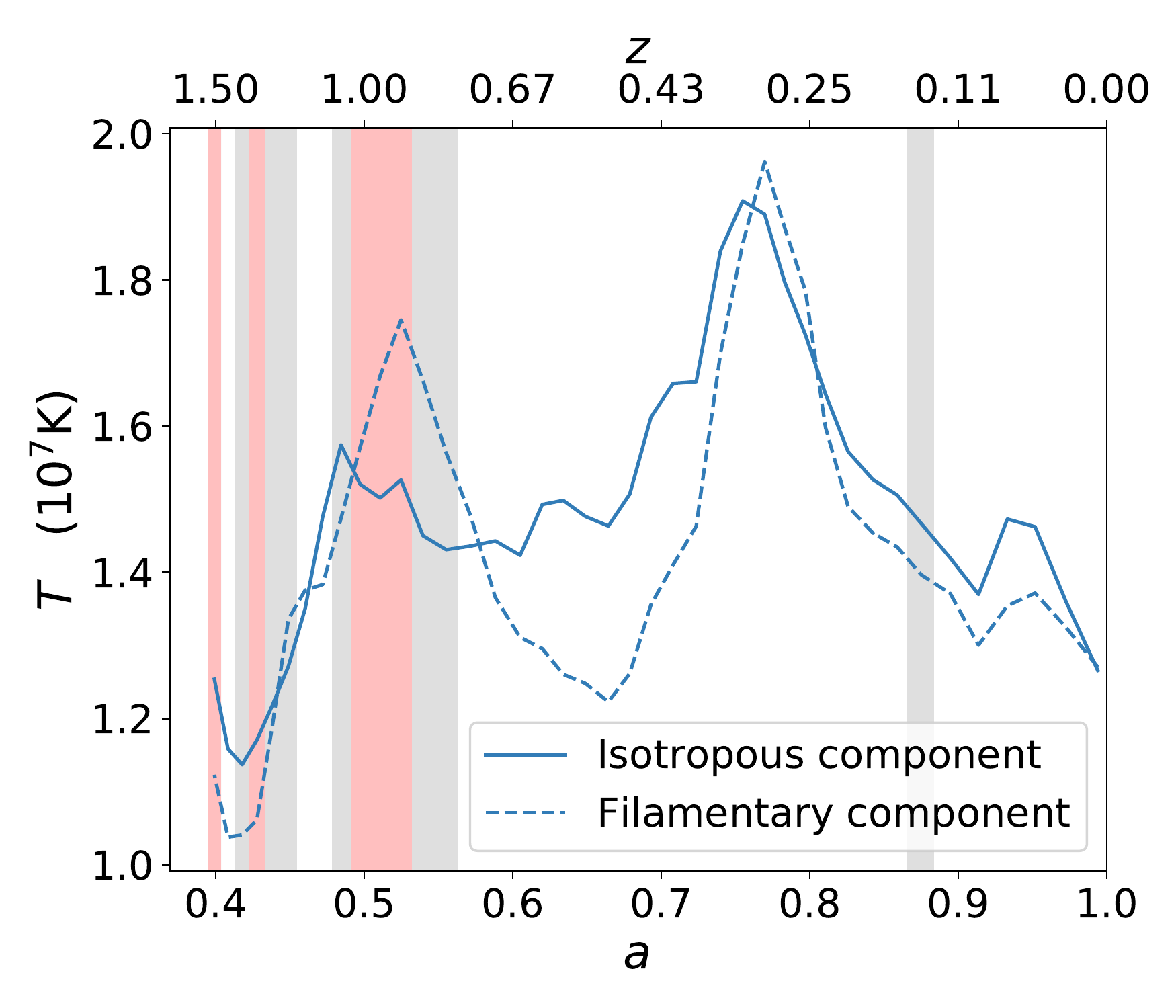}}~
	{\includegraphics[width = 0.33 \textwidth]{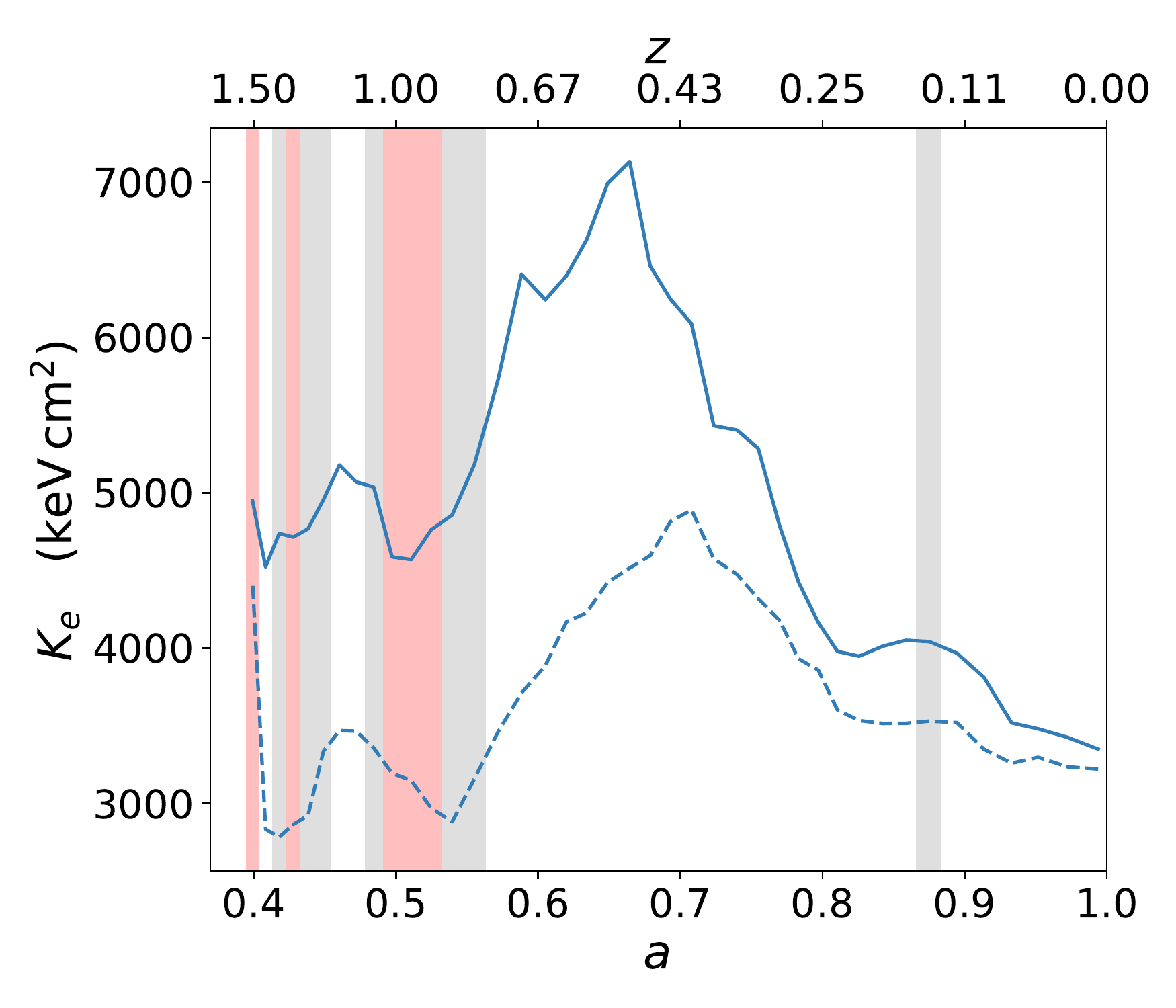}}~
	{\includegraphics[width = 0.33 \textwidth]{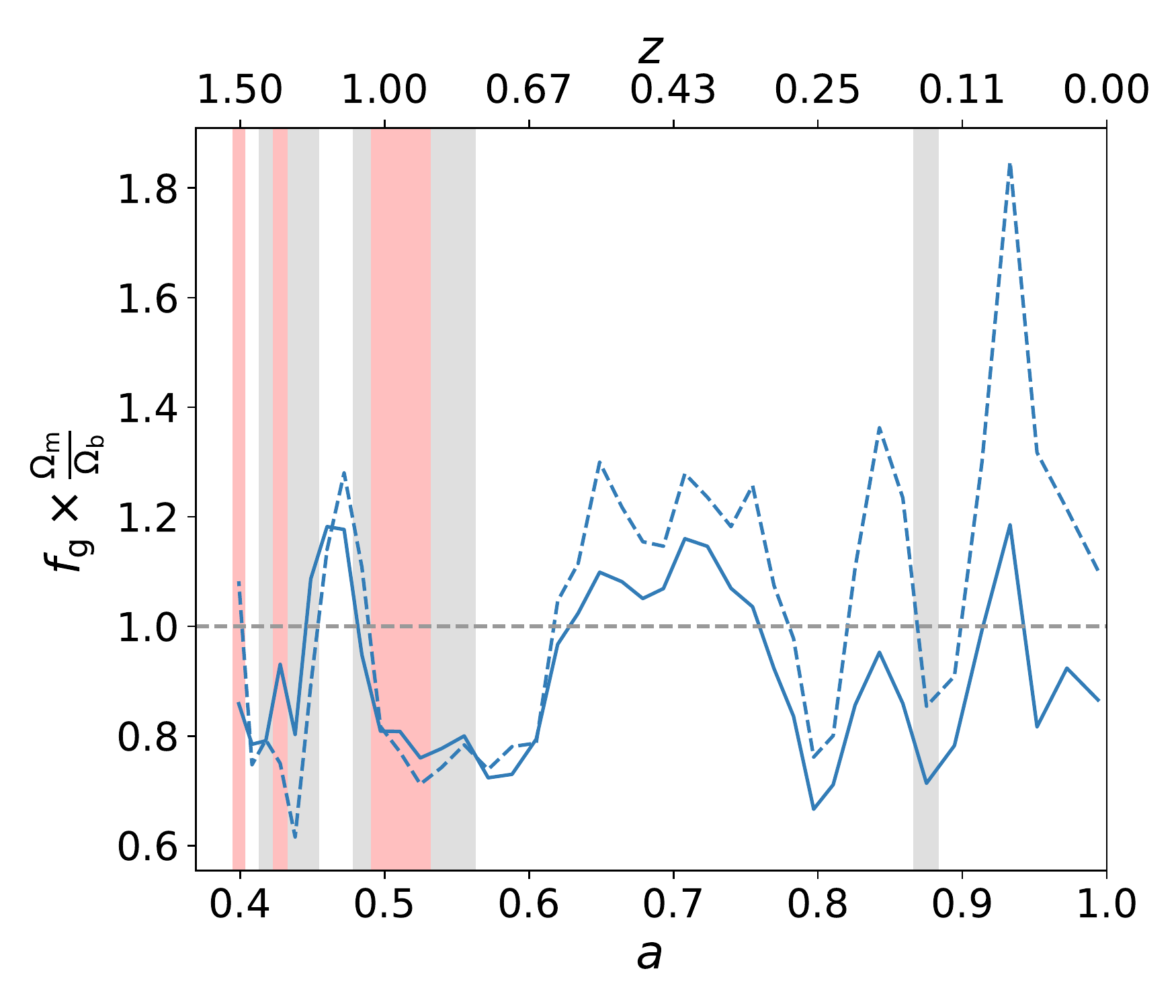}}
	\caption{The panels show different thermodynamical properties of the smooth or monopolar component (solid line) and the filamentary or aligned quadrupolar component (dashed line) of the mass flows. From left to right, panels show the temperature, the entropy and the gas fraction of the material being accreted to the cluster. Entropy displays the strongest differential behaviour, with the filamentary component having lower entropy. Shaded regions indicate the merging regimes as in Fig. \ref{fig:mergers}.}
	\label{fig:thermo_evolution}
\end{figure*}

The left and central panels in Figure \ref{fig:thermo_evolution} present, respectively, the evolution of gas temperature and gas entropy for cluster CL01. In all three panels in the figure, continuous lines represent the evolution of the magnitudes for the smooth component, while the dashed lines correspond to the component infalling through the filaments.

While temperatures of the monopolar and the quadrupolar components do not exhibit a clear distinctive behaviour, their entropies do. Gas being accreted through cosmic filaments tends to have systematically lower entropy than smoothly accreted gas, mainly as a consequence of its higher density. These effects appear to be larger at higher redshifts, while accretion rates are still high for this object (see Fig. \ref{fig:mergers}). The evolution of temperature or entropy of both accreting components does not seem to display a clear relation to merger regimes. However, in order to draw robust conclusions from the analyses of the evolution of thermal properties of the gas accreted smoothly and through filaments, larger samples of massive clusters need to be analysed.

As well as gas mass flows, the method described in Sec. \ref{s:results_angular.method} can also be applied to measure the angular distribution of the DM mass flux. The procedure is completely analogous to the one covered for the gas. From the gas and DM mass fluxes, the gas fraction of the accreting material of a given multipolar component, ($l, \, m$), is computed as:

\begin{equation}
	\left(f_g \right)_{lm} \equiv \frac{\int_{\tilde \Omega} \dd \Omega j_M(\theta, \phi) \mathcal{Y}_{lm}(\theta, \phi)}{\int_{\tilde \Omega} \dd \Omega \left[ j_M(\theta, \phi) + j_M^\mathrm{DM}(\theta, \phi) \right] \mathcal{Y}_{lm}(\theta, \phi)}.
	\label{eq:multipolar_component_gasfraction}
\end{equation}

The right panel of Fig. \ref{fig:thermo_evolution} shows the gas fractions of the matter infalling isotropically and through filaments. These fractions have been normalised to the cosmic baryon fraction (i.e., the gray dashed line in the plot). The gas fraction of both components fluctuates around the cosmic fraction, being slightly lower during the merger events (due to the collisional nature of gas). The two components show very similar qualitative behaviour, with the gas fractions of matter falling through the cosmic filaments slightly increased with respect to the matter being smoothly accreted. However, these differences are kept small and more statistics are needed in order to draw general conclusions.

\section{Discussion and conclusions}
\label{s:conclusion}

In this paper, we have analysed the results of an AMR hydrodynamical coupled to $N$-Body cosmological simulation of a small-volume domain, containing a central, massive galaxy cluster and several smaller systems. The main focus of this work has been placed on the quantification of the accretion phenomena, with a special focus on the dynamics of the gaseous component. The dynamical scenario depicted by the simulation is essentially dominated by the infall of matter onto the central cluster, which is also the best numerically resolved object. Consequently, most of the focus has been placed on this object, especially in Sec. \ref{s:results_angular}. Nevertheless, less-massive systems have turned out to be useful in assessing the differences in the MAHs of massive and less-massive clusters.

\subsection{Accretion rates}

Through these pages, several proxies for the MAR have been compared (see Figs. \ref{fig:alpha_vs_gamma} and \ref{fig:c00_vs_gamma}). The most widely adopted MAR proxy in numerical works, $\Gamma_\Delta^{[a_1, a_0]}$ \citep{Diemer_2014}, is difficult to relate to an actual observable, since it corresponds to the average accretion rate over several $\mathrm{Gyr}$ for typical choices of $a_1$. The instantaneous MAR proxies $\alpha_\Delta$ \citep{Lau_2015} and $\Gamma(a)$ (Eq. \ref{eq:mar_derivative}), however, could in principle be inferred from observations, and can moreover be determined for clusters at any redshift. 

\begin{itemize}
	\item As long as galaxies and gas bulk motions trace DM's, $\alpha_\Delta$ could be estimated, in the optical band, from the radial velocities of infalling galaxies in clusters' outskirts, or from the radial velocities of gas (from X-ray data, e.g. \citealp{Sanders_2020}; or from kinetic SZ observations, e.g. \citealp{Adam_2017}; see also \citealp{Simionescu_2019} for a recent review on the topic). In Sec. \ref{s:results_mar.velocity_profiles}, we have seen that gas systematically infalls with smaller radial velocities than DM, in consistency with \cite{Lau_2015}, who have computed the averaged radial velocity profiles for a sample of clusters. However, when looking at a specific cluster extracted from our simulation (as we have shown in Fig. \ref{fig:velocity_profile_CL01}), we see that the pattern for a single cluster can be fairly more complex than this average behaviour, with gas accreting at larger velocities than DM in some regions. Thus, even though the aforementioned behaviour generally holds, the details of the dynamics in the outskirts of each individual cluster introduces important uncertainties, which make unclear to which extent DM velocities can be inferred from gas or galactic velocities in particular objects, and consequently $\alpha_\Delta$ can be faithfully measured in observations. 
	\item $\Gamma(a)$ cannot be directly measured in observations, since it is defined as a rate of change of the mass. However, in Sec. \ref{s:results_mar.mergers_surroundings.correlation}, we have obtained a tantalising correlation between this quantity and the densities in the $1 \leq R/R_{200m} \leq 1.5$ in some massive clusters, suggesting that it might be possible to estimate $\Gamma(a)$ by probing the densities in these regions. However, more statistics are needed to confirm this correlation and to explore the possibility to relate these variables. When computing $\Gamma(a)$ in simulations, it is worth noting that, given that sampling the MAH of clusters introduces statistical noise, the particular differentiation scheme can bias the results: for instance, a gaussian smoothing of the numerical derivatives leads to overly flattened peaks. This effect can be prevented by using Savitzky-Golay filters.
\end{itemize}

When comparing $\alpha_{200m}$ to $\Gamma_{200m}^\mathrm{total}(a)$, we have found that these two quantities display a relatively tight anticorrelation (see Sec. \ref{s:results_mar.velocity_profiles.compare}). If gas velocities are used instead of dark matter velocities, as suggested above, the Spearman's rank correlation coefficient drops from $\rho_\mathrm{sp} = -0.832$ to $\rho_\mathrm{sp} = -0.649$, i.e., gas velocities are not always good indicators of dark matter velocities.  

Last, the coefficient $c_{00}$ presented in Sec. \ref{s:results_angular.results.monopolar} is a more direct measurement of the actual mass flux. We have found that, once the redshift and mass dependences are corrected, it correlates strongly to $\Gamma_{200m}^\mathrm{gas}(a)$, although the scatter continues being large. Even though this parameter is not easily derivable from observational data, it can serve in simulations as a reasonable estimate of the actual gas accretion rate, which can be compared to other proxies.

\subsection{Angular distribution of the mass flows}

In Sec. \ref{s:results_angular}, we have presented a novel and general approach to study the accretion phenomena in simulations, through the determination of the angular distribution of the mass flows through the clusters' spherical boundary. The method presented in Sec. \ref{s:results_angular.method} allows a simple determination of this quantity by treating cells as pseudo-Lagrangian fluid elements. 

The multipolar expansion of the mass flux in the principal axes system of the cluster has been used to extract the elementary contributions to the mass flux. In particular, we have payed special attention to the monopolar and the aligned quadrupolar components, which represent the average isotropic mass flux and the enhancement in the polar regions due to the preferential infall of mass through the cosmic filaments. We have defined a parameter, $\beta$, which quantifies the importance of the filamentary component of accretion, with respect to the isotropic component. Applying this analysis to the central, massive cluster in our simulated domain suggests that this geometry of accretion can vary importantly during the accretion history of the cluster, but we have not found any strong relation with merger events in this particular object.

This method can be extended to evaluate properties of the accreted gas, such as its temperature and its entropy. In particular, we have suggested a way to define these properties for each multipolar component. The most direct application of this procedure is assessing the difference between the gas accreted isotropically and the gas accreted through filaments. Our results point out that the gas being accreted through the cosmic filaments has systematically lower entropy than the isotropic component.

The analyses in this paper would importantly benefit from larger samples of clusters, in order to enhance the statistical significance of our results and, especially, to pursue further analysis, like relating the accretion phenomena studied in this work to the position of the cluster with respect to the SZ and X-ray mass-observable relations (see, e.g., \citealp{Yu_2015, Chen_2019} for recent studies in this line). Nevertheless, despite this lack of statistics, we have proposed some novel analyses which may offer a new insight into the study of accretion of gas in numerical simulations of cluster formation. With the new generation of X-ray and SZ instruments being able to systematically observe the outskirts of a larger number of galaxy clusters in the next decade, more light will be shed onto the physics happening in these regions.

\section*{Acknowledgements}
{We thank the anonymous referee for his/her constructive comments. This work has been supported by the Spanish Ministerio de Ciencia e Innovaci\'on (MICINN, grants AYA2016-77237-C3-3-P and PID2019-107427GB-C33) and by the Generalitat Valenciana (grant PROMETEO/2019/071). DVP acknowledges support from the Departament d'Astronomia i Astrof\'{i}sica through an \textit{Iniciaci\'{o} a la Investigaci\'{o}} scholarship. Simulations have been carried out using the supercomputer Llu\'is Vives at the Servei d'Inform\`atica of the Universitat de Val\`encia.}. This research has made use of the following open-source packages: \texttt{NumPy} \citep{Numpy}, \texttt{SciPy} \citep{Scipy} and \texttt{matplotlib} \citep{Matplotlib}. 

\section*{Data availability}
The data underlying this article will be shared on reasonable request to the corresponding author.

\bibliographystyle{mnras}
\bibliography{mnras_accretion} 

\begin{thebibliography}{}
\makeatletter
\relax
\def\mn@urlcharsother{\let\do\@makeother \do\$\do\&\do\#\do\^\do\_\do\%\do\~}
\def\mn@doi{\begingroup\mn@urlcharsother \@ifnextchar [ {\mn@doi@}
  {\mn@doi@[]}}
\def\mn@doi@[#1]#2{\def\@tempa{#1}\ifx\@tempa\@empty \href
  {http://dx.doi.org/#2} {doi:#2}\else \href {http://dx.doi.org/#2} {#1}\fi
  \endgroup}
\def\mn@eprint#1#2{\mn@eprint@#1:#2::\@nil}
\def\mn@eprint@arXiv#1{\href {http://arxiv.org/abs/#1} {{\tt arXiv:#1}}}
\def\mn@eprint@dblp#1{\href {http://dblp.uni-trier.de/rec/bibtex/#1.xml}
  {dblp:#1}}
\def\mn@eprint@#1:#2:#3:#4\@nil{\def\@tempa {#1}\def\@tempb {#2}\def\@tempc
  {#3}\ifx \@tempc \@empty \let \@tempc \@tempb \let \@tempb \@tempa \fi \ifx
  \@tempb \@empty \def\@tempb {arXiv}\fi \@ifundefined
  {mn@eprint@\@tempb}{\@tempb:\@tempc}{\expandafter \expandafter \csname
  mn@eprint@\@tempb\endcsname \expandafter{\@tempc}}}

\bibitem[\protect\citeauthoryear{{Adam} et~al.,}{{Adam}
  et~al.}{2017}]{Adam_2017}
{Adam} R.,  et~al., 2017, \mn@doi [\aap] {10.1051/0004-6361/201629182}, \href
  {https://ui.adsabs.harvard.edu/abs/2017A&A...598A.115A} {598, A115}

\bibitem[\protect\citeauthoryear{{Adhikari}, {Dalal}  \&
  {Chamberlain}}{{Adhikari} et~al.}{2014}]{Adhikari_2014}
{Adhikari} S.,  {Dalal} N.,   {Chamberlain} R.~T.,  2014, \mn@doi [\jcap]
  {10.1088/1475-7516/2014/11/019}, \href
  {https://ui.adsabs.harvard.edu/abs/2014JCAP...11..019A} {11, 19}

\bibitem[\protect\citeauthoryear{{Allen}, {Evrard}  \& {Mantz}}{{Allen}
  et~al.}{2011}]{Allen_2011}
{Allen} S.~W.,  {Evrard} A.~E.,   {Mantz} A.~B.,  2011, \mn@doi [\araa]
  {10.1146/annurev-astro-081710-102514}, \href
  {https://ui.adsabs.harvard.edu/abs/2011ARA&A..49..409A} {49, 409}

\bibitem[\protect\citeauthoryear{Arfken, Weber  \& Harris}{Arfken
  et~al.}{2013}]{Arfken_2013}
Arfken G.,  Weber H.,   Harris F.,  2013, Mathematical Methods for Physicists:
  A Comprehensive Guide.
Elsevier Science

\bibitem[\protect\citeauthoryear{{Biffi} et~al.,}{{Biffi}
  et~al.}{2016}]{Biffi_2016}
{Biffi} V.,  et~al., 2016, \mn@doi [\apj] {10.3847/0004-637X/827/2/112}, \href
  {https://ui.adsabs.harvard.edu/abs/2016ApJ...827..112B} {827, 112}

\bibitem[\protect\citeauthoryear{{Bryan} \& {Norman}}{{Bryan} \&
  {Norman}}{1998}]{Bryan_1998}
{Bryan} G.~L.,  {Norman} M.~L.,  1998, \mn@doi [\apj] {10.1086/305262}, \href
  {https://ui.adsabs.harvard.edu/abs/1998ApJ...495...80B} {495, 80}

\bibitem[\protect\citeauthoryear{{Cen}, {Roxana Pop}  \& {Bahcall}}{{Cen}
  et~al.}{2014}]{Cen_2014}
{Cen} R.,  {Roxana Pop} A.,   {Bahcall} N.~A.,  2014, \mn@doi [Proceedings of
  the National Academy of Science] {10.1073/pnas.1407300111}, \href
  {https://ui.adsabs.harvard.edu/abs/2014PNAS..111.7914C} {111, 7914}

\bibitem[\protect\citeauthoryear{{Chen}, {Avestruz}, {Kravtsov}, {Lau}  \&
  {Nagai}}{{Chen} et~al.}{2019}]{Chen_2019}
{Chen} H.,  {Avestruz} C.,  {Kravtsov} A.~V.,  {Lau} E.~T.,   {Nagai} D.,
  2019, \mn@doi [\mnras] {10.1093/mnras/stz2776}, \href
  {https://ui.adsabs.harvard.edu/abs/2019MNRAS.490.2380C} {490, 2380}

\bibitem[\protect\citeauthoryear{{Chen}, {Mo}, {Li}, {Wang}, {Yang}, {Zhang}
  \& {Wang}}{{Chen} et~al.}{2020}]{Chen_2020}
{Chen} Y.,  {Mo} H.~J.,  {Li} C.,  {Wang} H.,  {Yang} X.,  {Zhang} Y.,   {Wang}
  K.,  2020, arXiv e-prints, \href
  {https://ui.adsabs.harvard.edu/abs/2020arXiv200305137C} {p. arXiv:2003.05137}

\bibitem[\protect\citeauthoryear{{Cole} \& {Lacey}}{{Cole} \&
  {Lacey}}{1996}]{Cole_1996}
{Cole} S.,  {Lacey} C.,  1996, \mn@doi [\mnras] {10.1093/mnras/281.2.716},
  \href {https://ui.adsabs.harvard.edu/abs/1996MNRAS.281..716C} {281, 716}

\bibitem[\protect\citeauthoryear{{Cui} et~al.,}{{Cui} et~al.}{2016}]{Cui_2016}
{Cui} W.,  et~al., 2016, \mn@doi [\mnras] {10.1093/mnras/stv2839}, \href
  {https://ui.adsabs.harvard.edu/abs/2016MNRAS.456.2566C} {456, 2566}

\bibitem[\protect\citeauthoryear{{Diemer} \& {Kravtsov}}{{Diemer} \&
  {Kravtsov}}{2014}]{Diemer_2014}
{Diemer} B.,  {Kravtsov} A.~V.,  2014, \mn@doi [\apj]
  {10.1088/0004-637X/789/1/1}, \href
  {https://ui.adsabs.harvard.edu/abs/2014ApJ...789....1D} {789, 1}

\bibitem[\protect\citeauthoryear{{Diemer}, {Mansfield}, {Kravtsov}  \&
  {More}}{{Diemer} et~al.}{2017}]{Diemer_2017}
{Diemer} B.,  {Mansfield} P.,  {Kravtsov} A.~V.,   {More} S.,  2017, \mn@doi
  [\apj] {10.3847/1538-4357/aa79ab}, \href
  {https://ui.adsabs.harvard.edu/abs/2017ApJ...843..140D} {843, 140}

\bibitem[\protect\citeauthoryear{{Eisenstein} \& {Hu}}{{Eisenstein} \&
  {Hu}}{1998}]{Eisenstein_1998}
{Eisenstein} D.~J.,  {Hu} W.,  1998, \mn@doi [\apj] {10.1086/305424}, \href
  {http://ukads.nottingham.ac.uk/abs/1998ApJ...496..605E} {496, 605}

\bibitem[\protect\citeauthoryear{{Forero-Romero}, {Gottl{\"o}ber}  \&
  {Yepes}}{{Forero-Romero} et~al.}{2010}]{Forero-Romero_2010}
{Forero-Romero} J.~E.,  {Gottl{\"o}ber} S.,   {Yepes} G.,  2010, \mn@doi [\apj]
  {10.1088/0004-637X/725/1/598}, \href
  {https://ui.adsabs.harvard.edu/abs/2010ApJ...725..598F} {725, 598}

\bibitem[\protect\citeauthoryear{{Haardt} \& {Madau}}{{Haardt} \&
  {Madau}}{1996}]{Haardt_1996}
{Haardt} F.,  {Madau} P.,  1996, \mn@doi [\apj] {10.1086/177035}, \href
  {https://ui.adsabs.harvard.edu/abs/1996ApJ...461...20H} {461, 20}

\bibitem[\protect\citeauthoryear{{Hoffman} \& {Ribak}}{{Hoffman} \&
  {Ribak}}{1991}]{Hoffman_1991}
{Hoffman} Y.,  {Ribak} E.,  1991, \mn@doi [\apjl] {10.1086/186160}, \href
  {http://ukads.nottingham.ac.uk/abs/1991ApJ...380L...5H} {380, L5}

\bibitem[\protect\citeauthoryear{Hunter}{Hunter}{2007}]{Matplotlib}
Hunter J.~D.,  2007, \mn@doi [Computing in Science \& Engineering]
  {10.1109/MCSE.2007.55}, 9, 90

\bibitem[\protect\citeauthoryear{{Katz}, {Weinberg}  \& {Hernquist}}{{Katz}
  et~al.}{1996}]{Katz_1996}
{Katz} N.,  {Weinberg} D.~H.,   {Hernquist} L.,  1996, \mn@doi [\apjs]
  {10.1086/192305}, \href
  {https://ui.adsabs.harvard.edu/abs/1996ApJS..105...19K} {105, 19}

\bibitem[\protect\citeauthoryear{{Knebe}, {Libeskind}, {Knollmann}, {Yepes},
  {Gottl{\"o}ber}  \& {Hoffman}}{{Knebe} et~al.}{2010}]{Knebe_2010}
{Knebe} A.,  {Libeskind} N.~I.,  {Knollmann} S.~R.,  {Yepes} G.,
  {Gottl{\"o}ber} S.,   {Hoffman} Y.,  2010, \mn@doi [\mnras]
  {10.1111/j.1365-2966.2010.16514.x}, \href
  {https://ui.adsabs.harvard.edu/abs/2010MNRAS.405.1119K} {405, 1119}

\bibitem[\protect\citeauthoryear{{Knebe} et~al.,}{{Knebe}
  et~al.}{2011}]{Knebe_2011}
{Knebe} A.,  et~al., 2011, \mn@doi [\mnras] {10.1111/j.1365-2966.2011.18858.x},
  \href {https://ui.adsabs.harvard.edu/abs/2011MNRAS.415.2293K} {415, 2293}

\bibitem[\protect\citeauthoryear{{Kravtsov} \& {Borgani}}{{Kravtsov} \&
  {Borgani}}{2012}]{Kravtsov_2012}
{Kravtsov} A.~V.,  {Borgani} S.,  2012, \mn@doi [\araa]
  {10.1146/annurev-astro-081811-125502}, \href
  {https://ui.adsabs.harvard.edu/abs/2012ARA&A..50..353K} {50, 353}

\bibitem[\protect\citeauthoryear{{Lacey} \& {Cole}}{{Lacey} \&
  {Cole}}{1994}]{Lacey_1994}
{Lacey} C.,  {Cole} S.,  1994, \mn@doi [\mnras] {10.1093/mnras/271.3.676},
  \href {https://ui.adsabs.harvard.edu/abs/1994MNRAS.271..676L} {271, 676}

\bibitem[\protect\citeauthoryear{{Lau}, {Nagai}, {Avestruz}, {Nelson}  \&
  {Vikhlinin}}{{Lau} et~al.}{2015}]{Lau_2015}
{Lau} E.~T.,  {Nagai} D.,  {Avestruz} C.,  {Nelson} K.,   {Vikhlinin} A.,
  2015, \mn@doi [\apj] {10.1088/0004-637X/806/1/68}, \href
  {https://ui.adsabs.harvard.edu/abs/2015ApJ...806...68L} {806, 68}

\bibitem[\protect\citeauthoryear{{Lau}, {Hearin}, {Nagai}  \&
  {Cappelluti}}{{Lau} et~al.}{2020}]{Lau_2020}
{Lau} E.~T.,  {Hearin} A.~P.,  {Nagai} D.,   {Cappelluti} N.,  2020, arXiv
  e-prints, \href {https://ui.adsabs.harvard.edu/abs/2020arXiv200609420L} {p.
  arXiv:2006.09420}

\bibitem[\protect\citeauthoryear{{Lee} \& {Evrard}}{{Lee} \&
  {Evrard}}{2007}]{Lee_2007}
{Lee} J.,  {Evrard} A.~E.,  2007, \mn@doi [\apj] {10.1086/511003}, \href
  {https://ui.adsabs.harvard.edu/abs/2007ApJ...657...30L} {657, 30}

\bibitem[\protect\citeauthoryear{{Lee}, {Springel}, {Pen}  \& {Lemson}}{{Lee}
  et~al.}{2008}]{Lee_2008}
{Lee} J.,  {Springel} V.,  {Pen} U.-L.,   {Lemson} G.,  2008, \mn@doi [\mnras]
  {10.1111/j.1365-2966.2008.13624.x}, \href
  {https://ui.adsabs.harvard.edu/abs/2008MNRAS.389.1266L} {389, 1266}

\bibitem[\protect\citeauthoryear{{Mansfield}, {Kravtsov}  \&
  {Diemer}}{{Mansfield} et~al.}{2017}]{Mansfield_2017}
{Mansfield} P.,  {Kravtsov} A.~V.,   {Diemer} B.,  2017, \mn@doi [\apj]
  {10.3847/1538-4357/aa7047}, \href
  {https://ui.adsabs.harvard.edu/abs/2017ApJ...841...34M} {841, 34}

\bibitem[\protect\citeauthoryear{{Markevitch}, {Vikhlinin}  \&
  {Mazzotta}}{{Markevitch} et~al.}{2001}]{Markevitch_2001}
{Markevitch} M.,  {Vikhlinin} A.,   {Mazzotta} P.,  2001, \mn@doi [\apjl]
  {10.1086/337973}, \href
  {https://ui.adsabs.harvard.edu/abs/2001ApJ...562L.153M} {562, L153}

\bibitem[\protect\citeauthoryear{{More}, {Diemer}  \& {Kravtsov}}{{More}
  et~al.}{2015}]{More_2015}
{More} S.,  {Diemer} B.,   {Kravtsov} A.~V.,  2015, \mn@doi [\apj]
  {10.1088/0004-637X/810/1/36}, \href
  {https://ui.adsabs.harvard.edu/abs/2015ApJ...810...36M} {810, 36}

\bibitem[\protect\citeauthoryear{{Nagai}, {Vikhlinin}  \& {Kravtsov}}{{Nagai}
  et~al.}{2007}]{Nagai_2007}
{Nagai} D.,  {Vikhlinin} A.,   {Kravtsov} A.~V.,  2007, \mn@doi [\apj]
  {10.1086/509868}, \href
  {https://ui.adsabs.harvard.edu/abs/2007ApJ...655...98N} {655, 98}

\bibitem[\protect\citeauthoryear{{Navarro}, {Frenk}  \& {White}}{{Navarro}
  et~al.}{1997}]{NFW_1997}
{Navarro} J.~F.,  {Frenk} C.~S.,   {White} S. D.~M.,  1997, \mn@doi [\apj]
  {10.1086/304888}, \href
  {https://ui.adsabs.harvard.edu/abs/1997ApJ...490..493N} {490, 493}

\bibitem[\protect\citeauthoryear{Oliphant}{Oliphant}{2006}]{Numpy}
Oliphant T.~E.,  2006, A guide to NumPy.
 Vol. 1, Trelgol Publishing USA

\bibitem[\protect\citeauthoryear{{Planck Collaboration} et~al.,}{{Planck
  Collaboration} et~al.}{2018}]{Planck_2018}
{Planck Collaboration} et~al., 2018, arXiv e-prints, \href
  {https://ui.adsabs.harvard.edu/abs/2018arXiv180706209P} {p. arXiv:1807.06209}

\bibitem[\protect\citeauthoryear{{Planelles} \& {Quilis}}{{Planelles} \&
  {Quilis}}{2009}]{Planelles_2009}
{Planelles} S.,  {Quilis} V.,  2009, \mn@doi [\mnras]
  {10.1111/j.1365-2966.2009.15290.x}, \href
  {https://ui.adsabs.harvard.edu/abs/2009MNRAS.399..410P} {399, 410}

\bibitem[\protect\citeauthoryear{{Planelles} \& {Quilis}}{{Planelles} \&
  {Quilis}}{2010}]{Planelles_2010}
{Planelles} S.,  {Quilis} V.,  2010, \mn@doi [\aap]
  {10.1051/0004-6361/201014214}, \href
  {https://ui.adsabs.harvard.edu/abs/2010A&A...519A..94P} {519, A94}

\bibitem[\protect\citeauthoryear{{Planelles}, {Borgani}, {Dolag}, {Ettori},
  {Fabjan}, {Murante}  \& {Tornatore}}{{Planelles}
  et~al.}{2013}]{Planelles_2013}
{Planelles} S.,  {Borgani} S.,  {Dolag} K.,  {Ettori} S.,  {Fabjan} D.,
  {Murante} G.,   {Tornatore} L.,  2013, \mn@doi [\mnras]
  {10.1093/mnras/stt265}, \href
  {https://ui.adsabs.harvard.edu/abs/2013MNRAS.431.1487P} {431, 1487}

\bibitem[\protect\citeauthoryear{{Planelles}, {Borgani}, {Fabjan}, {Killedar},
  {Murante}, {Granato}, {Ragone-Figueroa}  \& {Dolag}}{{Planelles}
  et~al.}{2014}]{Planelles_2014}
{Planelles} S.,  {Borgani} S.,  {Fabjan} D.,  {Killedar} M.,  {Murante} G.,
  {Granato} G.~L.,  {Ragone-Figueroa} C.,   {Dolag} K.,  2014, \mn@doi [\mnras]
  {10.1093/mnras/stt2141}, \href
  {https://ui.adsabs.harvard.edu/abs/2014MNRAS.438..195P} {438, 195}

\bibitem[\protect\citeauthoryear{{Planelles}, {Schleicher}  \&
  {Bykov}}{{Planelles} et~al.}{2015}]{Planelles_2015}
{Planelles} S.,  {Schleicher} D.~R.~G.,   {Bykov} A.~M.,  2015, \mn@doi [\ssr]
  {10.1007/s11214-014-0045-7}, \href
  {https://ui.adsabs.harvard.edu/abs/2015SSRv..188...93P} {188, 93}

\bibitem[\protect\citeauthoryear{{Planelles} et~al.,}{{Planelles}
  et~al.}{2017}]{Planelles_2017}
{Planelles} S.,  et~al., 2017, \mn@doi [\mnras] {10.1093/mnras/stx318}, \href
  {https://ui.adsabs.harvard.edu/abs/2017MNRAS.467.3827P} {467, 3827}

\bibitem[\protect\citeauthoryear{{Planelles}, {Mimica}, {Quilis}  \&
  {Cuesta-Mart{\'{\i}}nez}}{{Planelles} et~al.}{2018}]{Planelles_2018}
{Planelles} S.,  {Mimica} P.,  {Quilis} V.,   {Cuesta-Mart{\'{\i}}nez} C.,
  2018, \mn@doi [\mnras] {10.1093/mnras/sty527}, \href
  {http://adsabs.harvard.edu/abs/2018MNRAS.476.4629P} {476, 4629}

\bibitem[\protect\citeauthoryear{{Pratt}, {Arnaud}, {Biviano}, {Eckert},
  {Ettori}, {Nagai}, {Okabe}  \& {Reiprich}}{{Pratt} et~al.}{2019}]{Pratt_2019}
{Pratt} G.~W.,  {Arnaud} M.,  {Biviano} A.,  {Eckert} D.,  {Ettori} S.,
  {Nagai} D.,  {Okabe} N.,   {Reiprich} T.~H.,  2019, \mn@doi [\ssr]
  {10.1007/s11214-019-0591-0}, \href
  {https://ui.adsabs.harvard.edu/abs/2019SSRv..215...25P} {215, 25}

\bibitem[\protect\citeauthoryear{{Press} \& {Teukolsky}}{{Press} \&
  {Teukolsky}}{1990}]{Press_1990}
{Press} W.~H.,  {Teukolsky} S.~A.,  1990, \mn@doi [Computers in Physics]
  {10.1063/1.4822961}, \href
  {https://ui.adsabs.harvard.edu/abs/1990ComPh...4..669P} {4, 669}

\bibitem[\protect\citeauthoryear{{Quilis}}{{Quilis}}{2004}]{Quilis04}
{Quilis} V.,  2004, \mn@doi [\mnras] {10.1111/j.1365-2966.2004.08040.x}, \href
  {http://adsabs.harvard.edu/abs/2004MNRAS.352.1426Q} {352, 1426}

\bibitem[\protect\citeauthoryear{{Quilis}, {Moore}  \& {Bower}}{{Quilis}
  et~al.}{2000}]{Quilis_2000}
{Quilis} V.,  {Moore} B.,   {Bower} R.,  2000, \mn@doi [Science]
  {10.1126/science.288.5471.1617}, \href
  {https://ui.adsabs.harvard.edu/abs/2000Sci...288.1617Q} {288, 1617}

\bibitem[\protect\citeauthoryear{{Quilis}, {Planelles}  \&
  {Ricciardelli}}{{Quilis} et~al.}{2017}]{Quilis_2017}
{Quilis} V.,  {Planelles} S.,   {Ricciardelli} E.,  2017, \mn@doi [\mnras]
  {10.1093/mnras/stx770}, \href
  {http://adsabs.harvard.edu/abs/2017MNRAS.469...80Q} {469, 80}

\bibitem[\protect\citeauthoryear{{Rasia} et~al.,}{{Rasia}
  et~al.}{2015}]{Rasia_2015}
{Rasia} E.,  et~al., 2015, \mn@doi [\apjl] {10.1088/2041-8205/813/1/L17}, \href
  {https://ui.adsabs.harvard.edu/abs/2015ApJ...813L..17R} {813, L17}

\bibitem[\protect\citeauthoryear{{Roediger}, {Br{\"u}ggen}, {Simionescu},
  {B{\"o}hringer}, {Churazov}  \& {Forman}}{{Roediger}
  et~al.}{2011}]{Roediger_2011}
{Roediger} E.,  {Br{\"u}ggen} M.,  {Simionescu} A.,  {B{\"o}hringer} H.,
  {Churazov} E.,   {Forman} W.~R.,  2011, \mn@doi [\mnras]
  {10.1111/j.1365-2966.2011.18279.x}, \href
  {https://ui.adsabs.harvard.edu/abs/2011MNRAS.413.2057R} {413, 2057}

\bibitem[\protect\citeauthoryear{{Sanders} et~al.,}{{Sanders}
  et~al.}{2020}]{Sanders_2020}
{Sanders} J.~S.,  et~al., 2020, \mn@doi [\aap] {10.1051/0004-6361/201936468},
  \href {https://ui.adsabs.harvard.edu/abs/2020A&A...633A..42S} {633, A42}

\bibitem[\protect\citeauthoryear{{Savitzky} \& {Golay}}{{Savitzky} \&
  {Golay}}{1964}]{Savitzky_1964}
{Savitzky} A.,  {Golay} M.~J.~E.,  1964, Analytical Chemistry, \href
  {https://ui.adsabs.harvard.edu/abs/1964AnaCh..36.1627S} {36, 1627}

\bibitem[\protect\citeauthoryear{{Shi}}{{Shi}}{2016a}]{Shi_2016}
{Shi} X.,  2016a, \mn@doi [\mnras] {10.1093/mnras/stw925}, \href
  {https://ui.adsabs.harvard.edu/abs/2016MNRAS.459.3711S} {459, 3711}

\bibitem[\protect\citeauthoryear{{Shi}}{{Shi}}{2016b}]{Shi_2016_shock}
{Shi} X.,  2016b, \mn@doi [\mnras] {10.1093/mnras/stw1418}, \href
  {https://ui.adsabs.harvard.edu/abs/2016MNRAS.461.1804S} {461, 1804}

\bibitem[\protect\citeauthoryear{{Simionescu} et~al.,}{{Simionescu}
  et~al.}{2019}]{Simionescu_2019}
{Simionescu} A.,  et~al., 2019, \mn@doi [\ssr] {10.1007/s11214-019-0590-1},
  \href {https://ui.adsabs.harvard.edu/abs/2019SSRv..215...24S} {215, 24}

\bibitem[\protect\citeauthoryear{{Springel} \& {Hernquist}}{{Springel} \&
  {Hernquist}}{2003}]{Springel_2003}
{Springel} V.,  {Hernquist} L.,  2003, \mn@doi [\mnras]
  {10.1046/j.1365-8711.2003.06206.x}, \href
  {https://ui.adsabs.harvard.edu/abs/2003MNRAS.339..289S} {339, 289}

\bibitem[\protect\citeauthoryear{{Sutherland} \& {Dopita}}{{Sutherland} \&
  {Dopita}}{1993}]{Sutherland_1993}
{Sutherland} R.~S.,  {Dopita} M.~A.,  1993, \mn@doi [\apjs] {10.1086/191823},
  \href {https://ui.adsabs.harvard.edu/abs/1993ApJS...88..253S} {88, 253}

\bibitem[\protect\citeauthoryear{{Tamura}, {Hayashida}, {Ueda}  \&
  {Nagai}}{{Tamura} et~al.}{2011}]{Tamura_2011}
{Tamura} T.,  {Hayashida} K.,  {Ueda} S.,   {Nagai} M.,  2011, \mn@doi [\pasj]
  {10.1093/pasj/63.sp3.S1009}, \href
  {https://ui.adsabs.harvard.edu/abs/2011PASJ...63S1009T} {63, S1009}

\bibitem[\protect\citeauthoryear{{Theuns}, {Leonard}, {Efstathiou}, {Pearce}
  \& {Thomas}}{{Theuns} et~al.}{1998}]{Theuns_1998}
{Theuns} T.,  {Leonard} A.,  {Efstathiou} G.,  {Pearce} F.~R.,   {Thomas}
  P.~A.,  1998, \mn@doi [\mnras] {10.1046/j.1365-8711.1998.02040.x}, \href
  {https://ui.adsabs.harvard.edu/abs/1998MNRAS.301..478T} {301, 478}

\bibitem[\protect\citeauthoryear{{Tormen}, {Moscardini}  \& {Yoshida}}{{Tormen}
  et~al.}{2004}]{Tormen_2004}
{Tormen} G.,  {Moscardini} L.,   {Yoshida} N.,  2004, \mn@doi [\mnras]
  {10.1111/j.1365-2966.2004.07736.x}, \href
  {https://ui.adsabs.harvard.edu/abs/2004MNRAS.350.1397T} {350, 1397}

\bibitem[\protect\citeauthoryear{{Virtanen} et~al.,}{{Virtanen}
  et~al.}{2020}]{Scipy}
{Virtanen} P.,  et~al., 2020, \mn@doi [Nature Methods]
  {https://doi.org/10.1038/s41592-019-0686-2}, \href {https://rdcu.be/b08Wh}
  {17, 261}

\bibitem[\protect\citeauthoryear{{Voit}}{{Voit}}{2005}]{Voit_2005}
{Voit} G.~M.,  2005, \mn@doi [Reviews of Modern Physics]
  {10.1103/RevModPhys.77.207}, \href
  {https://ui.adsabs.harvard.edu/abs/2005RvMP...77..207V} {77, 207}

\bibitem[\protect\citeauthoryear{{Walker} et~al.,}{{Walker}
  et~al.}{2019}]{Walker_2019}
{Walker} S.,  et~al., 2019, \mn@doi [\ssr] {10.1007/s11214-018-0572-8}, \href
  {https://ui.adsabs.harvard.edu/abs/2019SSRv..215....7W} {215, 7}

\bibitem[\protect\citeauthoryear{{Yepes}, {Kates}, {Khokhlov}  \&
  {Klypin}}{{Yepes} et~al.}{1997}]{Yepes_1997}
{Yepes} G.,  {Kates} R.,  {Khokhlov} A.,   {Klypin} A.,  1997, \mn@doi [\mnras]
  {10.1093/mnras/284.1.235}, \href
  {https://ui.adsabs.harvard.edu/abs/1997MNRAS.284..235Y} {284, 235}

\bibitem[\protect\citeauthoryear{{Yu}, {Nelson}  \& {Nagai}}{{Yu}
  et~al.}{2015}]{Yu_2015}
{Yu} L.,  {Nelson} K.,   {Nagai} D.,  2015, \mn@doi [\apj]
  {10.1088/0004-637X/807/1/12}, \href
  {https://ui.adsabs.harvard.edu/abs/2015ApJ...807...12Y} {807, 12}

\bibitem[\protect\citeauthoryear{{Zemp}, {Gnedin}, {Gnedin}  \&
  {Kravtsov}}{{Zemp} et~al.}{2011}]{Zemp_2011}
{Zemp} M.,  {Gnedin} O.~Y.,  {Gnedin} N.~Y.,   {Kravtsov} A.~V.,  2011, \mn@doi
  [\apjs] {10.1088/0067-0049/197/2/30}, \href
  {https://ui.adsabs.harvard.edu/abs/2011ApJS..197...30Z} {197, 30}

\makeatother
\end{thebibliography}

\appendix

\section{The symmetric logarithmic scale}
\label{s:appendix.symlog}
When representing the mass fluxes in Sec. \ref{s:results_angular.results}, or in many other situations where data can be both positive and negative and, at the same time, span a broad range of orders of magnitude, it may be useful to employ a \textit{symmetric logarithmic scale}. Even though some software packages for data visualisation (e.g., Python's \texttt{matplotlib}, \citealp{Matplotlib}) include implementations of this scale, its usage is not quite widespread and we have considered covering it here for the interested reader.

The basic underlying idea of our particular implementation of the symmetric logarithmic scale relies on mapping any interval $\left[-x_\mathrm{max}, x_\mathrm{max}\right]$ to the interval $\left[-1,1\right]$ by performing the following continuous transformation:

\begin{equation}
	x \mapsto f(x) = 
	\begin{cases}
		\mathrm{sign}(x)\left[1 + \frac{1 - a}{\alpha}\log_{10}\left(\frac{x}{x_\mathrm{max}}\right)\right], & \frac{|x|}{x_\mathrm{max}} \geq 10^{-\alpha} \\
		\frac{a}{10^{-\alpha} x_\mathrm{max}}x, & \frac{|x|}{x_\mathrm{max}} \leq 10^{-\alpha}
	\end{cases}
\end{equation}

The parameters have been chosen in this way to provide the clearest interpretation. $x_\mathrm{max}$ controls the maximum of the scale, i.e., the value that will be mapped to 1. $\alpha$ represents the dynamical range of the representation (the number of orders of magnitude represented in logarithmic scale). Absolute values above $10^{-\alpha}$ are treated logarithmically; below this threshold, linearly. Last, $a = y\left(10^{-\alpha}x_\mathrm{max}\right)$ represents the visual extent of the linear scale. In order for the transformation to be differentiable, $a$ shall be set to $a = \left( 1 + \alpha \ln 10 \right)^{-1}$. However, in many cases differentiability is not necessary, and $a$ can be chosen to better suite the representation purposes. In the plots shown in this work we have used $\alpha = 3.5$, $a = \left( 1 + \alpha \ln 10 \right)^{-1}$.

\section{The real spherical harmonics basis}
\label{s:appendix.realspharm}

In Sec. \ref{s:results_angular.results}, we have used the real spherical harmonics basis to study the angular dependence of the mass accretion fluxes. As there is not a general consensus regarding phase and normalisation conventions, let us formally define these functions as they have been used in this work.

Let $Y_{lm}(\theta, \phi)$ be the complex spherical harmonic of degree $l$ and order $m$, with unit square-integral and using the Condon-Shortley phase convention (see, e.g., \citealp{Arfken_2013}). These functions can be used to expand any square-integrable complex function defined on the unit sphere. However, for real functions, only a half of the coefficients of such expansion are free. Thus, it can be simplified by defining the real spherical harmonics, $\mathcal{Y}_{lm}(\theta, \phi)$, as:

\begin{equation}
	\mathcal{Y}_{lm}(\theta, \phi) = \begin{cases}
		(-1)^m \sqrt{2} \; \mathrm{Im} \, \left[Y_{l,-m} (\theta, \phi)\right], & m < 0 \\
		Y_{l0} (\theta, \phi), & m = 0\\
		(-1)^m \sqrt{2} \; \mathrm{Re} \, \left[Y_{l,m} (\theta, \phi)\right], & m > 0
	\end{cases}
	\label{eq:real_spharm}
\end{equation}

From the orthonormality properties of the $Y_{lm}$ functions, it is easy to show that any real, square-integrable function defined on the unit sphere, $f(\theta, \phi)$, can be expanded in a series of real spherical harmonics with real coefficients, as 

\begin{equation}
	f(\theta, \phi) = \sum_{l=0}^\infty \sum_{m=-l}^{l} c_{lm} \mathcal{Y}_{lm}
\end{equation}

\noindent with $c_\mathrm{lm} = \oiint \dd \Omega \mathcal{Y}_\mathrm{lm}(\theta, \phi) f(\theta, \phi)$.

\label{lastpage}

\end{document}